\documentclass{article}

\PassOptionsToPackage{numbers, compress}{natbib}
\usepackage{booktabs}
\usepackage{tabularx}  % 需要在导言区添加

\usepackage[final]{neurips_2025}
\usepackage[utf8]{inputenc} % allow utf-8 input
\usepackage[T1]{fontenc}    % use 8-bit T1 fonts
\usepackage{url}            % simple URL typesetting
\usepackage{booktabs}       % professional-quality tables
\usepackage{amssymb}
\usepackage{amsfonts}       % blackboard math symbols
\usepackage{mathrsfs}       % for \mathscr
\usepackage{multirow} % for multirow and multicolumn in tables
\usepackage{etoc}           % for enhanced table of contents
\usepackage{nicefrac}       % compact symbols for 1/2, etc.
\usepackage{microtype, wrapfig}

\usepackage{lipsum} % just for dummy text

\usepackage{amsmath,amsthm}
\usepackage{graphicx}
\numberwithin{equation}{section}
\newtheorem{theorem}{Theorem}[section]

\newtheorem{lemma}[theorem]{Lemma}

\newtheorem{remark}[theorem]{Remark}
\newtheorem{condition}[theorem]{Condition}
\newtheorem{definition}[theorem]{Definition}
\newtheorem{example}[theorem]{Example}

\usepackage{caption}   % 提供 \captionof 命令
\usepackage{float}  % 放在导言区
\usepackage{bm}
\usepackage{graphicx} 
\usepackage{xcolor} 
\usepackage{float}
\usepackage{tikz}
\usetikzlibrary{arrows.meta, positioning, shapes.geometric}
\usepackage{multirow} 
\usepackage{comment}

\usepackage{hyperref}
\hypersetup{
    colorlinks=true,  
    linkcolor=blue,      
    citecolor=blue,        
    urlcolor=blue          
}
\usepackage{enumitem} % For customized enumerate
\newcommand{\Pa}{\text{Pa}} % Parent set
\newcommand{\Ch}{\text{Ch}} % Children set
\newcommand{\An}{\text{An}} % Ancestors
\newcommand{\De}{\text{De}} % Descendants
\newcommand{\ND}{\text{ND}} % Non-descendants
\newcommand{\Z}{\mathbf{Z}}
\newcommand{\M}{\mathbf{M}}
\newcommand{\V}{\mathbf{V}}
\newcommand{\E}{\mathbb{E}}
\newcommand{\X}{\mathbf{X}}
\newcommand{\C}{\mathbf{C}}
\newcommand{\W}{\mathbf{W}}
\renewcommand{\c}{\mathbf{c}}
\newcommand{\w}{\mathbf{w}}

\renewcommand{\b}{\mathbf{b}}
\newcommand{\g}{\mathbf{g}}
\newcommand{\G}{\mathbf{G}}
\newcommand{\B}{\mathbf{B}}
\newcommand{\z}{\mathbf{z}}
\newcommand{\WCDE}{\text{WCDE}}
\newcommand{\EE}{\mathbb{E}}

\newcommand{\Q}{\mathbf{Q}}
\newcommand{\Y}{\mathbf{Y}}
\newcommand{\bigsetminus}{\mathrel{\scalebox{1}{$\setminus$}}}
\title{Optimal Adjustment Sets for Nonparametric Estimation of Weighted Controlled Direct Effect}
\author{Ruiyang Lin \\
  The University of Science and Technology of China\\
  \texttt{lruiyang@wustl.edu} \\
  \And
  Yongyi Guo \\
  University of Wisconsin–Madison\\
  \texttt{guo98@wisc.edu } \\
  \AND
  Kyra Gan \\
  Cornell Tech, Cornell University\\
\texttt{kyragan@cornell.edu} \\
}

\begin{document}
\etocdepthtag.toc{mtmainpaper}

\maketitle
\begin{abstract}
The \emph{weighted controlled direct effect} (WCDE) generalizes the standard \emph{controlled direct effect} (CDE) 
by averaging over the mediator distribution, providing a robust estimate when treatment effects vary across mediator levels. This makes the WCDE especially relevant in fairness analysis, where it isolates the direct effect of an exposure on an outcome, independent of mediating pathways.
% by accounting for heterogeneity in causal effects across mediator values.
% Unlike the $\mathrm{CDE}$—which evaluates effects at fixed mediator levels—the $\WCDE$ averages over the mediator distribution, providing a more robust estimate when treatment effects vary across subpopulations. This parameter is particularly relevant in fairness analysis, where it quantifies the direct impact of an exposure on an outcome, independent of mediating pathways.
This work establishes three fundamental advances for WCDE in observational studies: First, we
% In this work, we focus on observational studies and
establish necessary and sufficient conditions for the identifiability of the WCDE, clarifying when it diverges from the CDE. 
Next, we consider nonparametric estimation of the WCDE and derive its influence function, focusing on the class of {regular and asymptotically linear} estimators.
%Next, we consider nonparametric estimation and derive the 
% efficient
%influence function for the $\WCDE$ and consider the class of \emph{regular and asymptotically linear} estimators. 
Lastly, we
characterize the optimal covariate adjustment set that minimizes the asymptotic variance, demonstrating how mediator-confounder interactions introduce distinct requirements compared to \emph{average treatment effect} (ATE) estimation. 
% highlighting differences from standard ATE settings due to mediator-confounder interactions.
Using synthetic and real-world data, we validate our theory numerically, showing that the proposed optimal valid adjustment set yields the lowest variance at practical sample sizes.
% Through simulation studies on synthetic data, we compare variances of asymptotically linear estimators under different adjustment sets.
Our results offer a principled framework for efficient estimation of direct effects in complex causal systems, with practical applications in fairness and mediation analysis.

% Next, we derive the efficient influence function for the $\WCDE$ and characterize the optimal adjustment set—defined as the covariate set that minimizes asymptotic variance for nonparametric asymptotically linear estimators. Notably, this set differs from standard average treatment effect (ATE) settings due to mediator-confounder interactions. To validate our theoretical results, we conduct simulation studies comparing the variance of a few asymptotically linear estimates of the $\WCDE$ under different adjustment sets, using synthetically generated data. Our findings provide a rigorous framework for efficient estimation of direct effects in complex causal systems, with immediate applications in fairness assessment and mediation analysis.

\end{abstract}

% \vspace{-5pt}
\section{Introduction}\label{sec:intro}
% \vspace{-5pt}
% \kyra{mention observational study in abstract and intro}
% \kyra{@yongyi, for the title should we use semiparametric or nonparametric? nonparametric would mean that we don't restrict the direction of perturbation in semiparametric model}
The \emph{controlled direct effect}  quantifies the effect of an exposure when intervening to set the mediator to a fixed level, which may differ from its natural observed value \citep{pearl2022direct, robins2003semantics, robins1992identifiability}.
While CDE has traditionally been used to analyze mediation effects, in this work, we employ the \emph{weighted controlled direct effect} to detect the presence of a direct effect between treatment A and outcome Y given observed variables \citep{maasch2025local}. Specifically, we intervene on \emph{all} observed mediators rather than an arbitrary subset. This approach is motivated by both pragmatic and theoretical considerations: e.g., from a fairness perspective, we seek to determine whether direct discrimination exists \cite{maasch2025local}, while methodologically, 1) the identifiability of CDE becomes unnecessarily complex when some mediators are non-fixed \citep{vanderweele2011controlled}, 2) CDE values can vary unreliably across mediator levels 
\citep{tchetgen2012semiparametric, vansteelandt2017interventional}, and 3) CDE estimates frequently vary across subpopulations defined by baseline covariates~\citep{jackson2018decomposition}, complicating generalization.
% \citep{vanderweele2015explanation}.

Unlike \emph{natural direct effects} (NDE) that require untestable \emph{cross-world independence assumptions} (i.e., the potential outcome under one treatment level is independent of the potential mediator under another treatment level)~\citep{andrews2021insights, avin2005identifiability, robins2010alternative}, the CDE, and by extension WCDE, relies only on experimentally verifiable interventions 
% i.e., it involves setting the mediator to specific values through intervention)
(where the mediator is set to specific values through intervention)--a crucial advantage noted by \cite{petersen2006estimation} and \cite{vanderweele2015explanation}.
% This fundamental causal parameter 
The WCDE averages CDEs over the mediator distribution, yielding a robust effect measure that automatically accounts for mediation interaction while maintaining the CDE's avoidance of cross-world assumptions. This
% The $\mathrm{CDE}$ 
provides critical insights into direct pathways across diverse domains, from evaluating medical treatments in epidemiology~\citep{goetgeluk2008estimation, vanderweele2015explanation} to assessing  fairness~\citep{maasch2025local, zhang2018fairness} and informing policy decisions~\citep{imai2010identification}.

% Despite its popularity, the $\mathrm{CDE}$ suffers from several practical limitations. First, it conditions on a fixed level of the mediator—often chosen arbitrarily—which can substantially influence conclusions and fail to capture treatment effects that vary with the mediator~\citep{vansteelandt2017interventional, tchetgen2012semiparametric}. Second, $\mathrm{CDE}$ estimates frequently vary across subpopulations defined by baseline covariates~\citep{jackson2018decomposition}, complicating generalization.
% These limitations motivate the need for the \emph{weighted controlled direct effect}~\citep{pearl2000causality}, which averages $\mathrm{CDE}$s over the mediator distribution,
% yielding a robust effect measure that automatically accounts for mediation interaction
% % to provide a more robust, population-level summary that naturally accounts for effect heterogeneity 
% while maintaining $\mathrm{CDE}$'s avoidance of cross-world assumptions.
% The \emph{weighted controlled direct effect} ($\WCDE$) mitigates these issues while retaining the $\mathrm{CDE}$’s interpretability. By averaging $\mathrm{CDE}$s over the mediator distribution, the $\WCDE$ captures treatment effect heterogeneity and yields a more robust, population-level summary.

WCDE raises two key statistical challenges in observational studies that should be resolved for valid inference: (1) What are the necessary and sufficient conditions on the adjustment set such that WCDE is uniquely identifiable from observed data? (2) Among all valid adjustment sets satisfying identifiability, which achieves the asymptotic efficiency bound for WCDE estimation?
While optimal adjustment sets for 
ATEs
% average treatment effects
are well-characterized \citep{henckel2022graphical, pearl2000causality, rotnitzky2020efficient}, 
understanding WCDE introduces unique challenges due to the mediator's potential role as both a collider (between the exposure and the confounder) and an effect modifier (between the confounder and the outcome):
\begin{itemize}[itemsep=0pt, topsep=-3pt, partopsep=0pt, parsep=0pt, leftmargin=*]
    \item  Standard CDE adjustment criteria \citep{pearl2022direct} only ensure identification at fixed mediator levels and fail to address the integration required for WCDE (Example~\ref{example:condition_unqiue_WCDE}).
    % \kyra{reference to the section that you will show an example of non-unique identifiability under $\mathrm{CDE}$ criterion}
    \item ATE adjustment theory ignores the efficiency implications of mediator-confounder dependencies, which critically affect WCDE estimation variance through the weighting scheme. 
    % ~\citep{vanderweele2015explanation}
\end{itemize}
This gap leads to potentially inefficient estimators and makes it unclear whether WCDE provides any theoretical advantages over conventional CDE in practice.
% they prove inadequate for $\WCDE$ estimation due to their failure to account for the complex mediator-confounder interactions inherent in direct effect estimation. This oversight can lead to both inefficient estimation and residual confounding, particularly when the mediator-confounder relationship modifies treatment effects~\citep{vanderweele2015explanation}.

\textbf{Contributions\;\;} We study the identification and nonparametric estimation of WCDE given a known \emph{directed acyclic graph} (DAG).
% encoding the underlying causal structure. 
Our contributions are \emph{fourfold}. First, we establish necessary and sufficient conditions for 
\emph{valid adjustment sets} (VASs) that guarantee WCDE's identifiability (Lemma~\ref{lemma:adjustment_criterion}), a prerequisite for meaningful estimation.
Second, we derive the
% efficient
\emph{influence function} (IF) for WCDE (Theorem~\ref{thm:IF}). 
Third, we prove that its optimal adjustment set--defined as the covariate collection that minimizes asymptotic variance among all \emph{regular and asymptotically linear} (RAL) estimates--necessarily differs from ATE adjustment sets due to the required integration over mediator distributions,
and typically includes
% . In particular, it typically includes 
all parents of the outcome variable excluding the treatment (Theorem~\ref{thm:optimal_adjustment}).
%\kyra{might want to modify this to make it more precise: describing the adjustment set is the parent of Y}
These theoretical advances enable more precise direct effect estimation in settings with mediator-confounder dependence.
Lastly, we provide an efficient estimator, \emph{augmented inverse probability weighting} (AIPW), for WCDE, and numerically validate our theoretical results. 
% through numerical studies. 
% , as we demonstrate through both asymptotic analysis and numerical experiments.

\textbf{Outline\;\;}  Section~\ref{sec:WCDE_VAS} introduces 
% graphical model
notation, defines WCDE, and presents criteria for VASs. Section~\ref{sec:IF} 
% introduces
reviews nonparametric estimation and 
% the class of
RAL
% regular and asymptotically linear
estimators, and derives the IF
% influence function
of WCDE.
% in Theorem~\ref{thm:IF}. 
The asymptotic variance of any RAL estimator
% asymptotically linear estimator 
is then determined by the variance of its IF,
% influence function,
which depends explicitly on the choice of the VAS. 
% be characterized by their influence functions as a function of the valid adjustment set.
% characterizing their asymptotic variance as a function of the valid adjustment set.
% % establishes the asymptotic efficiency bound 
% for the $\WCDE$ and derives its influence function in Theorem~\ref{thm:IF}. 
Section~\ref{sec:optimal_VAS} presents 
% our main result on 
the optimal VAS
% adjustment set 
for  WCDE.
% , with a proof sketch provided in Section~\ref{subsec:proof_sketch}. 
% Section~\ref{sec:discussions} concludes with some additional discussion.
Section~\ref{sec: experiments} contains AIPW for WCDE and numerical results. Section~\ref{sec:discussions} discusses connections to the unknown-graph setting.

% \paragraph{\textbf{Key Contributions}} The contribution of the paper is summarized as follows.
% \begin{enumerate}
%     \item We propose adjustment criteria for identifying valid sets for estimating the Weighted Controlled Direct Effect ($\WCDE$).
%     \item We derive the influence function for the $\WCDE$ estimator to facilitate asymptotic variance analysis.
%     \item We develop graphical criteria for selecting the optimal adjustment set and for comparing different valid adjustment sets.
% \end{enumerate}

\textbf{Related Works\;\;}
%\kyra{@ruiyang, think about different lines of work that you want to include here that have not appeared in the intro. make an outline and ping us before you write}
Causal effect estimation from observational data typically relies on covariate adjustment, with prior work examining variable selection through simulations
% Identifying and efficiently estimating causal effects from observational data typically involves covariate adjustment. Variable selection has long been a central focus in this context, with numerous simulation studies investigating its impact 
\citep{brookhart2006variable,lefebvre2008statistical}
and analyzing the minimum asymptotic variance
% , and several theoretical works analyzing the minimum achievable asymptotic variance
\citep{hahn2004functional,rotnitzky1995semiparametric,rotnitzky2010note}. 
Complete and sound graphical criteria for valid adjustment sets identifying total causal effects are well established
% Foundational work has established complete and sound graphical criteria for valid adjustment sets that identify total causal effects 
\citep{pearl2000causality, perkovic2018complete, shpitser2010validity}, with later work optimizing asymptotic efficiency for ATEs under linear \cite{henckel2022graphical} and nonparametric \cite{rotnitzky2020efficient} settings.
% . Building on this foundation, subsequent research has focused on selecting adjustment sets that optimize asymptotic efficiency under both linear and nonparametric models, particularly for average treatment effects (ATE).
% For example, \citet{henckel2022graphical} proposed graphical rules for efficient ATE estimation in linear models, and \citet{rotnitzky2020efficient} extended these results to nonparametric settings.
Recent work has extended to individualized treatment rules and models with hidden variables
% extensions have addressed individualized treatment rules and models with hidden variables 
\citep{smucler2021efficient}; establishing 
necessary and sufficient conditions for globally optimal adjustment sets \citep{runge2021necessary};
and developing a polynomial-time algorithm for finding minimum-cost adjustment sets
\cite{smucler2022note}.
% further incorporated variable-specific costs and proposed a polynomial-time algorithm to identify efficient minimum-cost adjustment sets.

Beyond graphical criteria, data-driven methods for selecting sufficient adjustment sets include: 
% In addition to graphical criteria, numerous data-driven strategies have been proposed to select sufficient adjustment sets.
1) 
% The
\textit{CovSel} 
% algorithm 
\citep{deLuna2011covariate}, which identifies minimal sets via conditional independence 
% and performs well 
under correct model specification,
2) 
\textit{Outcome-adaptive lasso} \citep{shortreed2017outcome}, which selects variables predictive of the outcome to improve efficiency, 3)
\textit{Double selection} \citep{belloni2014inference}, which combines variables selected from both treatment and outcome models to ensure robustness in high dimensions.
Estimator-driven approaches, such as \textit{Change-in-Estimate} \citep{greenland2016causal} and \textit{Focused Confounder Selection} \citep{vansteelandt2012model}, aim to minimize MSE but require parametric assumptions.
These methods offer flexible and scalable alternatives when the causal structure is partially unknown, but their validity often hinges on model assumptions and sample size.

However, existing approaches remain centered on ATE and overlook key challenges posed by the WCDE. The mediator’s dual role as a collider and effect modifier introduces unique obstacles to both identifiability and efficiency. While standard CDE-based criteria \citep{VanderWeeleVansteelandt2009} do not guarantee identifiability of WCDE, ATE-oriented strategies treat mediators as forbidden variables \citep{henckel2022graphical,rotnitzky2020efficient}, precluding their integration as required by WCDE. 
We fill this gap by deriving graphical criteria and asymptotic efficiency results tailored to the mediator-weighted structure of the WCDE.
% To our knowledge, no prior work has offered a complete characterization of valid adjustment sets or efficiency theory specifically tailored to $\WCDE$. We address this gap by developing new graphical criteria and asymptotic efficiency results that directly incorporate the mediator-weighted structure of $\WCDE$. 

% \vspace{-5pt}
\section{Problem Setup and Valid Adjustment Sets for WCDE}\label{sec:WCDE_VAS}
% \vspace{-5pt}
% \kyra{this section should start with graphical model notation and introduce the $\WCDE$ target parameter. With a clear explanation of the notations. Then it should establish the graphical criterion for value adjustment sets}

To formally define valid adjustment sets for the WCDE, we first introduce notation for graphical models. We then present the inferential framework and associated notation required to specify the estimation problem. Lastly, we classify variable types with respect to the exposure and outcome and provide a formal definition of VASs.
% \section{Problem Setup and Efficient Influence Function for WCDE}
% This section gives a brief overview of Weighted Controlled Direct Effect and influence function.

% \paragraph{\textbf{Weighted Controlled Direct Effect}}

% The $\mathrm{CDE}$ can be used to test for direct discrimination, as the true value is non-zero if and only if there is a direct path from the protected attribute to the outcome (Zhang and Bareinboim 2018). 
% \paragraph{Directed Acyclic Graphs}
\textbf{Structural Causal Model (SCM)\;\;} 
An SCM is represented by a DAG, $\mathcal G = (\V,\mathbf{E})$, a graph with directed edges and no directed cycles, where vertices $\V = \{X_1,\ldots,X_d\}$ represent random variables and edges $\mathbf{E}$ encode direct causal relationships. 
% Additionally, 
An SCM is equipped with a probability distribution $P$ over $\V$.
The \textit{parent set} of a vertex $X_j$, denoted $\Pa(X_j)$, consists of all vertices $X_i$ for which $X_i \to X_j \in \mathbf{E}$, representing direct causes. The \textit{children} $\Ch(X_i)$ are vertices $X_j$ with $X_i \to X_j \in \mathbf{E}$, representing direct effects. \textit{Ancestors} $\An(X_j)$ comprise all vertices connected to $X_j$ via \emph{directed paths} 
(Def.~\ref{def:path-types}), while \textit{descendants} $\De(X_i)$ are all vertices reachable from $X_i$ via \emph{directed paths}.

The \textit{d-separation} (Def.~\ref{def:d-separation}) criterion formally characterizes conditional independence relationships implied by the graph structure. 
% For disjoint vertex sets $\X$, $\Y$, and $\Z$, we say $\Z$ \textit{d-separates} $\X$ from $\Y$ if every path between $\X$ and $\Y$ is \textit{blocked} by $\Z$. A path is blocked by $\Z$ if it contains either:
% \begin{enumerate}[label=(\arabic*), itemsep=0pt, topsep=-3pt, partopsep=0pt, parsep=0pt]
%     \item A chain $X_i \to Z_k \to X_j$ or fork $X_i \leftarrow Z_k \to X_j$ with $Z_k \in \Z$, or
%     \item A collider $X_i \to Z_k \leftarrow X_j$ where $Z_k \notin Z$ and none of its descendants are in $\Z$, i.e., $\De(Z_k) \cap \Z = \emptyset$.
% \end{enumerate}
Under the assumption of \emph{causal sufficiency} (no unmeasured confounding), d-separation perfectly captures the conditional independencies in the joint distribution through the \textit{Markov property}: if $\Z$ d-separates $\X$ from $\Y$ in $\mathcal{G}$ (i.e., $\X \perp\!\!\!\perp_{\mathcal{G}} \Y \mid \Z$), then $\X \perp\!\!\!\perp \Y \mid \Z$ in
all distributions that are Markov with respect to 
$\mathcal G$.
% compatible distributions.
This Markov property implies two key consequences. First, the joint distribution factorizes as:
$P(X_1,\ldots,X_d) = \prod_{j=1}^d P(X_j \mid \Pa(X_j))$
where each component represents the conditional distribution of a variable given its direct causes. Second, it yields the \textit{local Markov property} that each variable is conditionally independent of its non-descendant non-parents given its parents:
\vspace{-5pt}
$$
X_j \perp\!\!\!\perp \ND(X_j) \bigsetminus \Pa(X_j) \mid \Pa(X_j),
$$
where $\ND(X_j)$ denotes the non-descendants of $X_j$. These properties form the foundation for deriving identifiability conditions for causal effects in the presence of mediator-confounder relationships.
% We present the formal definitions of directed, backdoor, and mediator paths with the standard counterfactual independence assumptions in Appendices~\ref{def:path-types} and~\ref{def:cde-id}.

\textbf{Weighted Controlled Direct Effect\;\;} Consider a DAG 
$\mathcal{G}=(\V,\mathbf{E})$ representing a structural causal model. 
We focus on estimating WCDE of a binary treatment variable $A\in \V$ with values $\{a,a^*\}$ (treatment vs. control) on an outcome 
$Y\in \V$. 
% The WCDE captures the controlled direct effect while accounting for mediator-confounder interactions through simple weighting.
% \Lry{I think this sentence is unnecessary — the phrase “accounting for mediator–confounder interactions through simple weighting” is unclear, maybe it can just be removed. The explanation appears later anyway.}

% Formally, 
% % a mediator $M \in \V$ of $A$ and $Y$ is defined through ancestral relations as $M \in \De(A) \cap \An(Y) \setminus \{A,Y\}$. Let $\M$ be a set of mediators.
% the set of all mediators $\M\subset \V$ of $A$ and $Y$ is defined through ancestral relations as $\M:= \De(A) \cap \An(Y) \setminus \{A,Y\}$. 
Let $\mathbf{M} \subset \mathbf{V}$ be the set of all observed mediators between $A$ and $Y$, defined through ancestral relations as $\mathbf{M}:= \mathrm{De}(A) \cap \mathrm{An}(Y) \setminus {A,Y}$. 
The CDE measures the expected change in outcome as the exposure changes when mediators $\mathbf{M}$ are uniformly fixed to a constant value $\mathbf{m}$ through intervention:
% Under the identifiability assumptions in Def.~\ref{def:cde-id},
% It can be expressed using the do-operator as:
% Similarly, the set of confounders $\bm{C}\subset V$ between $A$ and $Y$ is defined as 
% Let $A=a$ and $A=a^*$ be the exposure values corresponding to treatment and no treatment, respectively.
\begin{definition}[CDE, \citet{pearl2014interpretation}]\label{def:CDE} 
$
\mathrm{CDE}(\mathbf{m}):=\mathbb{E}[Y \mid d o(a, \mathbf{m})]-\mathbb{E}\left[Y \mid d o\left(a^*, \mathbf{m}\right)\right].
$
\end{definition}
We provide the \emph{identifiability condition} of CDE 
% Under the identifiability assumptions
in Def.~\ref{def:cde-id}.
When mediator-exposure interactions are present, CDE becomes mediator-dependent, taking different values across $\M$~\citep{pearl2014interpretation}. 
To obtain an unique population-level measure that admits valid adjustment, we define WCDE using the subset of mediators that are also direct parents of $Y$, ensuring that all mediator paths (Def.~\ref{def:path-types}) are blocked: 
% Let $\mathbf{M}' := \mathbf{M} \cap \mathrm{Pa}(Y)$ and let $\mathcal{M}'$ be the set of all possible joint values of the mediators in $\mathbf{M}'$. The weighted controlled direct effect is then defined as:
% To obtain a single population-level measure without making strong parametric assumptions, we define WCDE by averaging $\mathrm{CDE}(\mathbf{m})$ over the marginal distribution of mediators. 

% To address this without making strong parametric assumptions, \citet{pearl2000causality} introduced the $\WCDE$ as the marginal effect:\footnote{In the first edition of \emph{Causality} \citep[Chapter~4.5.4, p.~131]{pearl2000causality}, 
% Pearl introduced the definition of the average direct effect as 
% $\sum_{\Pa_Y \setminus X} \left( \E[Y \mid do(x), do(\Pa_Y \setminus X)] - \E[Y \mid do(x'), do(\Pa_Y \setminus X)] \right) P(\Pa_Y \setminus X)$. 
% This formulation requires fixing all other parents of $Y$. 
% Later, in his article \emph{Interpretation and Identification of Causal Mediation} \citep{pearl2014interpretation}, 
% Pearl emphasized that fixing mediators alone suffices to measure controlled direct effects. 
% Accordingly, let $\mathbf{M}' := \mathbf{M} \cap \Pa(Y)$ denote the subset of $\Pa_Y \setminus X$ consisting of mediators. 
% By replacing $\Pa_Y \setminus X$ with $\mathbf{M}'$, we arrive at our definition of WCDE in \eqref{eqution:WCDE}.} 

\begin{definition}[WCDE, \cite{maasch2025local, pearl2000causality}]\label{def:WCDE} 
% Let $\M$ be the set of mediators that block all mediator paths (Def.~\ref{def:path-types}).
% Let $\mathbf{M}' := \mathbf{M} \cap \text{Pa}(Y)$.
% Let $\mathcal{M}'$ be the set of all possible joint values of the mediators in $\M'$.
Given a DAG $\mathcal{G}$,
let $\mathbf{M}' := \mathbf{M} \cap \mathrm{Pa}(Y)$ and let $\mathcal{M}'$ be the set of all possible joint values of the mediators in $\mathbf{M}'$. 
The WCDE
% weighted controlled direct effect 
can be expressed using do-probabilities as:
\begin{equation}
\label{eqution:WCDE}
    \mathrm{WCDE}=
\sum_{\mathbf{m}'\in\mathcal{M}'}\left(\mathbb{E}\left[Y \mid \operatorname{do}\left(a, \mathbf{m}'\right)\right]-\mathbb{E}\left[Y \mid \operatorname{do}\left(a^*, \mathbf{m}'\right)\right]\right) p\left(\mathbf{m}'\right).
\end{equation}
% follows:
% Let $\mathbf{M}^{\prime} \subseteq \mathbf{M}$ be parents of $Y$,  
% \begin{equation}
% \label{eqution:WCDE}
%     \mathrm{WCDE}=
% \sum_{\mathbf{m}^{\prime}\in\M'}\left(\mathbb{E}\left[Y \mid \operatorname{do}\left(a, \mathbf{m}^{\prime}\right)\right]-\mathbb{E}\left[Y \mid \operatorname{do}\left(a^*, \mathbf{m}^{\prime}\right)\right]\right) p\left(\mathbf{m}^{\prime}\right).
% \end{equation}
\end{definition}
\vspace{-5pt}
Definition~\ref{def:WCDE}
% This formulation 
builds upon prior work by \citet{pearl2000causality, pearl2014interpretation},\footnote{In the \emph{first edition} of \emph{Causality} \citep[Chapter~4.5.4, p.~131]{pearl2000causality}, Pearl defined the average direct effect as $\sum_{\Pa_Y \setminus X} \left( \E[Y \mid do(x), do(\Pa_Y \setminus X)] - \E[Y \mid do(x'), do(\Pa_Y \setminus X)] \right) P(\Pa_Y \setminus X)$. Subsequent work \citep{pearl2014interpretation} clarified that only mediators need to be fixed. Accordingly, we can replace $\Pa_Y \setminus X$ with $\mathbf{M}' := \mathbf{M} \cap \Pa(Y)$.}
% let $\mathbf{M}' := \mathbf{M} \cap \Pa(Y)$ denote the subset of $\Pa_Y \setminus X$ consisting of mediators. 
% By replacing $\Pa_Y \setminus X$ with $\mathbf{M}'$, we arrive at our definition of WCDE in \eqref{eqution:WCDE}.}
who established that fixing only the mediators that are parents of $Y$ is sufficient for identifying direct effects.\footnote{In this work, we do not assume that \emph{all mediators} are observed. Rather, a nonzero WCDE value indicates that a direct effect between $A$ and $Y$ exists when accounting for all observed mediators.}
% where he implied that fixing only the mediators that are parents of $Y$ is sufficient.
As $\mathbf{M}'$ is uniquely determined given $\mathcal{G}$, for any given SCM, Eq.~\eqref{eqution:WCDE} admits an \emph{unique value}
assuming CDE's identifiability.
% because $\mathrm{CDE}(\mathbf{m}')$ is \emph{identifiable} and $\mathbf{M}'$ is uniquely determined by the DAG $\mathcal{G}$.
% due to the identifiability of
% CDE
% % the do-probabilities 
% and that $\mathbf{M}'$ is uniquely determined given DAG $\mathcal{G}$.
% through the exhaustive and mutually exclusive nature of the causal partitions (details in Fig.~\ref{fig:sample}, Tab.~\ref{tab:causal-partitions}, App.~\ref{app:causal_partition}, and \citet{maasch2024local}).
% \kyra{here, for identifiability, we need also standard positivity assumption, and so on?}
% So far, 
\begin{remark}[WCDE definition]
Let $\mathcal{C}$ be the set of confounders between $A$ and $Y$.
In defining WCDE, we preserve
% it is essential to maintain 
the conceptual distinction between mediators and confounders, rather than treating mediators as additional confounders through joint conditioning.
% Therefore, we weight by
% Our definition weights by
We weight the marginal distribution $P(\mathbf{m}')$ rather than the conditional $P(\mathbf{m}'|\mathcal{C})$ because
% to preserve this distinction.
% for important conceptual and methodological reasons. 
the $\mathrm{CDE}(\mathbf{m}')$ in Eq.~\eqref{eqution:WCDE} is already causally standardized over confounders $\mathcal{C}$ through the do-operator, representing the population-average direct effect at mediator level $\mathbf{m}'$.
% If we instead w
Weighting by $P(\mathbf{m}'|\mathcal{C})$
would inappropriately treat the mediator as a confounder by reintroducing the mediator-confounder relationship after causal standardization. This creates conceptual redundancy by effectively conditioning twice on
$\mathcal{C}$,
% twice—once in the causal standardization of CDE($\mathbf{m}'$) and again in the mediator weighting—
% creating
% circularity between the causal standardization and weighting steps. 
% Moreover, it would 
yielding stratum-specific effects conditional on
% a circular dependency that lacks clear causal interpretation. Second, this approach would yield stratum-specific effects conditional on 
$\mathcal{C}$ rather than a coherent population-level measure.
% , which would require
% % , requiring 
% further marginalization over $\mathcal{C}$ to obtain a single population effect.

The alternative factorization of the joint distribution, $P(\mathcal{C}|\mathbf{m}')P(\mathbf{m}')$, requires standardizing effects using $P(\mathcal{C}|\mathbf{m}')$, contradicting CDE's identification (Def. \ref{def:cde-id}).
By using $P(\mathbf{m}')$,
we maintain philosophical clarity by respecting the distinct roles of confounders (adjusted for in causal standardization) and mediators (the pathway of interest). 
% Weighting by $P(\mathbf{m}')$ 
This provides a clear population-level interpretation: the average direct effect under natural mediator variation.
\end{remark}

% Our analysis has thus far relied on do-probabilities--counterfactual quantities defined through interventions.  
% A central challenge lies in expressing the $\WCDE$  (Eq.~\eqref{eqution:WCDE}) 
% using 
% observational data.
% While Def.~\ref{def:WCDE} ensures its identifiability, it does not specify how to compute WCDE from actual data -- something that is achieved through defining valid adjustment sets. 
% We now address the question of which adjustment sets can recover this specific WCDE value defined in Eq.~\eqref{eqution:WCDE}.
While Def.~\ref{def:WCDE} ensures WCDE identifiability, it does not specify how to compute it from observational data. We now characterize valid adjustment sets that recover the specific WCDE value in Eq.~\eqref{eqution:WCDE}. 
% In observational studies, identifying $\WCDE$ requires finding adjustment sets that both: 1)
% satisfy counterfactual independence conditions for $\WCDE$, and 2) yield a unique estimand from the observed data distribution.
While complete criteria exist for ATE adjustment sets \citep{shpitser2012addendum, shpitser2010validity},
no analogous solution currently exists for WCDE,
% no analogous solution currently addresses the $\WCDE$’s dual requirement of adjustment validity and uniqueness—
a gap we now resolve.
% in the rest of the section. 
% Next, w
We define VAS for WCDE in Def.~\ref{def:VAS}, and provide a set of adjustment criteria in Condition~\ref{condition:adj}. We establish their sufficiency and necessity in Lemma~\ref{lemma:adjustment_criterion}.
% \Lry{For completeness, 
% Appendix~\ref{app:notation} provides the standard counterfactual independence assumptions and the formal definitions of directed, backdoor, and mediator paths.
% }

% Covariate adjustment is a widely used method for estimating the average treatment effect (ATE) from observational data, and sound and complete graphical criteria for valid adjustment in DAGs have been established~\citep{shpitser2010validity,shpitser2012addendum}. Analogous to the ATE case, we now 、introduce adjustment criteria for estimating the $\WCDE$.
% \Lry{
\begin{definition}[Valid Adjustment Set for WCDE]\label{def:VAS}
Let $\mathscr{M}$ be an SCM that induces causal DAG $\mathcal{G}$
with pairwise disjoint node sets $A$, $Y$, and $\mathbf{M}$,
% . Let 
% $A$, $Y$, and $\mathbf{Z}$ be pairwise disjoint node sets in 
% % a causal DAG
% $\mathcal{G}$, 
representing the treatment, outcome, and the set of all mediators (Tab.~\ref{tab:partitions}), respectively.
Let $\Z_1 = \Z \cap \M$,  $\Z_2 = \Z \setminus \M$, and
% where $A$ denotes the treatment, $Y$ denotes the outcome, and $\mathbf{M}$ denotes the set of mediators.
% We say that $\mathbf{Z}$ is a \emph{valid adjustment set} for identifying the $\WCDE$ with respect to $(A, Y)$
% if and only if the
% following 
% equality holds in the population defined by $\mathscr{M}$:
% holds:
\begin{equation}\label{eq:WCDE_identify}
    \mathrm{WCDE}_{\mathbf{Z}} =
\underbrace{\EE_{\mathbf{Z}_1} \left\{ \EE_{\mathbf{Z}_2} \left[ \mathbb{E}[Y \mid A = a, \mathbf{Z}_1, \mathbf{Z}_2] \right] \right\}}_{T_{a}(\Z)}
    - 
    \underbrace{\EE_{\mathbf{Z}_1} \left\{ \EE_{\mathbf{Z}_2} \left[ \mathbb{E}[Y \mid A = a^*, \mathbf{Z}_1, \mathbf{Z}_2] \right] \right\}}_{T_{a^*}(\Z)}.
\end{equation}
% where $\Z_1 = \Z \cap \M$ and $\Z_2 = \Z \setminus \M$.
We say that $\mathbf{Z}$ is a \emph{valid adjustment set} for identifying the $\WCDE$ with respect to $(A, Y)$
if and only if 
$\mathrm{WCDE}_{\mathbf{Z}}$ equals the interventional $\WCDE$  defined in Def.~\ref{def:WCDE} when both are applied to the population defined by 
$\mathscr{M}$.
% the
% following 
% equality holds in the population defined by $\mathscr{M}$:
% That is, $\Z$ is a VAS 
% if the adjustment formula in Eq.~\eqref{eq:WCDE_identify} recovers the unique $\WCDE$ value defined in Def.~\ref{def:WCDE}
% when applied to the population defined by 
% % the structural causal model
% $\mathscr{M}$.
% if it allows the $\WCDE$ to be identified by Equation~\eqref{eq:WCDE_identify}
% under the true distribution $P$.
\end{definition}

\begin{condition}[WCDE Adjustment Criteria]\label{condition:adj}
Given the notation and SCM from Definition~\ref{def:VAS}, a candidate adjustment set $\mathbf{Z}$
% Let $A$, $Y$ be the treatment and outcome variables, and let $\mathbf{Z}$ be a candidate adjustment set in $\mathcal{G}$.
% Denote $\mathbf{Z_1} = \Z \cap \M$ and $\mathbf{Z_2} = \Z \setminus \M$.
% The set $\mathbf{Z}$
satisfies the \emph{adjustment criterion} relative to $(A, Y)$ for $\WCDE$ in $\mathcal{G}$ if:
% the following conditions hold:
\begin{enumerate}[label=C\arabic*,noitemsep,topsep=0pt,partopsep=0pt,parsep=0pt,leftmargin=*]
    \item All \emph{mediator paths} (Def.~\ref{def:path-types}) for $\{A, Y\}$ are blocked by $\Z_1$; \label{condition:mediator_path}
    \item The subset $\Z_2$ satisfies the sufficient and necessary graphical conditions for identifying the CDE with mediator $\Z_1$, 
    which we further discuss
    % as defined 
    in Appendix~\ref{app:cde-id}; \label{condition:cde}
    \item $\{\mathbf{M}' \setminus \mathbf{Z_1}\} \perp\!\!\!\perp_{\mathcal{G}} A,\{\Pa(Y) \setminus (\mathbf{M}' \cup \{A\})\}\mid \mathbf{Z_1}$,
    ensuring an unique weighting scheme across mediator levels (illustrated in Fig.~\ref{fig:sample5}, Example~\ref{example:condition_unqiue_WCDE}). \label{condition:unique_WCDE}
\end{enumerate}
\end{condition}
% \vspace{-0.5em}
% Condition~\ref{condition:cde} corresponds to the identification of the controlled direct effect defined in Appendix~\ref{def:cde-id}, 
% while Condition~\ref{condition:unique_WCDE} further ensures the uniqueness of the weighted $\mathrm{CDE}$ by guaranteeing a well-defined weighting scheme across mediator levels 
% (illustrated in Fig.~\ref{fig:sample5}, Example~\ref{example:condition_unqiue_WCDE}).

\begin{lemma}[Sufficiency and Necessity of the Adjustment Criterion]\label{lemma:adjustment_criterion}
Let $\mathbf{Z}$ be any set that satisfies Condition~\ref{condition:adj} for $\WCDE$ with respect to $(A, Y)$ in $\mathcal{G}$.
Then $\mathbf{Z}$ is a VAS (Def.~\ref{def:VAS}).
% valid adjustment set for identifying a unique $\WCDE$ via Equation~\eqref{eq:WCDE_identify}.
Conversely, any adjustment set $\Z$ that identifies 
the WCDE in Eq.~\eqref{eqution:WCDE}
% a unique $\WCDE$ as in Equation~\eqref{eq:WCDE_identify}
must necessarily satisfy Condition~\ref{condition:adj}.
\end{lemma}

% }
% \Lry{It is worth noting 
% Lemma~\ref{lemma:adjustment_criterion} 
% establishes that the $\WCDE$ is a well-defined causal quantity, whose population value is invariant to the choice of valid adjustment set (under Condition~\ref{condition:adj}). 
Lemma~\ref{lemma:adjustment_criterion} establishes that the $\WCDE$, which is uniquely defined by Def.~\ref{def:WCDE}, can be identified through any VAS satisfying Condition~\ref{condition:adj}, with the same population value regardless of which valid set is chosen.
However, in finite samples, estimates can vary across sets due to differences in statistical properties like bias and variance. 
A proof is provided in Appendix~\ref{appendix:adj}, with a sketch below.
% indicates 
% that the $\WCDE$ is a well-defined causal quantity that yields the same value in the population regardless of which valid adjustment set is used (under Condition~\ref{condition:adj}). 
% Nevertheless, different adjustment sets can produce distinct finite-sample estimates of this quantity due to differences in their statistical properties (e.g., bias and variance).
% }
% The proof of Lemma~\ref{lemma:adjustment_criterion} is contained in Appendix~\ref{appendix:adj}.
% We provide a proof sketch below.
% Proof Sketch of Lemma~\ref{lemma:adjustment_criterion} (see Appendix~\ref{appendix:adj})

\textbf{Necessity.}  
We focus on the necessity of
% Criterion~
\ref{condition:unique_WCDE}, as \ref{condition:mediator_path} is inherent to the definition of WCDE, and  \ref{condition:cde} follows from CDE's identification. 
% , since the first three are already known to be necessary for identifying the $\mathrm{CDE}$~\citep{vanderweele2011controlled}.  
For a given DAG $\mathcal G$ and a adjustment set $\mathbf Z$, suppose that 
% Criterion 
% \ref{condition:unique_WCDE} is violated. Then, for any adjustment set 
$\mathbf Z$ satisfies 
\ref{condition:mediator_path} and \ref{condition:cde},
% the first three criteria 
but violates \ref{condition:unique_WCDE},
% the last, 
we show that there exists a joint distribution over the vertices $\mathbf V$ in $\mathcal G$ so that $\mathrm{WCDE}_{\Z}$ defined in Eq. \eqref{eq:WCDE_identify} does not match the $\WCDE$ in \eqref{eqution:WCDE}. Specifically, let $\mathbf{W}=\mathbf{M ^{\prime}}\setminus\mathbf{Z_1}$, and $ \mathbf{C}=\Pa(Y)\setminus \mathbf{(M ^{\prime}}\cup A)$. Because  \ref{condition:unique_WCDE} is violated, $\mathbf{W} \not\!\perp\!\!\!\perp_{\mathcal{G}} A, \mathbf{C} \mid \mathbf{Z_1}$. This conditional dependence results in a discrepancy between \eqref{eq:WCDE_identify} and \eqref{eqution:WCDE} due to a bias term involving
$p(\w \mid \z_1) - p(\w \mid \z_1, a, \c),$
which is non-zero whenever $\mathbf{W} \not\!\perp\!\!\!\perp_{\mathcal{G}} A, \mathbf{C} \mid \mathbf{Z_1}$ (see Example \ref{example:condition_unqiue_WCDE}).
% for a detailed example.

%This conditional dependence leads to bias in the estimated effect, implying that the adjustment set $\mathbf{Z}$ fails to identify the $\WCDE$ unless Criterion \ref{condition:unique_WCDE} is satisfied.

% \begin{figure}[t]
% \begin{wrapfigure}{r}{0.48\textwidth}
% \centering
% \includegraphics[width=0.45\textwidth]{fig1.png}
% \caption{A DAG illustrating Condition \ref{condition:adj}.}
% \label{fig:sample5}
% \end{wrapfigure}
% \begin{tikzpicture}[
%     node distance=0.7cm and 0.9cm,
%     every node/.style={draw, ellipse, align=center, minimum width=0.89cm, minimum height=0.6cm, font=\normalsize\rmfamily},
%     arrow/.style={-Latex, thick} 
% ]

% \node (A) {$A$};
% \node[right=of A] (B1) {$B_1$};
% \node[right=of B1] (G1) {$G_1$};
% \node[right=of G1] (Y) {$Y$}; 
% \node[above=of G1] (G2) {$G_2$}; 
% % 绘制边
% \draw[arrow] (A) -- (B1); 
% \draw[arrow] (B1) -- (G1); 
% \draw[arrow] (G1) -- (Y); 
% \draw[arrow] (G2) -- (B1);
% \draw[arrow] (G2) -- (A); 
% \draw[arrow] (G2) -- (Y);
% \draw[arrow] (G2) -- (G1);
% \draw[arrow, bend right=25] (A) to (Y);
% \end{tikzpicture}
% \caption{A DAG illustrating Condition \ref{condition:adj}.}
% \label{fig:sample5}
% \end{wrapfigure}
% \end{figure}

\textbf{Sufficiency.}  
We show that if an adjustment set $\mathbf{Z}$ satisfies all three criteria in Condition~\ref{condition:adj}, then Eq.~\eqref{eq:WCDE_identify} identifies the $\WCDE$. 
% \Lry{
We first decompose $\mathbf{Z}$ into mediators $\mathbf{Z}_1$ and non-mediators $\mathbf{Z}_2$, and then replace $\mathbf{Z}_2$ with $\C$, which continues to identify the CDE and therefore preserves the expectation.
Next add $\mathbf{W}=\mathbf{M}'\setminus\mathbf{Z}_1$ to $\mathbf{Z}_1$, which likewise leaves the expectation unchanged.  
% \Lry{
After removing non-parent variables, the remaining expectations are taken over $\C$ and $\mathbf{M}'$. 
Substituting $\Z_2=\C$ and $\Z_1=\mathbf{M}'$ into Eq. \eqref{eq:WCDE_identify}, 
and noting that $\C$ identifies the CDE for the mediator set $\mathbf{M}'$, 
we obtain the WCDE expression in Eq.~\eqref{eqution:WCDE}.
% }
% }

% \kyra{above two have places that need to be fixed}

Given the identification of all VASs, a natural question arises: when multiple such sets are available, which one should we choose? In this work, 
we address this question by comparing VASs through the lens of \emph{asymptotic variance}.
Specifically, we characterize how the choice of VAS affects the asymptotic variance across a general class of estimators. Our objective is to identify the optimal VAS that minimizes asymptotic variance for any estimator within this class.
% we compare valid adjustment sets based on the \emph{asymptotic variance} of a specific class of estimators, and this variance depends on the choice of adjustment set. Our goal is to identify the adjustment set that achieves the smallest asymptotic variance for any fixed estimator from this class of estimators.
% In Section \ref{sec:IF}, we first present these estimators in a general framework 
% and characterize their asymptotic variance. We then specialize these results to the $\WCDE$ setting, deriving the corresponding influence function for the $\WCDE$ estimator. 
% % using tools from statistical efficiency theory.
% In Section \ref{sec:optimal_VAS}, we characterize the \emph{optimal} valid adjustment set, and illustrate how one can obtain such an optimal adjustment from altering any given valid adjustment set.
% ---one that minimizes the asymptotic variance of the $\WCDE$ estimator. We further provide methods for obtaining it from any given valid adjustment set.

% \vspace{-5pt}
\section{RAL Estimators and Influence Function of WCDE}\label{sec:IF}
% \vspace{-5pt}
% \kyra{this section formally sets up the semiparametric efficiency of $\WCDE$ estimation, and introduce the result on $\WCDE$ influence function}

%\YG{Why do we need efficiency theory? Why is it relevant for minimizing asymptotic variance with respect to adjustment set?}

% We now review the key elements of efficient nonparametric parameter estimation that we will use throughout the paper.
To identify the optimal VAS, we begin by introducing \emph{regular and asymptotically linear} estimators---a class of estimators with well-characterized asymptotic variance through their influence functions. These estimators provide the foundation for our optimality analysis, which we develop fully in Section~\ref{sec:optimal_VAS} through a constructive variance-minimization procedure.
We then specialize this framework to $\WCDE$ estimation, presenting concrete examples of RAL estimators. Theorem~\ref{thm:IF} establishes the explicit form of the $\WCDE$ influence function as a function of the chosen VAS.
% Next, we provide examples of asymptotically linear estimators for $\WCDE$, and derive the influence function of $\WCDE$ as a function of the valid adjustment set in Theorem~\ref{thm:IF}. 
% these estimators in a general framework 
% and characterize their asymptotic variance. We then specialize these results to the $\WCDE$ setting, deriving the corresponding influence function for the $\WCDE$ estimator. 
% using tools from statistical efficiency theory.
% In Section \ref{sec:optimal_VAS}, we characterize the \emph{optimal} valid adjustment set, and illustrate how one can obtain such an optimal adjustment from altering any given valid adjustment set.

\textbf{RAL Estimators\;\;} We begin by introducing RAL estimators in a general statistical framework.  Given $n$ i.i.d.\ observations $O_1, \dots, O_n$ drawn from an \emph{unknown} distribution $P$ on 
a sample space $\Omega$. Our goal is to estimate a real-valued target parameter,  $T(P)$ , based on these observations. For example, in $\WCDE$ , for a given valid adjustment set $\Z$, the target parameter is given by Eq.~\eqref{eq:WCDE_identify}.  
% $T(P)$, where $T:$ is a mapping from $\mathcal{M}$ to $\mathbb{R}$.
% We write $O \sim P$ to denote a generic observation drawn from $P$.% 
We consider a fully nonparametric model class $\mathcal{M}$, i.e., $P \in \mathcal{M}$, which contains all candidate distributions defined on a measurable space $(\Omega, \mathcal{F})$ dominated by a common $\sigma$-finite measure $\nu$. Let $p$ denote the corresponding density of $P$ with respect to $\nu$.

% Below, for a given distribution $P$, we denote by $L_2^0(P)$ the Hilbert space of square-integrable, mean-zero functions with respect to $P$, i.e.,
% \[
% L_2^0(P) := \left\{ f : \mathbb{E}_P[f(O)^2] < \infty,\ \mathbb{E}_P[f(O)] = 0 \right\}.
% \]
% \paragraph{Asymptotically Linear Estimators}
% A central concept in efficient estimation is asymptotic linearity, defined as follows:

We first introduce the notion of regularity, which connects estimators to their stability under small model perturbations (Def.~\ref{def:regular}). Next, we define asymptotic linearity (Def.~\ref{def:asymplin}).
\begin{definition}[Regular estimator,~\citet{vanderVaart2000}]\label{def:regular}
%Let $\mathcal{M}$ be a collection of probability distributions for a random variable $\mathbf{V}$, and 
Let $T(P):\mathcal{M}\rightarrow \mathbb{R}^d$ be a target parameter of interest.
% defined for each $P \in \mathcal{M}$.  
An estimator $\widehat{T}_n$ is said to be \textbf{regular} in $\mathcal{M}$ at $P$ if its convergence to $T(P)$ is \textbf{locally uniform} in a neighborhood of $P$ in $\mathcal{M}$.
\end{definition}

\begin{definition}[Asymptotically linear estimator,~\citet{vanderVaart2000}]\label{def:asymplin}
% Let $T(P)$ be a parameter of interest. 
Given $O_1, \ldots, O_n\overset{\text{i.i.d.}}{\sim} P$ and target parameter $T(P)$, an estimator $\widehat{T}_n = \widehat{T}_n(O_1, \ldots, O_n)$ of $T(P)$ is said to be \emph{asymptotically linear} at $P$ if there exists a function $\psi_P \in L_2^0(P)$ such that
$
\sqrt{n}(\widehat{T}_n - T(P)) = \frac{1}{\sqrt{n}} \sum_{i=1}^n \psi_P(O_i) + o_p(1), 
$
where $L_2^0(P)$ is the set of all mean-zero, square-integrable functions under $P$, i.e., $L_2^0(P) := \left\{ f : \mathbb{E}_P[f(O)^2] < \infty,\ \mathbb{E}_P[f(O)] = 0 \right\}.$\footnote{Here, we denote $O$ as a generic random variable drawn from the distribution $P$.}
\end{definition}

Common RAL estimators include the ordinary least squares estimator~\citep{newey1990semiparametric}, inverse probability weighting~\citep{robins1994estimation}, AIPW estimator~\citep{glynn2010introduction}, doubly robust estimators~\citep{bang2005doubly, funk2011doubly}
%\kyra{add a few more citations on this}
, targeted maximum likelihood estimator~\citep{van2006targeted}, and double machine learning~\citep{chernozhukov2018double}.

\textbf{Influence Function\;\;}
The function $\psi_P$ in Definition \ref{def:asymplin} above is known as the \emph{influence function} (IF) of $\widehat{T}_n$ at $P$. By the Central Limit Theorem, we have:
\begin{equation}\label{eq:asymptotic_variance}
\sqrt{n}(\widehat{T}_n - T(P)) \xrightarrow{d} N\left(0, \operatorname{Var}_P\left[\psi_P(O)\right]\right).
\end{equation}
In fully nonparametric models, this IF is unique for all RAL
% regular and asymptotically linear 
estimators and coincides with the influence function of the target parameter $T(P)$.
At a high level, IF of the target parameter $T(P)$ measures the sensitivity of $T(P)$ under small perturbations of the distribution $P$~\citep{tsiatis2006semiparametric}. We provide a formal definition in Appendix~\ref{app:IF_dicussion}.
%In nonparametric models, this IF is unique and equals the \emph{efficient influence function} (EIF)~\citep{tsiatis2006semiparametric}. 
%Consequently, all asymptotically linear estimators of a given target parameter $T$ share the same influence function, which coincides with the influence function of $T$ itself.
%In this case, 
\begin{remark}[Influence Functions of $\WCDE_{\Z}$]\label{remark:IF_WCDE_Z}
We emphasize that Eq.~\eqref{eq:WCDE_identify} defines distinct target parameters for each adjustment set $\Z$. While these parameters coincide in value at the true distribution $P$ (yielding an unique causal estimand), different choices of the adjustment set $\Z$ would affect the sensitivity of the corresponding target parameter $\mathrm{WCDE}_{\Z}$ to local distributional perturbations (see concrete definitions of local distributional perturbation in Appendix~\ref{app:IF_dicussion}). Intuitively, $\mathrm{WCDE}_{\Z}$ is sensitive to perturbations in variables within $\Z$, so the choice of $\Z$ directly affects its sensitivity to different perturbations.
This leads to distinct influence functions for $\mathrm{WCDE}_{\Z}$ across different $\Z$. We formalize this argument in Appendix \ref{app:IF_dicussion}.
%\kyra{fix; this is an important argument, if we have the space, we should try to explain in the main body}
\end{remark}
\vspace{-3pt}

\textbf{Application to WCDE\;\;}
Given a valid adjustment set $\Z$,
% Suppose $\mathbf{Z}$ is a valid adjustment set, and
let $\mathbf{Z_1} = \mathbf{Z} \cap \mathbf{M}$ and $\mathbf{Z_2} = \mathbf{Z} \setminus \mathbf{M}$. 
We first provide the IF
% influence function 
corresponding to the first component of $\WCDE$ defined in Eq.~\eqref{eq:WCDE_identify}, denoted by $T_a(\Z)$ 
%(Theorem~\ref{thm:IF})
.
The IF of $T_{a^*}(\Z)$ can be similarly obtained by replacing $a$ with $a^*$ in the expression. 
% 
% Next, for a fixed valid adjustment set $\Z$, we derive the unique influence function of $T_a(\Z)$
% in Theorem~\ref{thm:IF}.
% Then the weighted controlled direct effect at treatment level $a \in \mathcal{A}$ is given by
% \begin{equation}\label{eq::WCDE-a-functional}
%     T_a(\mathbf Z)
%   :=
%   \mathbb{E}_{\mathbf{Z_1}}\!\bigl\{\,
%     \mathbb{E}_{\mathbf{Z_2}}\!\bigl[
%       \mathbb{E}_Y[Y\mid A=a,\mathbf{Z_2},\mathbf{Z_1}]
%     \bigr]
%   \bigr\}.
% \end{equation}

% Viewing $T_a(\mathbf Z)$ as the target parameter, we derive the unique influence function of any asymptotic linear estimator for this parameter in Theorem \ref{thm:IF} below.
% Here the uniqueness of the influence function is guaranteed by our fully nonparametric setting, and thus from this point onward, we refer to it as \emph{the} influence function.
% Note that the influence function derived depends on the choice of valid adjustment set, and so does the resulting variance. 

\begin{theorem}[IF of $\WCDE_\Z$]\label{thm:IF}
Given any VAS $\mathbf Z$, the influence function of $T_a(\Z)$ is given by
% the $\WCDE$ at treatment level $a \in \mathcal{A}$: functional \eqref{eq::WCDE-a-functional} is given by
    \begin{align*}
    \psi_{a}\left(Y, A, \mathbf{Z_1}, \mathbf{Z_2} ; P\right)  &= 
    \frac{ I_a(A) p(\mathbf{Z_1})  p(\mathbf{Z_2})}
    {p( \mathbf{Z_1}, \mathbf{Z_2},a)} \cdot \bigg( Y - \mathbb{E}_Y[Y | A=a,\mathbf{Z_1},\mathbf{Z_2}] \bigg) \notag \\
    & + \mathbb{E}_{\mathbf{Z_2}} \bigg[ \mathbb{E}_Y[Y | A=a,\mathbf{Z_1},\mathbf{Z_2}] \bigg]   
    + \mathbb{E}_{\mathbf{Z_1}} \bigg[ \mathbb{E}_Y[Y | A=a,\mathbf{Z_1},\mathbf{Z_2}] \bigg] - 2 T_a(\mathbf{Z}).
    \end{align*}
    Here,
    % $\mathbf{Z_1} = \mathbf{Z} \cap \mathbf{M}$, $\mathbf{Z_2} = \mathbf{Z} \setminus \mathbf{M}$, and
    \(\mathbf{I}\{\cdot\}\) denotes the indicator function, and \(T_{a}(\Z
    )\) is evaluated at
    % is the target functional evaluated at
    \(P\).
\end{theorem}

The proof of Theorem~\ref{thm:IF} 
%(Appendix~\ref{app:proof_thm_IF})
begins by substituting this functional $T_a(\Z)$ into the definition of pathwise differentiability (see Appendix~\ref{app:proof_thm_IF} for details). Leveraging the fact that the influence function is unique in a fully nonparametric model, we directly obtain an unique influence function associated with each valid adjustment set.

% \vspace{-5pt}
\section{Optimal Valid Adjustment Sets for WCDE}\label{sec:optimal_VAS}
% \vspace{-5pt}
% \vspace{-2pt}
Given that distinct valid adjustment sets typically yield different IFs (and thus different asymptotic variances), we now leverage the adjustment criterion from Section~\ref{sec:WCDE_VAS} and the IF derived in Section~\ref{sec:IF} to identify the optimal VAS---the one minimizing asymptotic variance for all RAL estimators.
% Different valid adjustment sets generally correspond to different influence functions, leading to different asymptotic variances for the $\WCDE$ estimator. With the adjustment criterion established in Section~\ref{sec:WCDE_VAS} and the influence function characterization derived in Section~\ref{sec:IF}, we are now prepared to evaluate how the choice of valid adjustment sets affects the asymptotic variance of $\WCDE$ estimators.
% In this section, we present results for identifying the 
% \emph{optimal valid adjustment set}, and we organize the proof sketch across several subsections.
% \kyra{this is too strong, you just used the language. say to facilitate presentation, we introduce ... and also give some intuition on why you are introducing this set of notations}

To facilitate the presentation of our results, we adopt
% Our methodology builds on
the causal‐partition taxonomy in Table 1~\citep{maasch2024local}. Given a causal DAG $\mathcal{G}$, let $(A,Y)\subseteq \V$ be an exposure-outcome pair, where $Y$ is a nondescendant of $A$.
% For any exposure-outcome pair $(A,Y)$ and arbitrary variable set $\mathbf{X}$, 
This local taxonomy decomposes the rest variables in the DAG, $\X = \V\setminus \{A,Y\}$,
% any arbitrary variable set $\mathbf{X}$ 
into eight \emph{exhaustive and mutually exclusive} subsets, each defined by the types of causal paths that it can share with $A$ and $Y$.  
% Below,
% \textbf{Notations}
When the union of multiple partitions is needed, we write, for example, $\mathbf{X}_{1,3}:=\mathbf{X}_1\cup\mathbf{X}_3$. Moreover, for any $k \in \{1,\dots,8\}$, we denote by $\mathbf{X}_{k \in \Pa(Y)} := \mathbf{X}_k \cap \Pa(Y)$ the subset of $\mathbf{X}_k$ that also belongs to the set of parents of $Y$.

\begin{figure}[t]
  \centering
  % ---- 左侧：图片 ----
  \begin{minipage}{0.45\linewidth}
    \centering
    \includegraphics[width=0.85\linewidth]{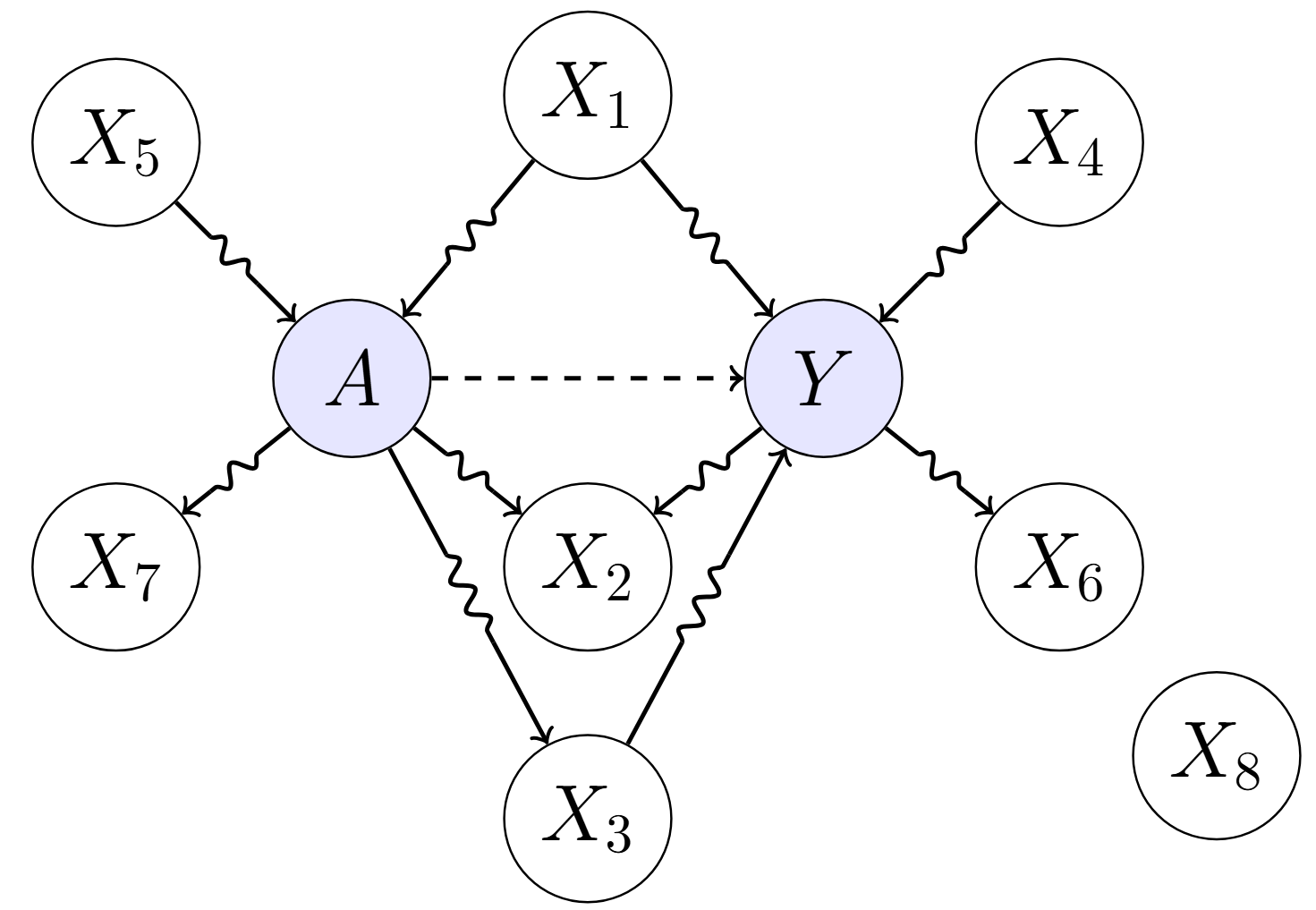}
    \vspace{-5pt}
    \caption{The partition visualization follows \citet{maasch2024local}, with directed squiggly edges indicating the presence of a directed causal path (not necessarily direct parent-child relationships). The dashed edge represents potential parent-child relationships. }        % ← 仍然是 Figure
    \label{fig:sample}
  \end{minipage}%
  \hfill
  % ---- 右侧：表格 ----
  \begin{minipage}{0.52\linewidth}
    \centering
    % 关键：用 \captionof{table} 令标题按 Table 编号
    \begin{tabularx}{\linewidth}{lX}
      \toprule
      \textbf{Group} & \textbf{Description} \\
      \midrule
      $\mathbf{X}_1$ & Confounders and their proxies. \\
      $\mathbf{X}_2$ & Colliders and their proxies. \\
      $\mathbf{X}_3$ & Mediators and their proxies. \\
      $\mathbf{X}_4$ & Non‑descendants of $Y$ where $\mathbf{X}_4 \!\perp\!\!\!\perp\! A$ and $\mathbf{X}_4 \not\perp\!\!\!\perp A \mid Y$. \\
      $\mathbf{X}_5$ & Instruments and their proxies. \\
      $\mathbf{X}_6$ & Descendants of $Y$ with $A$–$Y$ paths mediated by $Y$. \\
      $\mathbf{X}_7$ & Descendants of $A$. \\
      $\mathbf{X}_8$ & Nodes with no active paths to $A$ or $Y$. \\
      \bottomrule
    \end{tabularx}
    \captionof{table}{Exhaustive and Mutually Exclusive Causal Partitions \citep{maasch2024local}. Detailed definitions are included in Appendix~\ref{app:causal_partition}.}
    \label{tab:causal-partitions}
  \end{minipage}
  \vspace{-5pt}
\end{figure}

Under the causal partition taxonomy, we construct the set $\mathbf{O}$ as the union of all parents of $Y$ belonging to relevant partition groups in Definition~\ref{def::O}.

\begin{definition}[O-set]\label{def::O}
% Under the causal partition taxonomy, we define 
Let $\mathbf{O}(A,Y,\mathcal{G})$ denote the set of all non-treatment parents of $Y$ in $\mathcal{G}$. Formally,
under the causal partition taxonomy, these parents fall into three groups: confounders ($\mathbf{X}_1$), mediators ($\mathbf{X}_3$), and informative non-descendants ($\mathbf{X}_4$):
% all relevant parents of the outcome, including the confounding, mediating path, and other non-descendants of $Y$: 
% \kyra{what do you mean by relevant parents? }
\[
\mathbf{O}(A,Y,\mathcal{G}):= \mathbf{X}_{1 \in \Pa(Y)} \cup \mathbf{X}_{3 \in \Pa(Y)} \cup \mathbf{X}_{4 \in \Pa(Y)}.
\]
\end{definition}
% \vspace{-2pt}
Intuitively, $\mathbf{O}(A,Y,\mathcal{G})$ is constructed to maximize information on $Y$, while minimizing information on $A$ and preserving validity. Next, we formally define the notion of optimal VAS in Definition~\ref{def:optimal_VAS}.
Theorem~\ref{thm:optimal_adjustment} establishes the optimality of $\mathbf{O}(A,Y,\mathcal{G})$.

\begin{definition}[Optimal VAS]\label{def:optimal_VAS}
Given a DAG $\mathcal{G} = (\V, \mathbf{E})$ and an exposure–outcome pair $(A, Y) \subseteq \V$, let $\mathcal{Z}$ denote the collection of all VASs for identifying the $\WCDE$ of $A$ on $Y$.
We say that $\mathbf{O^*} \in \mathcal{Z}$ is an \emph{optimal valid adjustment set} if, for every $\mathbf{Z} \in \mathcal{Z}$,
$
\sigma_{\mathbf{Z}}^2(P) - \sigma_{\mathbf{O^*}}^2(P) \geq 0,
$
where  $\sigma_{\mathbf{Z}}^2(P)$  denotes the asymptotic variance of any RAL estimator of  $\WCDE_\Z$  based on adjustment set $\mathbf{Z}$ under distribution $P$, as characterized by Equation~\eqref{eq:asymptotic_variance}.
\end{definition}

%\kyra{this definition needs to be updated}

\begin{theorem}[O-set is Optimal]\label{thm:optimal_adjustment}
Given a DAG $\mathcal{G}=(\V,\mathbf{E})$,
% be a DAG with vertex set $\mathbf{V}$, 
and the exposure-outcome pair $(A,Y)\subseteq \V$,
% Let $\mathcal{G}$ be a DAG with vertex set $\mathbf{V}$, let $A \subset \mathbf{V}$ and $Y \in \mathbf{V} \backslash \mathbf{A}$, where $A$ is a random variable taking values on a finite set. Then
the set $\mathbf{O}(A,Y,\mathcal{G})$ in Definition \ref{def::O} is the \emph{optimal valid adjustment set} for the $\WCDE$ (\ref{eqution:WCDE}), satisfying Condition \ref{condition:adj} and producing minimal asymptotic variance for all RAL estimators.
\end{theorem}

% A detailed proof is provided in Appendix~\ref{appendix:sec4}. 
Below,
% In Sections~\ref{subsec:backdoor}-\ref{subsec:overadjustment_mediator},
we sketch the main proof idea (full proof in Appendix~\ref{appendix:sec4}).
% below we sketch the main idea. 
Starting from an arbitrary VAS $\mathbf{Z}$ with $\Z_1=\Z\cap\M$ and $\Z_2=\Z\setminus\M$, we iteratively transform it into $\mathbf{O}(A,Y,\mathcal{G})$ through four variance-reducing operations: (1) adding non-mediating parents of $Y$, (2) including mediating parents of $Y$, (3) removing redundant backdoor variables that do not contribute to $Y$, and (4) pruning mediators that offer no further gain. Each step preserves the validity of the adjustment set and reduces the asymptotic variance, ultimately converging to the same optimal set $\mathbf{O}$ regardless of the starting VAS. While this sequence of variance-reducing operations resembles the pruning procedure proposed by\,\cite{henckel2022graphical} and the graphical criteria introduced by \,\cite{rotnitzky2020efficient}, our work extends these ideas to a new functional that involves mediator-confounder interactions, introducing additional structural complexity.
%\kyra{you want to say from any $\Z$ you arrive at the same optimal O somewhere?} 
%\kyra{highlight the similar and different places with prior work in one or two sentences}
% The detailed arguments and supporting lemmas for each step are presented in the following subsections, while the full formal proofs of all supporting lemmas are provided in the Appendix\ref{appendix:sec4}.

\textbf{Step 1: Augmenting \texorpdfstring{$\mathbf{Z}_2$}. with Non-mediating Parents of \texorpdfstring{$Y$}.\hspace{0pt}.\;\;}
% \label{subsec:backdoor}
% \paragraph{Step 1: Augmenting $\mathbf{Z}_2$ with Non-mediating Parents of $Y$} 
% \kyra{you need to reference the Figure 3 somewhere here. Unclear how 3 is illustrating this lemma}
Lemma~\ref{lemma:1} (proof in Appendix~\ref{app:proof_lemma1}) establishes that including additional backdoor variables satisfying conditional independence with the treatment can reduce variance without losing validity, i.e., if we add $\mathbf{X}_{1 \in \Pa(Y)}$ and $\mathbf{X}_{4 \in \Pa(Y)}$ to $\mathbf{Z}_2$:
$$
\sigma^2_{(\mathbf{Z}_1,\, \mathbf{Z}_2 \cup \mathbf{X}_{1 \in \Pa(Y)} \cup \mathbf{X}_{4 \in \Pa(Y)})}(P) \leq \sigma^2_{(\mathbf{Z}_1, \mathbf{Z}_2)}(P).
$$
\begin{lemma}[Supplement of backdoor variables]\label{lemma:1}
Given a DAG $\mathcal{G}=(\V,\mathbf{E})$,
% be a DAG with vertex set $\mathbf{V}$, 
and the exposure-outcome pair $(A,Y)\subseteq \V$, 
% and
% Let $\mathcal{G}$ be a DAG with vertex set $\mathbf{V}$, let $A \subset \mathbf{V}$ and $Y \in \mathbf{V} \backslash \mathbf{A}$, where $A$ is a random variable taking values on a finite set.
suppose $(\G_1, \B_2) \subset \mathbf{V} \setminus \{A, Y\}$ is a
VAS,
% valid adjustment set relative
% to $(A, Y)$ in $\mathcal{G}$, 
% and suppose
with $\G_2$ 
% is
disjoint from both $\G_1$ and $\B_2$, satisfying
$
\G_2 \perp\!\!\!\perp_{\mathcal{G}} A, \G_1 \mid \B_2.
$
Then $(\G_1, \G_2, \B_2)$ is also a VAS, and for all $P \in \mathcal{M}$,
$
\sigma_{( \G_1, \B_2)}^2(P) - \sigma_{( \G_1,\G_2, \B_2)}^2(P) \geq 0.
$
\end{lemma}
%\YG{Proof of Lemma~\ref{lemma:1} is in Appendix ???}
\textbf{Proof Sketch of Lemma~\ref{lemma:1}.}
Figure~\ref{fig:sample1} 
illustrates
% provides a simplified illustration of
the conditions stated in Lemma~\ref{lemma:1}. For clarity, we focus on the first component of the $\WCDE$ in Eq.~\eqref{eq:WCDE_identify},  and let $\sigma_{a,\mathbf{Z}}^2(P)$ denote its asymptotic variance. The extension to the full $\WCDE$ is direct.
% straightforward.
We begin by decomposing the variance:
% \vspace{-3pt}
\begin{align*}
&\sigma^2_{a, ( \G_1, \B_2)}(P)
= \operatorname{Var}_P\left[\psi_{a}(Y,A,\G_1, \B_2; P)\right] = \operatorname{Var}_P\left[\psi_{a}(Y,A,\G_1, \G_2, \B_2; P)\right] \\
&+ \operatorname{Var}_P\left[\psi_{a}(Y,A,\G_1, \B_2; P) - \psi_{a}(Y,A,\G_1, \G_2, \B_2;P)\right] \\
& + 2\,\operatorname{Cov}_P\left(
\psi_{a}(Y,A,\G_1, \G_2, \B_2;P),
\psi_{a}(Y,A,\G_1, \B_2;P) - \psi_{a}(Y,A,\G_1, \G_2, \B_2;P)
\right).
\end{align*}
To analyze the covariance term, consider the difference in IFs evaluated at $(Y, A, \G_1, \G_2, \B_2)$: $$\psi_a(Y, A, \G_1, \B_2;P) - \psi_a(Y, A, \G_1, \G_2, \B_2;P).
$$
We construct a parametric submodel that perturbs only the conditional law $P(A, \G_1 \mid \G_2, \B_2)$, while leaving all other components unchanged. Under the assumption $\G_2 \perp\!\!\!\perp_{\mathcal{G}} A, \G_1 \mid \B_2$, the above difference is orthogonal to the IF of the larger adjustment set:
$$
\mathbb{E}_P\left[\psi_a(Y, A, \G_1, \G_2, \B_2;P) \cdot \left(
\psi_a(Y, A, \G_1, \B_2;P) - \psi_a(Y, A, \G_1, \G_2, \B_2;P)
\right)\right] = 0.
$$
Hence, the covariance term vanishes. Substituting this result back into the law of total variance yields:
\begin{align*}
\operatorname{Var}[\psi_a(Y, A, \G_1, \B_2;P)]
&= \operatorname{Var}[\psi_a(Y, A, \G_1, \G_2, \B_2;P)] 
\\& \quad + \operatorname{Var}[\psi_a(Y, A, \G_1, \B_2;P) - \psi_a(Y, A, \G_1, \G_2, \B_2;P)],   
\end{align*}
where the second term is nonnegative. This establishes
% Therefore, we have shown that
$\sigma^2_{a,(\G_1, \B_2)}(P) \geq \sigma^2_{a,(\G_1, \G_2, \B_2)}(P),$
showing that including $\G_2$ either improves or preserves estimation efficiency.

\begin{figure}[t]
\centering

% 左图
\begin{minipage}{0.48\linewidth}
\centering
\scalebox{0.9}[0.9]{%
\begin{tikzpicture}[
    node distance=0.7cm and 1cm,
    every node/.style={draw, ellipse, align=center, minimum width=0.88cm, minimum height=0.6cm, font=\normalsize},
    arrow/.style={-Latex, thick} 
]
\node (A) {A};
\node[right=of A] (G1) {$G_1$};
\node[right=of G1] (Y) {$Y$};
\node[above=of G1] (B2) {$B_2$};
\node[above=of Y] (G2) {$G_2$}; 

\draw[arrow] (A) -- (G1); 
\draw[arrow] (G1) -- (Y); 
\draw[arrow] (B2) -- (A);  
\draw[arrow] (B2) -- (G1);
\draw[arrow] (B2) -- (G2); 
\draw[arrow] (G2) -- (Y);
\draw[arrow, bend right=25] (A) to (Y);
\end{tikzpicture}
}
\caption{A DAG illustrating Lemmas \ref{lemma:1}, \ref{lemma:2}.}
\label{fig:sample1}
\end{minipage}
\hfill
% 右图
\begin{minipage}{0.48\linewidth}
\centering
\scalebox{0.8}[0.8]{%
\begin{tikzpicture}[
    node distance=0.5cm and 0.9cm,
    every node/.style={draw, ellipse, align=center, minimum width=0.88cm, minimum height=0.6cm, font=\normalsize},
    arrow/.style={-Latex, thick} 
]
\node (A) {$A$};
\node[right=of A] (B1) {$B_1$};
\node[right=of B1] (G1) {$G_1$};
\node[right=of G1] (Y) {$Y$}; 
\node[above=of G1] (G2) {$G_2$}; 

\draw[arrow] (A) -- (B1); 
\draw[arrow] (B1) -- (G1); 
\draw[arrow] (G1) -- (Y); 
\draw[arrow] (G2) -- (B1);
\draw[arrow] (G2) -- (A); 
\draw[arrow] (G2) -- (Y);
\draw[arrow, bend right=25] (A) to (Y);
\end{tikzpicture}
}
\caption{A DAG illustrating Lemmas \ref{lemma:3}, \ref{lemma:4}.}
\label{fig:sample2}
\end{minipage}
\vspace{-10pt}
\end{figure}

\textbf{Step 2: Augmenting \texorpdfstring{$\mathbf{Z}_1$}. with Mediating Parents of \texorpdfstring{$Y$}..\;\;}
% \label{subsec:mediator}
% \textbf{Step 2: Augmenting $\mathbf{Z}_1$ with Mediating Parents of $Y$\;\;}
% According to 
Next, Lemma~\ref{lemma:3} (proof in Appendix~\ref{app:proof_lemma3}) states that including mediators that are also parents of $Y$ leads to improved efficiency. Therefore, we augment $\mathbf{Z}_1$ with $\mathbf{X}_{3 \in \Pa(Y)}$:
% \vspace{-5pt}
$$\sigma^2_{(\mathbf{Z}_1 \cup \mathbf{X}_{3 \in \Pa(Y)},\, \mathbf{Z}_2 \cup \mathbf{X}_{1 \in \Pa(Y)} \cup \mathbf{X}_{4 \in \Pa(Y)})}(P) \leq \sigma^2_{(\mathbf{Z}_1,\, \mathbf{Z}_2 \cup \mathbf{X}_{1 \in \Pa(Y)} \cup \mathbf{X}_{4 \in \Pa(Y)})}(P).
$$
% \vspace{-2pt}
\begin{lemma}[Supplement of mediator variables]\label{lemma:3}
Given a DAG $\mathcal{G}=(\V,\mathbf{E})$,
% be a DAG with vertex set $\mathbf{V}$, 
and the exposure-outcome pair $(A,Y)\subseteq \V$,
% Let $\mathcal{G}$ be a DAG with vertex set $\mathbf{V}$, let $A \subset \mathbf{V}$ and $Y \in \mathbf{V} \backslash \mathbf{A}$, where $A$ is a random variable taking values on a finite set.  S
suppose $(\B_1, \G_2) \subset \mathbf{V} \backslash \{A, Y\}$ is a VAS,
% valid adjustment set 
% relative to $(A, Y)$ in $\mathcal{G}$,
% and suppose 
with $\G_1$ 
% is
disjoint from both $\B_1$ and $\G_2$, satisfying
$\G_1 \perp\!\!\!\perp_{\mathcal{G}} A, \G_2 \mid \B_1.
$
Then $(\G_1, \B_1, \G_2)$ is also a VAS, and for all $P \in \mathcal{M}$,
$
\sigma_{(\B_1, \G_2)}^2(P) - \sigma_{(\G_1, \B_1, \G_2)}^2(P) \geq 0.
$
\end{lemma}
% \vspace{-5pt}
Figure~\ref{fig:sample2} provides a simplified illustration of the conditions stated in Lemma~\ref{lemma:3}. The proof proceeds analogously to that of Lemma~\ref{lemma:1}, with $\G_1$ playing the same role as $\G_2$.

\textbf{Step 3: Pruning Redundant Backdoor Variables from \texorpdfstring{$\Z_2$. \;\;}
% Deletion of Overadjustment Backdoor Variables
}
% \label{subsec:overajustment_backdoor}
% \paragraph{Step 3: Pruning $\mathbf{Z}_2$ of Redundant Variables}
Lemma~\ref{lemma:2} (proof in Appendix~\ref{app:proof_lemma2}) implies that unnecessary variables in $\mathbf{Z}_2$ (those not in $\mathbf{X}_{1 \in \Pa(Y)}$ or $\mathbf{X}_{4 \in \Pa(Y)}$) can be removed without increasing variance:
% \vspace{-5pt}
$$
\sigma^2_{(\mathbf{Z}_1 \cup \mathbf{X}_{3 \in \Pa(Y)},\, \mathbf{X}_{1 \in \Pa(Y)} \cup \mathbf{X}_{4 \in \Pa(Y)})}(P) \leq \sigma^2_{(\mathbf{Z}_1 \cup \mathbf{X}_{3 \in \Pa(Y)},\, \mathbf{Z}_2 \cup \mathbf{X}_{1 \in \Pa(Y)} \cup \mathbf{X}_{4 \in \Pa(Y)})}(P).
$$

\begin{lemma}[Deletion of overadjustment backdoor variables]\label{lemma:2}
Given a DAG $\mathcal{G}=(\V,\mathbf{E})$,
% be a DAG with vertex set $\mathbf{V}$, 
and the exposure-outcome pair $(A,Y)\subseteq \V$, and
% Let $\mathcal{G}$ be a DAG with vertex set $\mathbf{V}$, let $A \subset \mathbf{V}$ and $Y \in \mathbf{V} \backslash \mathbf{A}$, where $A$ is a random variable taking values on a finite set.  S
suppose $(\G_1,\G_2 \cup \B_2) \subset \mathbf{V} \backslash \{A, Y\}$ is a 
VAS,
% valid adjustment set 
% relative to $(A, Y)$ in $\mathcal{G}$,
with $\G$ and $\B$ disjoint, and suppose
$
Y \perp\!\!\!\perp_{\mathcal{G}} \B_2 \mid \G_1, \G_2, A.
$
Then $(\G_1, \G_2)$ is also a VAS,
% valid adjustment set 
% relative to $(A, Y)$ in $\mathcal{G}$,
and for all $P \in \mathcal{M}$,
$\sigma_{(\G_1, \G_2, \B_2)}^2(P) - \sigma_{( \G_1, \G_2)}^2(P) \geq 0.$
\end{lemma}
% \vspace{-5pt}
\textbf{Proof Sketch of Lemma~\ref{lemma:2}.\;\;}
Figure~\ref{fig:sample1} provides a simplified illustration of the conditions stated in Lemma~\ref{lemma:2}. Again,  we focus on comparing the asymptotic variance of  $T_a(\Z)$.
We begin by applying the law of total variance to decompose the variance under the larger adjustment set $(\G_1, \G_2, \B_2)$:
$$
\begin{aligned}
\operatorname{Var}[\psi_a(Y, A, \G_1, \G_2, \B_2;P)] 
&= \operatorname{Var}[\mathbb{E}[\psi_a(Y, A, \G_1, \G_2, \B_2;P) \mid A, Y, \G_1, \G_2]] \\
&\quad + \mathbb{E}\left[\operatorname{Var}[\psi_a(Y, A, \G_1, \G_2, \B_2;P) \mid A, Y, \G_1, \G_2]\right].
\end{aligned}
$$
Now, under the conditional independence assumption  
$Y \perp\!\!\!\perp_{\mathcal{G}} \B_2 \mid \G_1, \G_2, A,$  
the IF based on the adjustment set $(\G_1, \G_2, \B_2)$ satisfies  
$
\mathbb{E}_P[\psi_a(Y, A, \G_1, \G_2, \B_2;P) \mid A, Y, \G_1, \G_2]
= \psi_a(Y, A, \G_1, \G_2;P).
$  
Substituting this into the variance decomposition yields:  
$$
\begin{aligned}
&\operatorname{Var}[\psi_a(Y, A, \G_1, \G_2, \B_2;P)]\\
=& \operatorname{Var}[\psi_a(Y, A, \G_1, \G_2;P)] + \mathbb{E}\left[\operatorname{Var}[\psi_a(Y, A, \G_1, \G_2, \B_2;P) \mid A, Y, \G_1, \G_2]\right].
\end{aligned}
$$  
Since the conditional variance term is nonnegative, we conclude that  
$
\sigma^2_{a,(\G_1, \G_2, \B_2)}(P) \geq \sigma^2_{a,(\G_1, \G_2)}(P),
$  
implying that $\B_2$ can be safely excluded without increasing the asymptotic variance.

\textbf{Step 4: Pruning Redundant Mediating Variables from \texorpdfstring{$\Z_1$}..\;\;}
% \label{subsec:overadjustment_mediator}
% \paragraph{Step 4: Pruning $\mathbf{Z}_1$ of Redundant Mediators}
Finally, Lemma~\ref{lemma:4} (proof in Appendix~\ref{app:proof_lemma4}) shows that variables in
$\mathbf{Z}_1 \setminus \mathbf{X}_{3 \in \Pa(Y)}$
% $\mathbf{Z}_1$ not included in $\mathbf{X}_{3 \in \Pa(Y)}$
can be removed:
\[
\sigma^2_{(\mathbf{X}_{3 \in \Pa(Y)},\, \mathbf{X}_{1 \in \Pa(Y)} \cup \mathbf{X}_{4 \in \Pa(Y)})}(P) \leq \sigma^2_{(\mathbf{Z}_1 \cup \mathbf{X}_{3 \in \Pa(Y)},\, \mathbf{X}_{1 \in \Pa(Y)} \cup \mathbf{X}_{4 \in \Pa(Y)})}(P).
\]
Therefore, the final adjustment set minimizing the asymptotic variance is given by
\[
\mathbf{O} := \mathbf{X}_{1 \in \Pa(Y)} \cup \mathbf{X}_{3 \in \Pa(Y)} \cup \mathbf{X}_{4 \in \Pa(Y)}.
\]

\begin{lemma}[Deletion of overadjustment mediator variables]\label{lemma:4}
Given a DAG $\mathcal{G}=(\V,\mathbf{E})$,
% be a DAG with vertex set $\mathbf{V}$, 
and the exposure-outcome pair $(A,Y)\subseteq \V$, 
% Let $\mathcal{G}$ be a DAG with vertex set $\mathbf{V}$, let $A \subset \mathbf{V}$ and $Y \in \mathbf{V} \backslash \mathbf{A}$, where $A$ is a random variable taking values on a finite set.  S
suppose $(\G_1 \cup \B_1, \G_2) \subset \mathbf{V} \backslash \{A, Y\}$ is a 
VAS
% valid adjustment set 
% relative to $(A, Y)$ in $\mathcal{G}$
with $\G$ and $\B$ disjoint, and suppose
$Y \perp\!\!\!\perp_{\mathcal{G}} \B_1 \mid \G_1, \G_2, A.$
Then $(\G_1, \G_2)$ is also a VAS,
% valid adjustment set relative to $(A, Y)$ in $\mathcal{G}$,
and for all $P \in \mathcal{M}$,
$\sigma_{(\G_1, \B_1, \G_2)}^2(P) - \sigma_{(\G_1, \G_2)}^2(P) \geq 0.$
\end{lemma}

Figure~\ref{fig:sample2} provides a simplified illustration of the conditions stated in Lemma~\ref{lemma:4}. The proof follows the same argument as Lemma~\ref{lemma:2}, with $\B_1$ playing the same role as $\B_2$.

% \vspace{-5pt}
\section{Experiments}\label{sec: experiments}
% \vspace{-5pt}

% To validate the robustness of our procedure under finite samples, 
To evaluate the robustness of our method under finite-sample conditions, we conduct synthetic experiments to verify 
Lemmas~\ref{lemma:1}-\ref{lemma:4}, corresponding to Figures~\ref{fig:sample1} and \ref{fig:sample2}, and further demonstrate the practical stability of the O-set using real-world data.\footnote{Code available at \url{https://github.com/Lin-Ruiyang/WCDE-Simulation}}

\textbf{AIPW Estimator\;\;}
In our experiments, we implemented the AIPW estimator, constructed using the plug-in components of the influence function: 
\begin{align}
&\widehat{\mathrm{WCDE}} = \frac{1}{2n} \sum_{i=1}^n \Bigg\{ 
\underbrace{
\frac{ \mathbb{I}\{A_i = a\} \cdot \hat{p}(\mathbf{Z}_{1i}) \cdot \hat{p}(\mathbf{Z}_{2i})}
{\hat{p}(\mathbf{Z}_{1i}, \mathbf{Z}_{2i}, a)} 
\cdot \left( Y_i - \hat{\mu}(a, \mathbf{Z}_{1i}, \mathbf{Z}_{2i}) \right)
}_{\text{IPW term at level } a}  + 
\underbrace{
\sum_{\mathbf{z}_2} \hat{\mu}(a, \mathbf{Z}_{1i}, \mathbf{z}_2) \cdot \hat{p}(\mathbf{z}_2)
}_{\text{Marginal over } \mathbf{Z}_2 \text{ at } a}
\notag\\
&\quad +
\underbrace{
\sum_{\mathbf{z}_1} \hat{\mu}(a, \mathbf{z}_1, \mathbf{Z}_{2i}) \cdot \hat{p}(\mathbf{z}_1)
}_{\text{Marginal over } \mathbf{Z}_1 \text{ at } a}
\Bigg\} 
- \frac{1}{2n} \sum_{i=1}^n \Bigg\{ 
\underbrace{
\frac{ \mathbb{I}\{A_i = a^*\} \cdot \hat{p}(\mathbf{Z}_{1i}) \cdot \hat{p}(\mathbf{Z}_{2i})}
{\hat{p}(\mathbf{Z}_{1i}, \mathbf{Z}_{2i}, a^*)} 
\cdot \left( Y_i - \hat{\mu}(a^*, \mathbf{Z}_{1i}, \mathbf{Z}_{2i}) \right)
}_{\text{IPW term at level } a^*} \notag\\
&\quad + 
\underbrace{
\sum_{\mathbf{z}_2} \hat{\mu}(a^*, \mathbf{Z}_{1i}, \mathbf{z}_2) \cdot \hat{p}(\mathbf{z}_2)
}_{\text{Marginal over } \mathbf{Z}_2 \text{ at } a^*}
+
\underbrace{
\sum_{\mathbf{z}_1} \hat{\mu}(a^*, \mathbf{z}_1, \mathbf{Z}_{2i}) \cdot \hat{p}(\mathbf{z}_1)
}_{\text{Marginal over } \mathbf{Z}_1 \text{ at } a^*}
\Bigg\}.\label{eq:aipw}
\end{align}
To estimate the conditional expectations $\hat{\mu}(a, \mathbf{Z}_1, \mathbf{Z}_2)$, we fit linear regressions with spline-transformed features (degree 5, 10 knots) for continuous outcomes,
% \Lry{, 
and use multinomial logistic regression for discrete outcomes.  
The treatment probabilities $\hat{p}(A=a \mid \mathbf{Z}_1, \mathbf{Z}_2)$ are obtained from logistic regressions. 
% \Lry
{The marginal and joint densities are estimated nonparametrically, using empirical frequencies for discrete variables and Gaussian kernel density estimation for continuous ones.}

\textbf{Synthetic Experiments\;\;}
Under Figures~\ref{fig:sample1} and \ref{fig:sample2}, data are generated from a nonlinear structural equation model with random edge coefficients, nonlinear transformations, and additive Gaussian noise. Appendix~\ref{app:synthetic_DGP} contains full details of the data-generating process and estimator implementation.
% To evaluate the robustness of WCDE estimation, we repeat the data generation process over 100 independent sets of randomly sampled coefficients.
% we conducted additional experiments and will add them to our revised paper. We construct two 7-node DAGs with multiple causal pathways, corresponding to the scenarios in Lemmas~1--4 (Figures~3 and~4). Our data are generated as follows:
% 
% \paragraph{Estimation Method.}
% We implement an AIPW estimator with:
% \begin{itemize}
%   \item \textbf{Q-model}: linear regression with spline-transformed features (degree 4, 20 knots),
%   \item \textbf{g-model}: logistic regression for the treatment model.
% \end{itemize}
% The detailed form of the AIPW is included in the response to Reviewer Nm1A (Question~3).
% 
% \paragraph{Evaluation Metrics.}
For each adjustment set, we report average variance and average MSE across 50 replications. Table~\ref{table:combined_synthetic} top and bottom respectively 
% corresponds to Figure~\ref{fig:sample1} and
illustrate the results associated with Lemmas~\ref{lemma:1} and~\ref{lemma:2}, and  
% Table~\ref{table:synthetic2} 
% % corresponds to Figure~\ref{fig:sample2}, which
% supports the theoretical findings in 
Lemmas~\ref{lemma:3} and~\ref{lemma:4}.
Our proposed O-set $\{G_1, G_2\}$ consistently achieves the lowest variance and MSE across all sample sizes, ranging from
% as small as
$n=250$ to 
% as large as
$n=2000$,
demonstrating its finite-sample robustness despite our theoretical guarantees being asymptotic. 
\vspace{-15pt}  
\begin{table}[H]
\centering
\caption{Average Variance and MSE for Figures~\ref{fig:sample1} and~\ref{fig:sample2}. Optimal VAS are highlighted in bold.}
\label{table:combined_synthetic}
\resizebox{\textwidth}{!}{
\begin{tabular}{llcccccccc}
\toprule
& & \multicolumn{2}{c}{$n=250$} & \multicolumn{2}{c}{$n=500$} & \multicolumn{2}{c}{$n=1000$} & \multicolumn{2}{c}{$n=2000$} \\
\cmidrule(lr){3-4} \cmidrule(lr){5-6} \cmidrule(lr){7-8} \cmidrule(lr){9-10}
\textbf{Figure} & \textbf{Adjust. Set} & Var & MSE & Var & MSE & Var & MSE & Var & MSE \\
\midrule
\multirow{3}{*}{Fig~\ref{fig:sample1}}
& $\mathbf{G_1, G_2}$ & \textbf{0.0029} & \textbf{0.0030} & \textbf{0.0016} & \textbf{0.0016} & \textbf{0.0009} & \textbf{0.0009} & \textbf{0.0006} & \textbf{0.0006} \\
& $G_1, B_2$ & 0.0102 & 0.0103 & 0.0110 & 0.0112 & 0.0101 & 0.0102 & 0.0075 & 0.0076 \\
& $G_1, G_2, B_2$ & 0.0050 & 0.0051 & 0.0028 & 0.0029 & 0.0021 & 0.0022 & 0.0016 & 0.0016 \\
\midrule
\multirow{3}{*}{Fig~\ref{fig:sample2}}
& $\mathbf{G_1, G_2}$ & \textbf{0.0027} & \textbf{0.0028} & \textbf{0.0015} & \textbf{0.0015} & \textbf{0.0010} & \textbf{0.0010} & \textbf{0.0006} & \textbf{0.0006} \\
& $B_1, G_2$ & 0.0100 & 0.0101 & 0.0072 & 0.0073 & 0.0066 & 0.0067 & 0.0050 & 0.0050 \\
& $G_1, B_1, G_2$ & 0.0059 & 0.0059 & 0.0040 & 0.0041 & 0.0040 & 0.0040 & 0.0029 & 0.0029 \\
\bottomrule
\end{tabular}
}
\vspace{-\baselineskip}
\end{table}
\vspace{-20pt}
\begin{table}[H]
\centering
\caption{Variance and MSE of WCDE estimates on the ASIA network under different sample sizes. Variable names are abbreviated as follows: $b$: bronc, $s$: smoke, $l$: lung.}
\label{tab:asia_combined}
\resizebox{\textwidth}{!}{
\begin{tabular}{lccccccccccc}
\toprule
& & \multicolumn{2}{c}{$n=250$} & \multicolumn{2}{c}{$n=500$} & \multicolumn{2}{c}{$n=1000$} & \multicolumn{2}{c}{$n=4000$} & \multicolumn{2}{c}{$n=10000$} \\
\cmidrule(lr){3-4} \cmidrule(lr){5-6} \cmidrule(lr){7-8} \cmidrule(lr){9-10} \cmidrule(lr){11-12}
\textbf{Adj. Set} & & Var & MSE & Var & MSE & Var & MSE & Var & MSE & Var & MSE \\
\midrule
{[}$\mathbf{b}${]} & & \textbf{0.0143} & \textbf{0.0158} & \textbf{0.0049} & \textbf{0.0052} & \textbf{0.00266} & \textbf{0.00292} & \textbf{0.00062} & \textbf{0.00067} & \textbf{0.00027} & \textbf{0.00029} \\
{[}$s${]} & & 0.0210 & 0.0238 & 0.00862 & 0.00911 & 0.00408 & 0.00454 & 0.00101 & 0.00104 & 0.00046 & 0.00054 \\
{[}$b$, $s${]} & & 0.0176 & 0.0229 & 0.00628 & 0.00728 & 0.00419 & 0.00480 & 0.00087 & 0.00091 & 0.00044 & 0.00054 \\
{[}$b$, $l$, $s${]} & & 0.0296 & 0.0528 & 0.0174 & 0.0231 & 0.0141 & 0.0153 & 0.00333 & 0.00335 & 0.00177 & 0.00194 \\
{[}$b$, $l${]} & & 0.0360 & 0.0540 & 0.0207 & 0.0248 & 0.0133 & 0.0139 & 0.00340 & 0.00342 & 0.00170 & 0.00183 \\
{[}$l${]} & & 0.0532 & 0.0624 & 0.0205 & 0.0215 & 0.0133 & 0.0137 & 0.00346 & 0.00347 & 0.00184 & 0.00193 \\
{[}$l$, $s${]} & & 0.0458 & 0.0613 & 0.0218 & 0.0240 & 0.0151 & 0.0157 & 0.00367 & 0.00368 & 0.00189 & 0.00204 \\
\bottomrule
\end{tabular}
}
\vspace{-\baselineskip}
\end{table}

% \vspace{-0.5em}  

% The current tables present preliminary simulation results, with more extensive experiments planned for the final version to strengthen empirical validation and provide a more complete set of results.

\textbf{Real-World Experiments\;\;}  
% We additionally conducted simulations using three widely adopted semi-synthetic Bayesian networks from the \texttt{bnlearn} repository: \textbf{ASIA}, \textbf{SIGNALING}, and \textbf{MILDEW} networks \cite{Lauritzen1988local, Sachs2005causal, Jensen1996midas}, which are commonly used for benchmarking causal inference algorithms due to their realistic structures and domain relevance.
We evaluate
% further evaluated 
our method on three widely used semi-synthetic Bayesian networks from  \texttt{bnlearn}:
% repository: 
\textbf{ASIA}, \textbf{SIGNALING}, and \textbf{MILDEW}, which serve as standard benchmarks for causal inference due to their realistic structures and domain relevance \cite{Jensen1996midas, Lauritzen1988local, Sachs2005causal}.
For each dataset, we validated our theoretical results through a five-step process: (1) selecting a treatment and outcome variable (for multi-valued treatments, the two most frequent levels are used to define a binary treatment); (2) identifying all valid adjustment sets for small/medium DAGs and a sufficiently large and representative subset for large DAGs; (3) estimating the WCDE via the AIPW estimator (Eq. \eqref{eq:aipw}); (4) simulating data from the underlying causal model across sample sizes
% For each dataset, we validate our theoretical results by performing the following 5 steps:
% Specifically, for each DAG, 
% (1) we selected a treatment and outcome variable of interest; (2) identified all valid adjustment sets satisfying the identification criterion; (3) estimated the WCDE using the Augmented Inverse Probability Weighting estimator; (4) simulated data from the underlying causal model under sample sizes 
$n = 250$, $500$, $1000$, $4000$, and $10000$; and (5)
computing the empirical variance and MSE for each adjustment set.
% computed the empirical variance and MSE of the estimated WCDE under each adjustment set.

% We applied this procedure to the following network:
The empirical validation across three distinct networks consistently affirms the efficacy of our identification criterion. In the \textbf{ASIA network} (an 8-node 8-edge Bayesian network for medical diagnosis), our criterion identified \textbf{bronc} as the optimal set (O-set) for estimating the effect of a single composite variable---indicating the presence of either tuberculosis or lung cancer---on \textbf{dysp} (shortness of breath). 
As shown in Table~\ref{tab:asia_combined}, \textbf{bronc} attains the lowest variance and MSE across all sample sizes, consistent with our asymptotic theory and demonstrating robust finite-sample performance.
% As shown in Table~\ref{tab:asia_combined}, \textbf{bronc} achieves the lowest variance and MSE across all sample sizes,
% % sample sizes of $500$ and above, 
% aligning with
% % when $n = 500, 1000, 4000,$ and $10000$, consistent with 
% our asymptotic theoretical guarantees and also illustrates its robust performance in finite samples. 
% Even at a small sample size ($n=250$), where its variance is not the smallest, \textbf{bronc} achieves the l
% When $n = 250$, although its variance is not the smallest among all adjustment sets, it still yields one of the lowest MSE values, likely due to its relatively low bias compared to other sets, highlighting its robustness even in small-sample settings. 
In the \textbf{SIGNALING network} (a 11-node 17-edge DAG on protein signaling cascades), our criterion identifies 
% , a biological pathway DAG representing protein signaling cascades with 11 nodes and 17 edges, our method selects 
\textbf{\{PKA, Erk\}} as the O-set for estimating the effect of \textbf{PKC} on \textbf{Akt}. 
% As shown in 
Table~\ref{tab:signaling_combined} 
% The empirical results 
illustrates a nuanced bias-variance trade-off. For larger sample sizes ($n \geq 1000$), our O-set achieved the smallest variance among all adjustment sets, though its MSE was not the absolute lowest. When the sample size was $500$ or smaller, the O-set, while not achieving the very lowest variance or MSE, delivered performance that was highly comparable to the best-performing sets, demonstrating its practical utility and robustness even in finite-sample regimes where strict optimality is not guaranteed.
% this O-set performs very well for $n =1000, 4000,$ and $10000$, while remaining competitive and close to the best-performing adjustment set even when $n = 250, 500$, where the asymptotic optimality may not strictly hold. 
Similar to the ASIA network,
in \textbf{MILDEW} (a 35-node 46-edge DAG on crop disease management), the identified O-set (\textbf{\{meldug\_3, middel\_3, mikro\_3\}} for estimating the effect of \textbf{mikro\_1} on \textbf{meldug\_4}) 
% \Lry{
achieves the smallest variance and MSE when $n \ge 500$, 
and remains among the best-performing sets for smaller samples (Tables~\ref{tab:mildew_combined} and \ref{tab:mildew_top50_combined_allabbr} in Appendix~\ref{app:realworld_DAG}). 

\vspace{-1.5em} 

\begin{table}[H]
\centering
\caption{Variance and MSE of WCDE estimates on the SIGNALING network under different sample sizes. Variable names are abbreviated as follows: $e$: Erk, $p$: PKA, $m$: Mek, $r$: Raf.}
\label{tab:signaling_combined}
\resizebox{\textwidth}{!}{
\begin{tabular}{lccccccccccc}
\toprule
& & \multicolumn{2}{c}{$n=250$} & \multicolumn{2}{c}{$n=500$} & \multicolumn{2}{c}{$n=1000$} & \multicolumn{2}{c}{$n=4000$} & \multicolumn{2}{c}{$n=10000$} \\
\cmidrule(lr){3-4} \cmidrule(lr){5-6} \cmidrule(lr){7-8} \cmidrule(lr){9-10} \cmidrule(lr){11-12}
\textbf{Adj. Set} & & Var & MSE & Var & MSE & Var & MSE & Var & MSE & Var & MSE \\
\midrule
{[}$\mathbf{e}$, $\mathbf{p}${]} & & 0.00491 & 0.00618 & 0.00269 & 0.00328 & \textbf{0.00125} & \textbf{0.00134} & \textbf{0.00033} & \textbf{0.00036} & \textbf{0.00012} & \textbf{0.00012} \\
{[}$e$, $p$, $r${]} & & 0.00483 & 0.00569 & 0.00293 & 0.00344 & 0.00156 & 0.00164 & 0.00041 & 0.00044 & 0.00019 & 0.00019 \\
{[}$e$, $m$, $p${]} & & \textbf{0.00429} & \textbf{0.00476} & \textbf{0.00238} & \textbf{0.00260} & 0.00142 & 0.00146 & 0.00062 & 0.00064 & 0.00020 & 0.00020 \\
{[}$e$, $m$, $p$, $r${]} & & 0.00456 & 0.00490 & 0.00272 & 0.00289 & 0.00159 & 0.00160 & 0.00065 & 0.00067 & 0.00020 & 0.00020 \\
{[}$m$, $p$, $r${]} & & 0.00843 & 0.01024 & 0.00486 & 0.00586 & 0.00383 & 0.00407 & 0.00205 & 0.00217 & 0.00105 & 0.00105 \\
{[}$m$, $p${]} & & 0.00752 & 0.00941 & 0.00408 & 0.00534 & 0.00351 & 0.00386 & 0.00201 & 0.00213 & 0.00108 & 0.00108 \\
\bottomrule
\end{tabular}
}
\end{table}

\vspace{-2em}

% \vspace{-2.5em} 
\vspace{-5pt}
\section{Discussion}\label{sec:discussions}
\vspace{-5pt} 
We provide the first comprehensive framework for identifying and estimating the WCDE in observational settings, where
% which we define as the weighted CDE while intervening to fix
\emph{all} observed mediators are fixed. We leave the analysis of more complex WCDE parameters, where only a subset of mediators is fixed, to future work. WCDE introduces unique challenges stemming from the integration over mediator distributions and the dual role of mediators as both colliders and effect modifiers. To address these challenges, we first establish a graphical criterion that characterizes VASs for WCDE. Second, we derive the unique IF of WCDE under nonparametric models, enabling principled comparison of valid adjustment strategies. Third, we identify the optimal VAS that minimizes asymptotic variance, showing that it systematically differs from those used for ATE estimation due to mediator–confounder interactions. Together, our identification results and efficiency analysis lay the foundation for extending WCDE estimation to more complex settings
% , including fairness-aware decision-making and high-dimensional causal inference,
where careful handling of mediator–confounder structure is essential.

{While we assume that the underlying DAG is known and accurate, in practice, the DAG must often be learned from data, a challenge actively studied in the causal discovery literature. Global discovery algorithms can, in principle, recover DAGs from observational data, but they often suffer from poor finite-sample performance or exponential runtime in the worst case \cite{he2008active, kalisch2007estimating}. As a computationally efficient alternative, local discovery algorithms focus on the relative relationships between exposure and outcome, bypassing the need to recover the full DAG \cite{maasch2025local, maasch2024local}. \citet{maasch2025local} proposes a local discovery method tailored for WCDE estimation under suitable assumptions, showing polynomial runtime while allowing for unobserved confounders that do not directly affect the outcome. Our results further establish that the set returned by their procedure is asymptotically optimal in their problem setting. Another promising approach is to assume an \emph{additive noise model}, under which global discovery algorithms are provably consistent in polynomial time \cite{hiremath2025losam,hiremath2024hybrid,hoyer2008nonlinear}, potentially enabling the efficient identification of adjustment sets. While these alternatives offer practical avenues for learning the O-set from data, a key open challenge is to understand how errors from causal discovery propagate into WCDE estimation. Developing robust frameworks that integrate structure learning with optimal adjustment set selection remains an important direction for future work.}

\section*{Acknowledgments}
The authors wish to thank Ezequiel Smucler and 
Stephan Poppe
for insightful discussions that helped improve this work. Y.G. is supported by NSF DMS-2515285. This work was supported in part by an AWS Cloud Credit Grant from Cornell's Center for Data Science for Enterprise and Society.

% Methodologically, we are conducting simulation studies to evaluate the finite-sample performance of our estimators under various data-generating scenarios, including high-dimensional and nonlinear settings. 
% On the applied side, we are applying our framework to fairness-aware policy evaluation and mediation-based discrimination testing, where isolating direct effects is essential for understanding and mitigating structural bias in decision-making systems.

% \section*{References}
\bibliography{citation}
\medskip

{
\small

}

%%%%%%%%%%%%%%%%%%%%%%%%%%%%%%%%%%%%%%%%%%%%%%%%%%%%%%%%%%%%
\newpage
\setcounter{figure}{0} 
\renewcommand{\thefigure}{\Alph{section}.\arabic{figure}}
\setcounter{table}{0} 
\renewcommand{\thetable}{\Alph{section}.\arabic{table}}
\appendix
% \section*{Appendix}
\begin{center}
	\Large \textbf{Appendix}
\end{center}

\etocdepthtag.toc{mtappendix}
\etocsettagdepth{mtmainpaper}{none}
\etocsettagdepth{mtappendix}{subsubsection}
\begingroup
\parindent=0em
\begin{center}
\large\textbf{Table of contents}
\vspace{-15pt}
\end{center}
\etocsettocstyle{\rule{\linewidth}{1pt}\vskip0.5\baselineskip}{\rule{\linewidth}{1pt}}
%\etocsettocstyle{}{}
\tableofcontents 
\endgroup

% \noindent\textbf{Appendix Contents}
% \vspace{0.5em}

% \begin{itemize}
%   \item \hyperref[appendix:sec3]{Section~\ref*{appendix:sec3}: Proofs and Extended Discussions for Sections 2 and 3}
%   \begin{itemize}
%     \item \hyperref[app:criteria]{Section~\ref*{app:criteria}: Proof of Lemma \ref{lemma:adjustment_criterion}}
%     \item \hyperref[app:IF_dicussion]{Section~\ref*{app:IF_dicussion}: Discussions on Influence Functions}
%     \item \hyperref[app:proof_thm_IF]{Section~\ref*{app:proof_thm_IF}: Proof of Theorem \ref{thm:IF}}
%   \end{itemize}
%   \item \hyperref[appendix:sec4]{Section~\ref*{appendix:sec4}: Proofs of Results in Section 4}
%     \begin{itemize}
%     \item \hyperref[app:proof_lemma1]{Section~\ref*{app:proof_lemma1}: Proof of Lemma \ref{lemma:1}}
%     \item \hyperref[app:proof_lemma2]{Section~\ref*{app:proof_lemma2}: Proof of Lemma \ref{lemma:2}}
%     \item \hyperref[app:proof_lemma3]{Section~\ref*{app:proof_lemma3}: Proof of Lemma \ref{lemma:3}}
%     \item \hyperref[app:proof_lemma4]{Section~\ref*{app:proof_lemma4}: Proof of Lemma \ref{lemma:4}}
%     \item \hyperref[app:proof_thm_optimal]{Section~\ref*{app:proof_thm_optimal}: Proof of Theorem \ref{thm:optimal_adjustment}}
%   \end{itemize}
% \end{itemize}
% \section{Main Proofs}
% Technical appendices with additional results, figures, graphs and proofs may be submitted with the paper submission before the full submission deadline (see above), or as a separate PDF in the ZIP file below before the supplementary material deadline. There is no page limit for the technical appendices.
\section{Definition, Notation and Graph Terminology}
\label{app:notation}
\begin{definition}[d-separation]\label{def:d-separation}
    For disjoint vertex sets $\X$, $\Y$, and $\Z$, we say $\Z$ \textit{d-separates} $\X$ from $\Y$ if every path between $\X$ and $\Y$ is \textit{blocked} by $\Z$. A path is blocked by $\Z$ if it contains either:
\begin{enumerate}[label=(\arabic*), itemsep=0pt, topsep=-3pt, partopsep=0pt, parsep=0pt]
    \item A chain $X_i \to Z_k \to X_j$ or fork $X_i \leftarrow Z_k \to X_j$ with $Z_k \in \Z$, or
    \item A collider $X_i \to Z_k \leftarrow X_j$ where $Z_k \notin \Z$ and none of its descendants are in $\Z$, i.e., $\De(Z_k) \cap \Z = \emptyset$.
\end{enumerate}
\end{definition}

\begin{definition}[Path types]\label{def:path-types}
To ensure consistency with the standard causal inference literature 
\citep{pearl2000causality,perkovic2018complete}, 
we define seven types of paths in a directed acyclic graph (DAG) 
$\mathcal{G}=(\mathcal{V},\mathbf{E})$ connecting treatment $A$ and outcome $Y$:

\begin{enumerate}[label=(\arabic*), itemsep=3pt, topsep=3pt, parsep=0pt, partopsep=0pt]

    \item \textbf{Directed path:}  
    A path in which all arrows point in the same direction, from the starting node to the end node.  
    Example: $A \to B \to C \to Y$.

    \item \textbf{Backdoor path:}  
    Any undirected path between $A$ and $Y$ that starts with an arrow pointing into $A$.  
    Example: $A \leftarrow Z \rightarrow Y$.

    \item \textbf{Mediator path:}  
    Any directed path from $A$ to $Y$ that passes through at least one mediator variable $M\in\mathbf{M}$.  
    Example: $A \to M \to Y$.
% \Lry{
    \item \textbf{Causal path:}  
    A path from $A$ to $Y$ in which all arrows are oriented away from $A$ and toward $Y$.

    \item \textbf{Proper causal path and causal nodes:}  
    A causal path is \emph{proper} (with respect to $A$) if only its first node lies in $A$.  
    The set of all nodes lying on proper causal paths from $A$ to $Y$, excluding $A$ itself,  
    is called the set of \emph{causal nodes} and denoted by $\mathrm{cn}(A,Y,\mathcal{G})$.

    \item \textbf{Non-causal path:}  
    Any path between $A$ and $Y$ that is not causal, i.e., that starts with an arrow pointing into $A$.
% }
\end{enumerate}

Intuitively, directed and causal paths transmit causal influence from $A$ to $Y$,
whereas backdoor and other non-causal paths represent spurious associations 
that must be blocked for unbiased causal estimation.
\end{definition}

% \subsection{Identification of CDE}\label{app:CDE_identification}
\begin{definition}[Identification of the controlled direct effect]\label{def:cde-id}
Following the notation of \citet{VanderWeeleVansteelandt2009}, 
let $Y(a,m)$ denote the potential outcome that would have been observed
had the treatment been set to $A=a$ and the mediator to $M=m$. 
% \Lry{
Let $\C$ denote a (possibly multivariate) set of pre-exposure covariates.
% }
Identification of the CDE relies on the following counterfactual independence assumptions:
\begin{enumerate}[label=(\arabic*), itemsep=0pt, topsep=-3pt, partopsep=0pt, parsep=0pt]
    \item No unmeasured confounding of the treatment–outcome relationship:
    \[
    Y(a,m) \perp\!\!\!\perp A \mid C,
    \]
    \item No unmeasured confounding of the mediator–outcome relationship:
    \[
    Y(a,m) \perp\!\!\!\perp M \mid A,C.
    \] 
    % \label{condition:cde-2}
\end{enumerate}
Under these assumptions, the controlled direct effect of changing $A$ 
from $a$ to $a^\ast$ while holding $M$ fixed at $m$ is identified as
\begin{align}
\mathrm{CDE}(a,a^\ast; m)
&\;\equiv\; \mathbb{E}[Y(a,m) - Y(a^\ast,m)] \notag\\
&=\; \int \!\Big(
\mathbb{E}[Y \mid A=a,\, M=m,\, C=c]
- \mathbb{E}[Y \mid A=a^\ast,\, M=m,\, C=c]
\Big)\, dF_C(c). \label{eq:cde-id}
\end{align}
\end{definition}

% \Lry{
\subsection{Sufficient and Necessary Conditions for CDE identification \ref{condition:cde}}\label{app:cde-id}

% \Lry{
To characterize the identification of the controlled direct effect (CDE), 
we first revisit the general conditions under which a post-intervention density 
involving the $do$-operator can be expressed as a function of observable conditional densities.  
These general graphical concepts---including adjustment sets, forbidden sets, 
and the adjustment criterion---provide the foundational framework for identifying
causal effects such as the CDE.  
\subsubsection{Necessary and Sufficient Adjustment Criteria for Identifying do-probability}
Following \citet{pearl2000causality} and \citet{perkovic2018complete},
we recall the key definitions and the associated soundness and completeness theorem.

\begin{definition}[Adjustment set {\citep[Def.~54]{pearl2000causality, perkovic2018complete}}]\label{def:adjustment-set}
Let $\X$, $\Y$, and $\Z$ be pairwise disjoint node sets in a causal DAG $\mathcal{G}$.
Then $Z$ is an \emph{adjustment set} relative to $(\X,\Y)$ in $\mathcal{G}$ if, for any density
$f$ consistent with $\mathcal{G}$,
\[
f(\mathbf{y} \mid do(\mathbf{x})) =
\begin{cases}
f(\mathbf{y} \mid \mathbf{x}), & \text{if } \mathbf{Z} = \emptyset,\\[6pt]
\displaystyle \int f(\mathbf{y} \mid \mathbf{x}, \mathbf{z})\,f(\mathbf{z})\,d\mathbf{z}, & \text{otherwise.}
\end{cases}
\]
\end{definition}

\begin{definition}[Forbidden set {\citep{perkovic2018complete}}]\label{def:forbidden-set}
Let $\X$ and $\Y$ be disjoint node sets in a causal DAG $\mathcal{G}=(\V,\mathbf{E})$.  
The \emph{forbidden set} relative to $(\X,\Y)$ is defined as
\[
\mathrm{Forb}(\X,\Y,\mathcal{G})
= \bigl\{\, W' \in \V :
W' \in \mathrm{De}(W,\mathcal{G})
\text{ for some } W \in \mathrm{cn}(\X,\Y,\mathcal{G}) \,\bigr\}.
\]
That is, $\mathrm{Forb}(\X,\Y,\mathcal{G})$ contains all descendants of nodes 
that lie on proper causal paths from $\X$ to $\Y$.  
Variables in this set must \emph{not} be adjusted for, 
since conditioning on them may introduce post-treatment bias.
\end{definition}

\begin{definition}[Adjustment criterion {\citep[Def.~55]{perkovic2018complete}}]\label{def:adjustment-criterion}
Let $\X$, $\Y$, and $\Z$ be pairwise disjoint node sets in a DAG $\mathcal{G}$.
Let $\mathrm{Forb}(\X,\Y,\mathcal{G})$ denote the set of all descendants in $\mathcal{G}$
of any $W \notin \X$ that lies on a proper causal path from $\X$ to $\Y$.
Then $\Z$ satisfies the \emph{adjustment criterion} relative to $(\X,\Y)$ in $\mathcal{G}$
if the following two conditions hold:
\begin{enumerate}[label=(A\arabic*), itemsep=3pt]
    \item \textbf{Forbidden set condition:}
    $\Z \cap \mathrm{Forb}(\X,\Y,\mathcal{G}) = \emptyset$;
    \label{condition:adj-forbidden}
    \item \textbf{Blocking condition:}
    all proper non-causal paths from $\X$ to $\Y$ in $\mathcal{G}$ are blocked by $\Z$.
    \label{condition:adj-blocking}
\end{enumerate}
These two conditions together ensure that adjusting for $\Z$ yields an unbiased estimate
of the causal effect of $\X$ on $\Y$.
\end{definition}

\begin{theorem}[Soundness and completeness of the adjustment criterion {\citep[Thm.~56]{perkovic2018complete}}]\label{thm:adjustment-dag}
Let $\X$, $\Y$, and $\Z$ be pairwise disjoint node sets in a causal DAG 
$\mathcal{G}=(\V,\mathbf{E})$.  
Then $\Z$ satisfies the adjustment criterion (Definition~\ref{def:adjustment-criterion})
if and only if $\Z$ is an adjustment set (Definition~\ref{def:adjustment-set}).
\end{theorem}

These definitions and Theorem~\ref{thm:adjustment-dag} jointly provide a 
complete graphical characterization for when a post-intervention quantity such as 
$f(\mathbf{y}\mid do(\mathbf{x}))$ can be expressed through integration over observed densities.  
That is, the adjustment criterion is both \emph{necessary and sufficient} 
for identifying a single interventional distribution in a causal DAG.

\subsubsection{CDE Adjustment Criteria}
In the context of the controlled direct effect, 
this criterion (Def.~\ref{def:adjustment-criterion}) remains \emph{sufficient} for identification.  
Specifically, by letting $\X$ in Definition~\ref{def:adjustment-set} 
denote the union of the treatment and mediator sets, 
$\X = (A,\M)$, and by choosing $\Z$ that satisfies the conditions in 
Definition~\ref{def:adjustment-criterion}, 
we can, according to Theorem~\ref{thm:adjustment-dag}, 
identify the post-intervention distribution 
\[
f(y\mid do(a,\mathbf{m})) 
= \int f(y\mid a,\mathbf{m},\z)\, f(\z)\,d\z,
\]
where the integration is performed over the adjustment variables $\Z$.  
Hence, each component 
$\E[Y(a,\mathbf{m})] = \int y\, f(y\mid do(a,\mathbf{m}))\,dy$
is identifiable, and consequently, the difference 
$\E[Y(a,\mathbf{m})] - \E[Y(a^\ast,\mathbf{m})]$ 
that defines the CDE is also identifiable.

Nevertheless, it is important to note that the CDE involves the 
\emph{difference between two interventional expectations}, 
$\E[Y(a,\mathbf{m})] - \E[Y(a^\ast,\mathbf{m})]$.  
Because of this differencing structure, the adjustment criterion serves as a 
\emph{sufficient but not necessary} condition:  
in certain causal graphs, neither $f(y\mid do(a,\mathbf{m}))$ 
nor $f(y\mid do(a^\ast,\mathbf{m}))$ 
is individually identifiable, 
yet their difference---the CDE---can still be identified 
because non-identifiable components cancel out when taking the contrast. We will illustrate such cases in Figures~\ref{fig:cde-id-counter-example-1} 
and~\ref{fig:cde-id-counter-example-2}.

\vspace{6pt}
In parallel, \citet{vanderweele2011controlled} proposed another set of 
\emph{sufficient} conditions specifically tailored to the identification of the CDE.  
Unlike the adjustment criterion, which applies to arbitrary causal effects defined by a single $do$-operator, 
these conditions explicitly target the treatment--mediator--outcome structure 
and are formulated in terms of conditional independence among potential outcomes:

% }
% \Lry{
\begin{enumerate}[label=(B\arabic*), itemsep=0pt, topsep=-3pt, partopsep=0pt, parsep=0pt]
    \item The potential outcomes $Y(a,m)$ are conditionally independent of $A$ given $\Z$:
    \[
    Y(a,m) \perp\!\!\!\perp A \mid \Z, \quad \text{for all } a,m;
    \]
    \label{condition:backdoor}
    \item The potential outcomes $Y(a,m)$ are conditionally independent of $M$ given $(A,\Z)$:
    \[
    Y(a,m) \perp\!\!\!\perp M \mid A,\Z, \quad \text{for all } a,m.
    \]
    \label{condition:mediator_backdoor}
\end{enumerate}
%\begin{enumerate}[label=(A\arabic*), itemsep=0pt, topsep=-3pt, %partopsep=0pt, parsep=0pt]
%    \item All \emph{backdoor paths} (Def.~\ref{def:path-types}) for $\{A, Y\}$ are blocked by $\Z$; \label{condition:backdoor}
%    \item All backdoor paths for $\{\mathbf{M}, Y\}$ are blocked by $\Z$;\label{condition:mediator_backdoor}
%\end{enumerate}
% We note that Definition~\ref{def:cde-id} provides only a \emph{sufficient} set of conditions for the identification of the controlled direct effect \cite{vanderweele2011controlled}.  
% Specifically, 

These conditions ensure that the CDE, defined as 
$\E[Y(a,m)] - \E[Y(a^*,m)]$, 
can be consistently estimated from observed data.  
Nevertheless, similar to the adjustment criterion, 
they only guarantee sufficiency rather than necessity.

\paragraph{Counterexamples to Necessity.}
Next, we provide counterexamples to illustrate that 
% Indeed, for CDE identification,
Conditions~\ref{condition:adj-forbidden}, \ref{condition:adj-blocking}, 
and~\ref{condition:mediator_backdoor} are all \emph{not necessary} for CDE's identification.  
Consider the DAGs in Figure~\ref{fig:cde-id-counter-example-1} 
and Figure~\ref{fig:cde-id-counter-example-2}.  
In the first DAG, the empty set $\emptyset$ identifies $\mathrm{CDE}(m)$ even though it does not contain $X_1$, which lies on a backdoor path between $M$ and $Y$, thus violating Conditions~\ref{condition:adj-blocking} and~\ref{condition:mediator_backdoor}. 
% \Lry{
Since there is no direct arrow from $A$ to $Y$, the controlled direct effect is zero, i.e., $\mathrm{CDE}(m)=0$.  
When the adjustment set is empty, Equation~\ref{eq:cde-id} reduces to
\[
E(Y\mid A=a,m)-E(Y\mid A=a^\ast,m)
=E(Y\mid m)-E(Y\mid m)=0,
\]
which exactly matches the true value of $\mathrm{CDE}(m)$.  
Hence, the empty set is sufficient to identify $\mathrm{CDE}(m)$ in this graph.
% }
%\kyra{need to explain why this is valid. This is not obvious}

In the second DAG, the set $\{X_2\}$ identifies $\mathrm{CDE}(m)$ even though 
$X_2 \in \mathrm{Forb}(A,Y,\mathcal{G})$, violating Condition~\ref{condition:adj-forbidden}.  
% \Lry{
Since there is again no direct arrow from $A$ to $Y$, the controlled direct effect equals zero, i.e., $\mathrm{CDE}(m)=0$.  
When adjusting for $X_2$, Equation~\ref{eq:cde-id} becomes
\[
E(Y\mid A=a,m,X_2)-E(Y\mid A=a^\ast,m,X_2)
=E(Y\mid m,X_2)-E(Y\mid m,X_2)=0,
\]
which coincides with the true value of $\mathrm{CDE}(m)$.  
Therefore, the set $\{X_2\}$ is also sufficient to identify $\mathrm{CDE}(m)$ in this graph.
% }

\begin{figure}[h]
\centering
\begin{minipage}[b]{0.48\linewidth}
\centering
\caption{Example illustrating Condition \ref{condition:adj-blocking} and \ref{condition:mediator_backdoor} is not necessary for CDE identification.}
\label{fig:cde-id-counter-example-1}
\begin{tikzpicture}[
    node distance=0.7cm and 1cm,
    every node/.style={draw, ellipse, align=center, minimum width=0.88cm, minimum height=0.6cm, font=\normalsize},
    arrow/.style={-Latex, thick} 
]
\node (A) {A};
\node[right=of A] (M) {$M$};
\node[right=of M] (Y) {$Y$};
\node[below=of M] (X1) {$X_1$}; 

\draw[arrow] (A) -- (M); 
\draw[arrow] (M) -- (Y); 
\draw[arrow] (X1) -- (M);
\draw[arrow] (X1) -- (Y);
\end{tikzpicture}
\end{minipage}
\hfill
\begin{minipage}[b]{0.48\linewidth}
\centering
\caption{Example illustrating Condition \ref{condition:adj-forbidden} is not necessary for CDE identification.}
\label{fig:cde-id-counter-example-2}
\begin{tikzpicture}[
    node distance=0.7cm and 1cm,
    every node/.style={draw, ellipse, align=center, minimum width=0.88cm, minimum height=0.6cm, font=\normalsize},
    arrow/.style={-Latex, thick} 
]
\node (A) {A};
\node[right=of A] (M) {$M$};
\node[right=of M] (Y) {$Y$};
\node[below=of M] (X2) {$X_2$}; 

\draw[arrow] (A) -- (M); 
\draw[arrow] (M) -- (Y); 
\draw[arrow] (M) -- (X2);
\draw[arrow] (Y) -- (X2);
\end{tikzpicture}
\end{minipage}
\end{figure}

These examples demonstrate that both frameworks—the adjustment criterion for the $do$-operator 
and VanderWeele’s conditions for the CDE—are \emph{sufficient but not necessary}.  
A complete graphical characterization that is both necessary and sufficient for CDE identification 
has yet to be established.  
% \Lry{
Given that our primary focus is on the WCDE, we leave this theoretical development for future work.
% }

\vspace{6pt}

\subsection{Definitions of Causal Partition Introduced by Maasch et al. [32]
% \citet{maasch2024local}
}\label{app:causal_partition}
A complete description of the eight causal partitions is included in Table~\ref{tab:partitions}.
\begin{table}[H]
    \centering
    % \begin{adjustbox}{max width=0.47\textwidth}
    \begin{tabular}{p{0.3cm} p{8.5cm}}
    % \begin{tabular}{c | l}
    \toprule
    \multicolumn{2}{c}{\fontfamily{cmr}\textsc{Exhaustive and Mutually Exclusive Partitions}} \\
    \midrule
        $\X_1$ &  \textit{Confounders and their proxies}: Non-descendants of $X$ that lie on an active backdoor path between $A$ and $Y$ (Definition \ref{def:path-types}), and their proxies (Definition \ref{def:proxy}). \\
        $\X_2$ &  \textit{Colliders and their proxies}: Non-ancestors of $\{A,Y\}$ with at least one active path to $X$ not mediated by $Y$ and at least one active path to $Y$ not mediated by $A$.\\
        $\X_3$ &  \textit{Mediators and their proxies}: Descendants of $A$ that are ancestors of $Y$, and their proxies (Definition \ref{def:proxy}). \\
        $\X_4$ &  Non-descendants of $Y$ that are marginally dependent on $Y$ but marginally independent of $A$  (Definition \ref{def:z4}). \\
        $\X_5$ &  \textit{Instruments and their proxies}: Non-descendants of $A$ whose causal effect on $Y$ is fully mediated by $A$, and that share no confounders with $Y$ (Definition \ref{def:z5}). \\
        $\X_6$ &  Descendants of $Y$ where all active paths shared with $A$ are mediated by $Y$. \\
        $\X_7$ &  Descendants of $A$ where all active paths shared with $Y$ are mediated by $A$. \\
        $\X_8$ &  All nodes that share no active paths with $A$ nor $Y$. \\
    \midrule
    \end{tabular}
    % \end{adjustbox}
    \caption{{Partitions are formally defined by the path combinations enumerated in Table 3 of \citet{maasch2024local}.}}
    \label{tab:partitions}
    % \vspace{-10pt}
\end{table}

\begin{definition}[Proxy variables in $\X_1$, $\X_2$, and $\X_3$, \citealt{maasch2024local}] \label{def:proxy}
    A proxy variable for $\X_1$, $\X_2$, or $\X_3$ is a member of these partitions that is an ancestor or descendant of another member of its respective partition, such that the proxy is not strictly a confounder, mediator, or collider, but still satisfies the allowable path types for its respective partition (as defined in \citet[Table 3]{maasch2024local}). This includes $\X_3$ that are descended from $\X_3$ that lie on mediator chains, $\X_1$ that are ancestral to $\X_1$ on backdoor paths, etc.
    (\citep[Figure B.3]{maasch2024local}).
\end{definition}

\begin{definition}[Partition $\X_4$, \citealt{maasch2024local}] Partition $\X_4$ encompasses all non-descendants of $Y$ that are marginally dependent on $Y$ but marginally independent of $A$ (Table \ref{tab:partitions}). Given this definition, we observe that any $X_4$ participates in a $v$-structure $A \cdots \rightarrow Y \leftarrow \cdots X_4$. This implies the following:
    \begin{enumerate}[noitemsep,topsep=0pt]
    \item $A$ cannot share active paths with any $X_4$. Thus, $A$ can share no common causes with any $X_4$.
    \item $\X_4$ is conditionally dependent on $A$ given $Y$. This implicitly requires that $A$ and $Y$ are marginally dependent, though they may not be directly adjacent in $\mathcal{G}$.
\end{enumerate}
\label{def:z4}
\end{definition}

\begin{definition} [Instrumental variable, \citet{lousdal_introduction_2018}] Any instrument $X_5$ meets the following criteria: \label{def:z5}
\begin{enumerate}[noitemsep,topsep=0pt]
    \item \textit{Relevance assumption}: $X_5$ is causal for exposure $A$.
    \item \textit{Exclusion restriction}: The effect of instrument $X_5$ on outcome $Y$ is fully mediated by $A$.
    \item \textit{Exchangeability assumption}: $X_5$ and $Y$ do not share a common cause. 
\end{enumerate}
\end{definition}

\section{Proofs and Extended Discussions for Sections 2 and 3}\label{appendix:sec3}
In this section, we provide the technical details and proofs omitted from Sections \ref{sec:WCDE_VAS} and \ref{sec:IF} of the main text. We begin by establishing the sufficiency and necessity of the graphical adjustment criteria for identifying the $\WCDE$, as formalized in Lemma~\ref{lemma:adjustment_criterion}. We then proceed to elaborate on semiparametric foundations, including regular submodels, score functions, tangent spaces, and pathwise differentiability, which lay the groundwork for deriving the influence function of $\WCDE$. The section concludes with a detailed proof of Theorem~\ref{thm:IF}, which characterizes the influence function of $\WCDE$ in terms of its identifying functional.

\subsection{Proof of Lemma \ref{lemma:adjustment_criterion}}\label{app:criteria}
\begin{proof}[Proof of Lemma \ref{lemma:adjustment_criterion}]\label{appendix:adj}

\textbf{Sufficiency of the Graphical Criteria.} We demonstrate that any valid adjustment set $\mathbf{Z}$ satisfying Condition~\ref{condition:adj} identifies the same WCDE (Equation~\eqref{eqution:WCDE}) through two expectation-preserving transformations.  The key insight is that the conditional independence in Criteria~\ref{condition:adj} allows us to safely incorporate the target mediators $\mathbf{M}'$ into the adjustment set without changing the identified effect.
% We first prove that the adjustment criteria are sufficient. Specifically, we show that for any adjustment set $\mathbf{Z}$ satisfying the four adjustment criteria, the expression for $\mathrm{WCDE}_{\mathbf{Z}}$ in Equation~\eqref{eq:WCDE_identify} can be simplified and is equal to Equation~\eqref{eqution:WCDE}.
Let $\mathbf{Z_1}=\mathbf{Z}\cap \mathbf{M}, \mathbf{Z_2}=\mathbf{Z}\setminus \mathbf{M}$, $\C = \Pa(Y) \setminus (\mathbf{M}' \cup \{A\})$.

% \mathbf{Z_1  ^{\prime}}= \mathbf{Z_1}\cup \mathbf{M  ^{\prime}}, \mathbf{Z_2  ^{\prime}}=\mathbf{Z_2}\cup \Pa(Y) \setminus {(\mathbf{M ^{\prime}}\cup A)}$. 

% \Lry{
\paragraph{Step 1: Replacing $\mathbf{Z}_2$ with $\C$ Including Non-Mediating Parents of $Y$} 

Since $\C$ satisfies the graphical conditions for identifying the CDE with mediator $\mathbf{Z}_1$ (Definition~\ref{def:cde-id}), yielding:
\begin{align*}
& \E_{\mathbf{Z}_1}\!\left\{\E_{\mathbf{Z}_2}\!\left[\E_Y(Y\mid A=a,\mathbf{Z}_1,\mathbf{Z}_2)\right]\!\right\}
- \E_{\mathbf{Z}_1}\!\left\{\E_{\mathbf{Z}_2}\!\left[\E_Y(Y\mid A=a^*,\mathbf{Z}_1,\mathbf{Z}_2)\right]\!\right\}\\
&= \E_{\mathbf{Z}_1}\!\Big(
\E[Y\mid\operatorname{do}(A=a,\mathbf{Z}_1)] 
- \E[Y\mid\operatorname{do}(A=a^*,\mathbf{Z}_1)]
\Big)\\
&= \E_{\mathbf{Z}_1}\!\left\{\E_{\C}\!\left[\E_Y(Y\mid A=a,\mathbf{Z}_1,\C)\right]\!\right\}
- \E_{\mathbf{Z}_1}\!\left\{\E_{\C}\!\left[\E_Y(Y\mid A=a^*,\mathbf{Z}_1,\C)\right]\!\right\}.
\end{align*}

\paragraph{Step 2: Incorporating Mediators in $\M'$} Define $\mathbf{Z}_1^{\prime} = \mathbf{Z}_1 \cup \mathbf{M}'$.
% By the fourth condition 
We now show that augmenting the conditioning set with $\mathbf{M}' \setminus \mathbf{Z}_1$ does not change the outer expectation over $\mathbf{Z}_1$. Let $\mathbf{W} = \mathbf{M}' \setminus \mathbf{Z}_1$. so we have:
%The key observation is that both $\mathbf{Z}_1 \cup \C$ and $\mathbf{Z}_1^{\prime} \cup \C$ both satisfy Condition \ref{condition:adj}. This is because:
%1) $\mathbf{W} \subseteq \mathbf{M}' \subseteq \Pa(Y)$ and consists of non-descendants of $A$,
%2) Adding $\mathbf{W}$ cannot open new paths (only helps block mediator paths),
%3) Both sets satisfy Criterion \ref{condition:backdoor}-\ref{condition:unique_WCDE}.
%Since both are valid adjustment sets, they must both yield the same interventional expectation:
% , we can augment $\mathbf{Z}_1$ with $\mathbf{M}^{\prime}$ without affecting the outer expectation:

\begin{align*}
&E_{\mathbf{Z}_1} \!\left\{ \E_{\C} \!\left[ \E_{Y}(Y \mid A=a, \mathbf{Z}_1, \C) \right] \right\}
    - \E_{\mathbf{Z}_1} \!\left\{ \E_{\C} \!\left[ \E_{Y}(Y \mid A=a^*, \mathbf{Z}_1, \C) \right] \right\}\\
    & =\E_{\C} \!\left\{ \E_{\mathbf{Z}_1}   \!\big( \E_{Y}(Y \mid A=a, \mathbf{Z}_1, \C) \big) \right\}  
    - \E_{\C} \!\left\{ \E_{\mathbf{Z}_1} \!\big( \E_{Y}(Y \mid A=a^*, \mathbf{Z}_1, \C) \big)  \right\} \\
    &= \E_{\C} \!\left\{ \E_{\mathbf{Z}_1} \!\left[ \E_{\mathbf{W}\mid \mathbf{Z}_1,\C,A} \!\big( \E_{Y}(Y \mid A=a, \mathbf{Z}_1, \mathbf{W}, \C) \big) \right] \right\}  
    - \E_{\C} \!\left\{ \E_{\mathbf{Z}_1} \!\left[ \E_{\mathbf{W}\mid \mathbf{Z}_1,\C,A} \!\big( \E_{Y}(Y \mid A=a^*, \mathbf{Z}_1, \mathbf{W}, \C) \big) \right] \right\} \\
    &= \E_{\C} \!\left\{ \E_{\mathbf{Z}_1} \!\left[ \E_{\mathbf{W}\mid \mathbf{Z}_1} \!\big( \E_{Y}(Y \mid A=a, \mathbf{Z}_1, \mathbf{W}, \C) \big) \right] \right\}  
    - \E_{\C} \!\left\{ \E_{\mathbf{Z}_1} \!\left[ \E_{\mathbf{W}\mid \mathbf{Z}_1} \!\big( \E_{Y}(Y \mid A=a^*, \mathbf{Z}_1, \mathbf{W}, \C) \big) \right] \right\} \\
    &= \E_{\C} \!\left\{ \E_{\mathbf{Z}_1, \W} \!\left[ \E_{Y}(Y \mid A=a, \mathbf{Z}_1, \W, \C) \right] \right\}
    - \E_{\C} \!\left\{ \E_{\mathbf{Z}_1, \W} \!\left[ \E_{Y}(Y \mid A=a^*, \mathbf{Z}_1, \W, \C) \right] \right\} \\
    &= \E_{\mathbf{Z}_1, \W} \!\left\{ \E_{\C} \!\left[ \E_{Y}(Y \mid A=a, \mathbf{Z}_1, \W, \C) \right] \right\}
    - \E_{\mathbf{Z}_1, \W} \!\left\{ \E_{\C} \!\left[ \E_{Y}(Y \mid A=a^*, \mathbf{Z}_1, \W, \C) \right] \right\} \\
    &= \E_{\mathbf{Z}_1^{\prime}} \!\left\{ \E_{\C} \!\left[ \E_{Y}(Y \mid A=a, \mathbf{Z}_1^{\prime}, \C) \right] \right\}
    - \E_{\mathbf{Z}_1^{\prime}} \!\left\{ \E_{\C} \!\left[ \E_{Y}(Y \mid A=a^*, \mathbf{Z}_1^{\prime}, \C) \right] \right\}.
\end{align*}
The third equality follows from 
$\mathbf{W}\!\perp\!\!\!\perp_{\mathcal G}\!(\C, A)\mid\mathbf{Z}_1$ 
in Criterion~\ref{condition:unique_WCDE} of Condition~\ref{condition:adj}, 
which implies 
$p(\mathbf{w}\mid\mathbf{Z}_1,\C,A)=p(\mathbf{w}\mid\mathbf{Z}_1)$;
the fourth uses the tower property, and the last merges 
$\mathbf{Z}_1$ and $\mathbf{W}$ into 
$\mathbf{Z}_1'=\mathbf{Z}_1\cup\mathbf{W}$.

\paragraph{Step 3: Simplification to WCDE}
Then we remove elements in $\mathbf{Z}_1^{\prime}$  that are not in $\Pa(Y)$ and show that it is equal to Equation~\eqref{eqution:WCDE}. 
\begin{align*}
& \mathbb{E}_{\mathbf{Z}_1^{\prime}} \left\{ \mathbb{E}_{\C} \left[ \mathbb{E}_{Y}(Y \mid A=a, \mathbf{Z}_1^{\prime}, \C) \right] \right\}
- \mathbb{E}_{\mathbf{Z}_1^{\prime}} \left\{ \mathbb{E}_{\C} \left[ \mathbb{E}_{Y}(Y \mid A=a^*, \mathbf{Z}_1^{\prime}, \C) \right] \right\}\\
&= \mathbb{E}_{\C} \left\{ \mathbb{E}_{\mathbf{Z}_1 \cup \mathbf{M}'} \left[ \mathbb{E}_{Y}(Y \mid A=a, \mathbf{Z}_1 \cup \mathbf{M}', \C) \right] \right\}
- \mathbb{E}_{\C} \left\{ \mathbb{E}_{\mathbf{Z}_1 \cup \mathbf{M}'} \left[ \mathbb{E}_{Y}(Y \mid A=a^*, \mathbf{Z}_1 \cup \mathbf{M}', \C) \right] \right\}\\
&= \mathbb{E}_{\C} \left\{ \mathbb{E}_{\mathbf{Z}_1 \cup \mathbf{M}'} \left[ \mathbb{E}_{Y}(Y \mid A=a, \mathbf{M}', \C) \right] \right\}
- \mathbb{E}_{\C} \left\{ \mathbb{E}_{\mathbf{Z}_1 \cup \mathbf{M}'} \left[ \mathbb{E}_{Y}(Y \mid A=a^*, \mathbf{M}', \C) \right] \right\}\\
&= \mathbb{E}_{\C} \left\{ \mathbb{E}_{\mathbf{M}'} \left[ \mathbb{E}_{Y}(Y \mid A=a, \mathbf{M}', \C) \right] \right\}
- \mathbb{E}_{\C} \left\{ \mathbb{E}_{\mathbf{M}'} \left[ \mathbb{E}_{Y}(Y \mid A=a^*, \mathbf{M}', \C) \right] \right\}\\
&= \mathbb{E}_{\mathbf{M}'} \left\{ \mathbb{E}_{\C} \left[ \mathbb{E}_{Y}(Y \mid A=a, \mathbf{M}', \C) \right] \right\}
- \mathbb{E}_{\mathbf{M}'} \left\{ \mathbb{E}_{\C} \left[ \mathbb{E}_{Y}(Y \mid A=a^*, \mathbf{M}', \C) \right] \right\}\\
&= \E_{\mathbf{M}'} \!\Big( 
    \E\!\left[Y \mid \operatorname{do}(A=a, \mathbf{M}')\right]
    - \E\!\left[Y \mid \operatorname{do}(A=a^*, \mathbf{M}')\right]
\Big)\\
&=\sum_{\mathbf{m}^{\prime}\in\M'}\left(\mathbb{E}\left[Y \mid \operatorname{do}\left(a, \mathbf{m}^{\prime}\right)\right]-\mathbb{E}\left[Y \mid \operatorname{do}\left(a^*, \mathbf{m}^{\prime}\right)\right]\right) p\left(\mathbf{m}^{\prime}\right)\\
&= \mathrm{WCDE}
\end{align*}
The first equality expands the conditioning sets to include $\mathbf{M}'$; 
the second removes non-parent variables in $\mathbf{Z}_1'$ without changing the expectation (all parents of $Y$ are already conditioned on); 
the third marginalizes out $\mathbf{Z}_1$; 
the fourth swaps the order of integration over $\C$ and $\mathbf{M}'$ ; 
and the fifth uses the CDE identification with mediator $\mathbf{M}'$: since $\C$ satisfies the graphical conditions for identifying the CDE with mediator $\mathbf{M}'$.
This yields the WCDE in Equation~\eqref{eqution:WCDE}.
% }
% \kyra{I am still a bit confused with the logic of this proof. If the goal is to show that all VAS reduce to 2.1, then we don't need the first augmentation, we can just do the second augmentation and argue that they have the same expectation because they are all valid VAS for ATE. If the point is to establish that they all admit the same expression, then an argument on why $\Pa(Y)\setminus A$ is a VAS is missing. I guess i am asking whether step 1 is necessary}

\paragraph{Necessity of the Graphical Criteria}
% \Lry{
We now show that the adjustment criteria are not only sufficient but also necessary. 
By definition, 
The necessity of
Condition~\ref{condition:mediator_path} follows directly from the definition of the CDE, 
which requires blocking all mediator paths between $A$ and $Y$. 
By definition,
% The necessity of
Condition~\ref{condition:cde} 
is necessary
(we discuss specific conditions that could lead to necessary and sufficient VAS for CDE in Section~\ref{app:cde-id}). 
% as it ensures the satisfaction of the graphical requirements for CDE identification. 
We therefore focus on demonstrating that the \textbf{third adjustment criterion} is also necessary 
for a given set $\mathbf{Z}$ to qualify as a valid adjustment set for the WCDE.

Assume we have an adjustment set $\mathbf{Z}$ that satisfies the first three criteria, but violates the third criterion with respect to $(A,Y)$ in graph $\mathcal{G}$. Define the following subsets:
\[
\mathbf{Z}_1 = \mathbf{Z} \cap \mathbf{M}, \quad
\mathbf{Z}_2 = \mathbf{Z} \setminus \mathbf{M}, \quad
\mathbf{W} = \mathbf{M}^{\prime} \setminus \mathbf{Z}_1, \quad
\mathbf{C} = \Pa(Y) \setminus (\mathbf{M}^{\prime} \cup \{A\}).
\]
Note that $\Pa(Y) \setminus \{A\}$ is a valid adjustment set for identifying $\WCDE$. We compare $\mathrm{WCDE}_{\Pa(Y)\setminus A}$ with $\mathrm{WCDE}_{\mathbf{Z}}$ and consider their difference:
\[
\mathrm{WCDE}_{\Pa(Y)\setminus A} - \mathrm{WCDE}_{\mathbf{Z}} = \left( T_{a}(\Pa(Y)\setminus A) - T_{a}(\mathbf{Z}) \right) - \left( T_{a^*}(\Pa(Y)\setminus A) - T_{a^*}(\mathbf{Z}) \right).
\]
We first analyze the term $T_{a}(\Pa(Y) \setminus \{A\}) - T_{a}(\mathbf{Z})$:
\begin{align*}
&T_{a}(\Pa(Y)\setminus A)-T_{a}(\Z)\\
&=\E_{\mathbf{M}^{\prime}}\{ \E_{\Pa(Y)\setminus (\M ^{\prime}\cup A)}\left\{\E_{{Y}}(Y \mid A=a,\M^{\prime}, \Pa(Y)\setminus (\M ^{\prime}\cup A))\right\}\} \\
&-\E_{\mathbf{ Z}_1}\{\E_{\Z_2}\left\{\E_{{Y}}(Y \mid A=a, \Z_1, \Z_2)\right\}\}\\
&=\E_{\mathbf{M}^{\prime}}\{ \E_{\Pa(Y)\setminus (\M ^{\prime}\cup A)}\left\{\E_{{Y}}(Y \mid A=a,\M^{\prime},\Pa(Y)\setminus (\M ^{\prime}\cup A))\right\}\}\\
&-\E_{\mathbf{ Z}_1}\{ \E_{\Pa(Y)\setminus (\M ^{\prime}\cup A)}\left\{\E_{{Y}}(Y \mid A=a, \Z_1, \Pa(Y)\setminus (\M ^{\prime}\cup A))\right\}\}\\
&=\E_{\mathbf{M}^{\prime}\cup \Z_1}\{ \E_{\Pa(Y)\setminus \mathbf{(M ^{\prime}}\cup A)}\left\{\E_{{Y}}(Y \mid A=a,\M^{\prime}\cup \Z_1,\Pa(Y)\setminus (\M ^{\prime}\cup A))\right\}\}\\
&-\E_{\mathbf{ Z_1}}\{ \E_{\Pa(Y)\setminus (\M ^{\prime}\cup A)}\left\{\E_{{Y}}(Y \mid A=a, \Z_1, \Pa(Y)\setminus (\M ^{\prime}\cup A))\right\}\}\\
\end{align*}
The second equality holds because both $\Pa(Y)\setminus (\M^{\prime}\cup \{A\})$ and $\Z_2$ satisfy the graphical conditions for identifying the CDE with mediator $\mathbf{M}'$.
 The third equality follows from the fact that all parents of $Y$ are already included in the conditioning set; thus, adding elements from $\Z_1$ does not affect the conditional expectation of $Y$.

To simplify further, we translate these expectations into integrals:
\begin{align*}
&\E_{\mathbf{M}^{\prime}\cup \Z_1}\{ \E_{\Pa(Y)\setminus (\M ^{\prime}\cup A)}\left\{\E_{{Y}}(Y \mid A=a,\M^{\prime}\cup \Z_1,\Pa(Y)\setminus \mathbf{(M ^{\prime}}\cup A))\right\}\}\\
&-\E_{\mathbf{ Z_1}}\{ \E_{\Pa(Y)\setminus \mathbf{(M ^{\prime}}\cup A)}\left\{\E_{{Y}}(Y \mid A=a, \Z_1, \Pa(Y)\setminus \mathbf{(M ^{\prime}}\cup A))\right\}\}\\
&= \iint_{\mathbf{W}, \mathbf{Z}_1} \left( \int_{\mathbf{C}} \left( \int y \, p(y \mid A=a,\c, \mathbf{w}, \mathbf{z}_1) \, dy \right) \, p(\mathbf{c})d\c \right) p(\mathbf{w}\mid \mathbf{z_1})p(\z_1)d\w d\z_1 \\
&\quad - \int_{\mathbf{Z_1}} \left( \int_{\mathbf{C}} \left( \int y \, p(y \mid A=a, \c, \mathbf{z}_1) \, dy \right) \, p(\mathbf{c} )d\c \right) p(\mathbf{z}_1)d\z_1 \\
&= \int_{ \mathbf{Z}_1} \left( \int_{\mathbf{C}} \left( \int y \left(  \int  p(y \mid A=a,\c, \mathbf{w}, \mathbf{z}_1)p(\mathbf{w}\mid \mathbf{z_1})d\w\right) \, dy\right) \, p(\mathbf{c})d\c \right) p(\z_1)d\z_1\\
&\quad - \int_{\mathbf{Z_1}} \left( \int_{\mathbf{C}} \left( \int y \, p(y \mid A=a, \mathbf{c}, \mathbf{z}_1) \, dy \right) \, p(\mathbf{c} )d\mathbf{c} \right) p(\mathbf{z}_1)d\z_1 \\
&= \int_{ \mathbf{Z}_1} \left( \int_{\mathbf{C}} \left( \int y \left(  \int  p(y \mid A=a,\mathbf{c}, \mathbf{w}, \z_1)[p(\mathbf{w}\mid \mathbf{z}_1)- p(\mathbf{w}\mid A=a, \mathbf{c},\z_1)]d\w\right) \, dy\right) \, p(\mathbf{c})d\c \right) p(\mathbf{z}_1)d\z_1\\
&= \int y \int_{ \mathbf{Z}_1} \left( \int_{\mathbf{C}} \left( \int  p(y \mid A=a,\mathbf{c}, \mathbf{w}, \mathbf{z_1})[p(\mathbf{w}\mid \mathbf{z}_1)- p(\mathbf{w}\mid A=a, \z_1, \mathbf{c})] p(\mathbf{c})p(\mathbf{z}_1)d\w\right)  \,d\mathbf{c} \right) d\z_1 dy \\
\end{align*}
The first equality expands the nested expectations using the law of total expectation, with integration over \( p(\mathbf{w}|\mathbf{z}_1)p(\mathbf{z}_1) \) and \( p(\mathbf{c}) \). The second applies Fubini’s theorem to switch the order of integration. The third rewrites the inner integral to highlight the difference. Similarly,
\begin{align*}
&T_{a^*}(\Pa(Y)\setminus A)-T_{a^*}(\Z)\\
&= \int y \int_{ \mathbf{Z_1}} \left( \int_{\mathbf{C}} \left( \int  p(y \mid A=a^*,\mathbf{c}, \mathbf{w}, \mathbf{z_1})[p(\mathbf{w}\mid \mathbf{z_1})- p(\mathbf{w}\mid A=a^*,\mathbf{z_1,c})] p(\mathbf{c})p(\mathbf{z_1})d\w\right)  \,d\c \right) d\z_1 dy \\
\end{align*}
Subtracting the two:
\begin{align*}
&\mathrm{WCDE}_{\Pa(Y)\setminus A}-\mathrm{WCDE}_{\Z}=(T_{a}(\Pa(Y)\setminus A)-T_{a}(\Z))-(T_{a^*}(\Pa(Y)\setminus A)-T_{a^*}(\Z))\\
&=\int y \int_{ \mathbf{Z}_1} \left( \int_{\mathbf{C}} \left( \int  p(y \mid A=a,\mathbf{c}, \mathbf{w}, \mathbf{z_1})[p(\mathbf{w}\mid \mathbf{z}_1)- p(\mathbf{w}\mid A=a, \z_1, \mathbf{c})] p(\mathbf{c})p(\mathbf{z}_1)d\w\right)  \,d\c \right) d\z_1 dy \\
&- \int y \int_{ \mathbf{Z_1}} \left( \int_{\mathbf{C}} \left( \int  p(y \mid A=a^*,\mathbf{c}, \mathbf{w}, \mathbf{z_1})[p(\mathbf{w}\mid \mathbf{z_1})- p(\mathbf{w}\mid A=a^*,\mathbf{z_1,c})] p(\mathbf{c})p(\mathbf{z_1})d\w\right)  \,d\c \right) d\z_1 dy
\end{align*}
We can view 
$\int p(y \mid A=a,\c,\w,\z_1)\,d\w\,d\c\,d\z_1$ 
and 
$\int p(y \mid A=a^*,\c,\w,\z_1)\,d\w\,d\c\,d\z_1$
as two operators that map probability distributions over 
$(\C,\W,\Z_1)$ onto probability distributions over $Y$.
Arrange 
$p(y \mid A=a,\c,\w,\z_1)$ 
and 
$p(y \mid A=a^*,\c,\w,\z_1)$ 
so that this mapping is one-to-one. 
Then, to ensure equivalence between the induced distributions, 
we would require
\[
p(\w\mid\z_1)p(\c)p(\z_1)
= p(\w\mid A=a,\z_1,\c)p(\c)p(\z_1)
= p(\w\mid A=a^*,\z_1,\c)p(\c)p(\z_1)
\]
to hold for all models consistently with $\G$.
However, this equality generally fails because 
$\W \not\!\perp\!\!\!\perp_{\mathcal{G}} (A,\C)\mid \Z_1$ 
when the third adjustment criterion is violated.
Consequently, the integrand does not vanish in general, 
and the resulting functional $\WCDE_Z$ deviates from the true WCDE. 
We therefore conclude that the third adjustment criterion 
is necessary for a given $\Z$ to qualify as a valid adjustment set.

\end{proof}

\subsection{Illustration of Criterion~\ref{condition:unique_WCDE}}\label{app:criterion_uniqueness}
% The first three conditions are
\begin{figure}[h]
% \vspace{-15pt}
\centering
\includegraphics[width=0.5\textwidth]{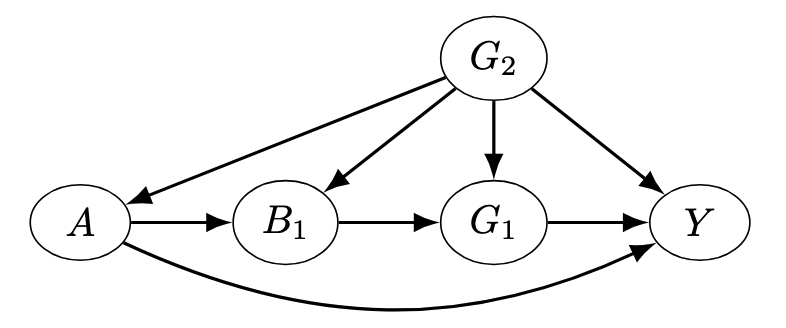}
% \vspace{-5pt}
\caption{A DAG illustrating Condition \ref{condition:adj}.}
\label{fig:sample5}
\end{figure}

\begin{example}[Illustration of Criterion~\ref{condition:unique_WCDE}]\label{example:condition_unqiue_WCDE}
In Figure~\ref{fig:sample5}, $\mathbf{M}=\{G_1,B_1\}, \Pa(Y)=\{A, G_1, G_2\}$, and $\mathbf{M}'=\{G_1\}$. One can verify that the sets $\{G_1, G_2\}$ and $\{G_1, B_1, G_2\}$ both satisfy Condition~\ref{condition:adj}, and are therefore valid adjustment sets.

In contrast, the set $\Z=\{B_1, G_2\}$ satisfies Criteria~\ref{condition:mediator_path}-\ref{condition:cde}, but violates the Criterion~\ref{condition:unique_WCDE}. 
In this case,
% , given $\mathbf{Z}=\{B_1, G_2\}$,
$\mathbf{Z}_1=\{B_1\}$, $\mathbf{Z}_2 = \{G_2\}$, $\M'\setminus \Z_1=\{G_1\}$, and $\Pa(Y)\setminus \M' = \{A, G_2\}$. 
In Figure~\ref{fig:sample5}, we observe that when conditioned on $G_1$, $B_1$ is dependent on $G_2$ due to the direct edge between them; similarly for $A$.
Thus
${G_1} \not\!\perp\!\!\!\perp_{\mathcal{G}} {G_2}, A \mid {B_1}$, violating Criterion~\ref{condition:unique_WCDE}. One could further verify that $\WCDE_{\{B_1, G_2\}}\neq \WCDE^* = \WCDE_{\{G_1, G_2\}}$ as defined by Equation~\eqref{eq:WCDE_identify}. 
% Below, we illustrate this point in detail.
% \kyra{need to explicitly talk about which nodes are Z1 and which are Z2}
To see this, we compare the difference between the first term in $\WCDE_{\{B_1, G_2\}}$ and that in $\WCDE_{\{G_1, G_2\}}$:
% compare the functional corresponding to $\{B_1, G_2\}$ with the one defined in Definition~\ref{def:WCDE}:
% \kyra{revise below}
\begin{align*}
&T_a(\{B_1, G_2\}) - T_a(\{G_1, G_2\})\\
% & E_{{B_1}}\left\{ E_{{G_2}}\left\{ E_{{Y}}\left(Y \mid A=a, B_1,G_2\right) \right\} \right\} 
% - E_{{G_1}}\left\{ E_{{G_2}}\left\{ E_{{Y}}\left(Y \mid A=a, G_1, B_1, G_2\right) \right\} \right\} \\
&= \iint \left( \int y \left[ \int  p(y \mid A=a, g_1, b_1, g_2)(p(g_1| b_1,g_2,a)-p(g_1| b_1))dg_1\right] dy \right) p(g_2) \, dg_2 \, p(b_1) \, db_1. 
\end{align*}
Because
${G_1} \not\!\perp\!\!\!\perp_{\mathcal{G}} {G_2,A} \mid {B_1}$, there exists a data generating process under which $p(g_1 | b_1, g_2, a) \neq p(g_1 | b_1)$
% \kyra{here}
the integrand does not necessarily vanish (detailed construction in proof of Lemma~\ref{lemma:adjustment_criterion}), and the resulting functional based on $\{B_1, G_2\}$ may differ from the true $\WCDE$ as defined. A similar argument can be made on the difference of the second term $T_{a^*}(\{B_1, G_2\}) - T_{a^*}(\{G_1, G_2\})$.
Therefore, the $\WCDE$ cannot be identified using $\mathbf{Z} = \{B_1, G_2\}$.
\end{example}

\subsection{Discussions on Influence Functions}\label{app:IF_dicussion}
For a more detailed review of influence functions and their role in causal inference, we refer the reader to \citet{hines2022demystifying}, \citet{kennedy2022semiparametric}, and \citet{tsiatis2006semiparametric}. Here, we focus on definitions relevant to our derivations in Section~\ref{sec:IF}.

To formally introduce nonparametric estimation, we begin by considering a general statistical model $\mathcal{M}$, which is a collection of candidate distributions defined on a measurable space $(\Omega, \mathcal{F})$. Each element $P \in \mathcal{M}$ admits a Radon–Nikodym derivative $p = dP/d\nu$ with respect to a dominating $\sigma$-finite measure $\nu$, satisfying $p \geq 0$ $\nu$-a.e.\ and $\int_{\Omega} p \, d\nu = 1$. With slight abuse of notation, we use $P$ and $p$ interchangeably to refer to the distribution and its density.

To prepare for the proof of Theorem~\ref{thm:IF}, we begin by recalling several foundational concepts from semiparametric theory, including regular parametric submodels, score functions, tangent spaces, and pathwise differentiability. These concepts provide a principled framework for defining and deriving influence functions. To rigorously understand and formalize the sensitivity discussed in Remark~\ref{remark:IF_WCDE_Z}, we also introduce the definition of influence functions via local distributional perturbations. This alternative characterization is particularly useful for interpreting the robustness and local sensitivity of statistical functionals to contamination at individual points.

%\kyra{connection to remark 3.3}
\textbf{Tangent Space, Pathwise Differentiability, and Influence Function.}

We now introduce key concepts from the local geometry of the model $\mathcal{M}$ that play a central role in semiparametric estimation theory. We proceed step-by-step, starting with the notion of regular parametric submodels, which formalize how a nonparametric distribution can be locally perturbed in a smooth manner.

\begin{definition}[Regular One-dimensional Submodel, \citealt{bickel1993efficient}]
Let $\mathcal{M}$ be a nonparametric model. A \emph{regular one-dimensional submodel} $\{P_t\}_{t \in (-\varepsilon, \varepsilon)} \subset \mathcal{M}$ passing through $P$ at $t = 0$ is a smooth parametric path such that $P_0 = P$ and the corresponding density $p_t$ with respect to a dominating measure $\nu$ is differentiable in $t$ at $t = 0$, with the derivative satisfying
\[
\left.\frac{\partial}{\partial t} \log p_t(O)\right|_{t=0} \in L_2(P).
\]
\end{definition}
The derivative of the log-density along a submodel is referred to as the \emph{score function} and represents the direction of infinitesimal perturbation in the model:

\begin{definition}[Score Function, \citealt{vanderVaart2000}]
For a regular one-dimensional submodel $\{P_t\}$ with density $p_t$, the \emph{score function} at $P$ is defined as
\[
S(O) := \left.\frac{\partial}{\partial t} \log p_t(O)\right|_{t=0},
\]
which describes the direction of local perturbation of $P$ along the submodel.
\end{definition}

\medskip

A commonly used example of a regular submodel is constructed by directly specifying a score function $S$ and perturbing the baseline density $p$ as follows:
\[
p_t(o) = (1 + t S(o)) p(o),
\]
where the score function satisfies $\int S(o)\, p(o) \,d\nu(o) = 0$ and $\int S^2(o)\, p(o) \,d\nu(o) < \infty$. One can verify that this path satisfies $P_0 = P$ and that its score function is
\[
\left.\frac{\partial}{\partial t} \log p_t(o)\right|_{t=0} = \left.\frac{\partial}{\partial t} \log (1 + t S(o)) \right|_{t=0} = S(o).
\]
Such submodels are particularly useful for constructing influence functions and deriving efficiency bounds in semiparametric models.

The set of all such score functions defines the tangent space, which describes the directions in which the model $\mathcal{M}$ can be perturbed:
\begin{definition}[Tangent Space, \citealt{vanderVaart2000}]
The \emph{tangent space} at $P$ is the closed linear subspace of $L_2(P)$ spanned by all score functions $S(O)$ generated by regular one-dimensional submodels through $P$. It characterizes the set of directions in which the distribution $P$ can be locally perturbed.
\end{definition}
We now turn to a central concept in semiparametric theory: pathwise differentiability. This notion formalizes when a parameter of interest can be linearly approximated along such perturbations, and leads directly to the definition of the influence function:
\begin{definition}[Pathwise Differentiability and Influence Function, \citealt{tsiatis2006semiparametric}]\label{def:pathdiff}
A parameter $T$ is said to be \emph{pathwise differentiable} at $P$ if there exists a function $\psi_P \in L_2(P)$ with $\mathbb{E}_P[\psi_P(O)] = 0$ such that, for every regular submodel $\{P_t\}$ with score function $S(O)$,
\begin{equation}\label{equation:pathwise}
    \left.\frac{d}{dt} T(P_t)\right|_{t=0} = \mathbb{E}_P[\psi_P(O) \cdot S(O)]
\end{equation}
Any such function $\psi_P$ is called an \emph{influence function} (IF) of $T$ at $P$.
\end{definition}
Equation~\ref{equation:pathwise} provides a constructive method for deriving influence functions, which we elaborate upon in the next subsection.

In addition to submodel-based definitions, influence functions can be equivalently understood via local distributional perturbations. This perspective, which originates in robust statistics, is often more intuitive:
\begin{definition}[Influence Function via Local Distributional Perturbations, \citealt{hampel1974influence, huber2009robust}]
Let $\delta_o$ denote the Dirac measure at observation $o$, and consider a perturbed distribution
\[
    P_\varepsilon = (1 - \varepsilon) P + \varepsilon \delta_{o}.
\]
Then the \emph{influence function} of a statistical functional $T$ at $P$ evaluated at $o$ is defined as
\[
\mathrm{IF}(o; T, P) := \left. \frac{d}{d\varepsilon} T(P_\varepsilon) \right|_{\varepsilon = 0}.
\]
\end{definition}
This definition captures the first-order sensitivity of $T$ to contamination at point $o$, and naturally connects to the notions of robustness and sensitivity analysis. It corroborates our observation in Remark~\ref{remark:IF_WCDE_Z} that $\mathrm{WCDE}_{\Z}$ is sensitive to perturbations in variables within $\Z$, implying that the choice of $\Z$ directly impacts the sensitivity of the functional.

\medskip

A fundamental result in semiparametric theory links these definitions to statistical estimation. Specifically, if $\hat{T}$ is an asymptotically linear estimator of $T(P)$ at $P$ with influence function $\psi_P$, then $\hat{T}$ is regular at $P$ in model $\mathcal{M}$ if and only if $T(P)$ is pathwise differentiable and $\psi_P$ is the influence function of $T$. See Theorem 2.2 of \citet{newey1990semiparametric}. This explains why the influence function defined in Definition~\ref{def:asymplin} coincides with the influence function of the target parameter $T(P)$.

\subsection{Proof of Theorem \ref{thm:IF}}
\label{app:proof_thm_IF}

\begin{lemma}~\citep{levy2024deriving}\label{lemma:levy}
Let $O = (O_1, O_2, \dots, O_d)$ be a random vector with joint density $p_t(o)$ under a smooth parametric submodel $\{P_t\}$. Assume the density factorizes as
\[
p_t(o) = \prod_{i=1}^d p_{O_i, t}(o_i \mid \bar{o}_{i-1}),
\]
where $\bar{o}_{i-1} = (o_1, \dots, o_{i-1})$, and $p_{O_i, t}(o_i \mid \bar{o}_{i-1}) = p_{O_i}(o_i \mid \bar{o}_{i-1})$ at $t=0$. Then, the pathwise derivative of each conditional density satisfies:
\[
\left.\frac{\partial}{\partial t} p_{O_i, t}(o_i \mid \bar{o}_{i-1})\right|_{t=0} = p_{O_i}(o_i \mid \bar{o}_{i-1}) \cdot \left.\frac{\partial}{\partial t} \log p_{O_i, t}(o_i \mid \bar{o}_{i-1})\right|_{t=0}.
\]
Define the conditional score function $S_{O_i}(\bar{o}_i)$ as the derivative of the log-density:
\[
S_{O_i}(\bar{o}_i) := \left.\frac{\partial}{\partial t} \log p_{O_i, t}(o_i \mid \bar{o}_{i-1})\right|_{t=0}.
\]
Then:
\[
\left.\frac{\partial}{\partial t} p_{O_i, t}(o_i \mid \bar{o}_{i-1})\right|_{t=0} = S_{O_i}(\bar{o}_i) p_{O_i}(o_i \mid \bar{o}_{i-1}).
\]
Moreover, under the chain rule for scores, this can be written in terms of the full-data score function $S(O)$ as:
\[
\left.\frac{\partial}{\partial t} p_{O_i, t}(o_i \mid \bar{o}_{i-1})\right|_{t=0} =
\left( \mathbb{E}[S(O) \mid \bar{O}_i = \bar{o}_i] - \mathbb{E}[S(O) \mid \bar{O}_{i-1} = \bar{o}_{i-1}] \right) p_{O_i}(o_i \mid \bar{o}_{i-1}).
\]
\end{lemma}
This lemma allows for expressing the derivative of a marginal or conditional density in terms of conditional expectations of the full-data score function and is essential for decomposing pathwise derivatives in IF calculations.
\begin{proof}(Proof of Theorem \ref{thm:IF})\label{appendix:IF}

For the $\WCDE$ functional:
\[
T_{a}(\Z) = {\EE_{\mathbf{Z}_1} \left\{ \EE_{\mathbf{Z}_2} \left[ \mathbb{E}[Y \mid A = a, \mathbf{Z}_1, \mathbf{Z}_2] \right] \right\}}
\]
Recall the definition of the pathwise derivative:
\begin{equation*}
\left.\frac{\partial}{\partial t} T\left(P_t\right)\right|_{t=0}=\int \psi(o; P) S(o)p(o)do.
\end{equation*}
where the score function $S(o)$ is defined as
\[
S(o)=\left. \frac{\partial}{\partial t} \log p_t(o)\right |_{t=0}
\]
for any smooth parametric submodel $\{P_t\}$.

We start by expressing $T_a(P_t)$ as:
\begin{align*}
T_{a}\left(P_t\right) & =\int y p_t(y \mid a, \z_1,\z_2) p_t(\z_1)p_t(\z_2) d y d \z_2d\z_1\\
& =\int y \frac{p_t(y, \z_1,\z_2,a) p_t(\z_1)p_t(\z_2)}{p_t(\z_1,\z_2,a)} d y d \z_2d\z_1
\end{align*}
Taking the derivative w.r.t.\ $t$ at $t=0$, and applying the product and chain rules, we get:
\begin{align*}
& \left.\frac{\partial T_a\left(P_t\right)}{\partial t}\right|_{t=0} \\
&= \int y \bigg\{ 
    \underbrace{
        \left.\frac{p\left(\z_1\right) p\left(\z_2\right)}{p\left(a, \z_1, \z_2\right)} 
        \frac{\partial}{\partial t} p_t\left(y, a, \z_1, \z_2\right)\right|_{t=0}
    }_{\text{(i) first term}} +
    \underbrace{
        \left.\frac{p\left(y, a, \z_1, \z_2\right) p\left(\z_1\right)}{p\left(a, \z_1, \z_2\right)} 
        \frac{\partial}{\partial t} p_t\left(\z_2\right)\right|_{t=0}
    }_{\text{(ii) second term}} \\
&\quad +
    \underbrace{
        \left.\frac{p\left(y, a, \z_1, \z_2\right) p\left(\z_2\right)}{p\left(a, \z_1, \z_2\right)} 
        \frac{\partial}{\partial t} p_t\left(\z_1\right)\right|_{t=0}
    }_{\text{(iii) third term}}-
    \underbrace{
        \left.\frac{p\left(y, a, \z_1, \z_2\right) p\left(\z_1\right) p\left(\z_2\right)}{p\left(a, \z_2, \z_1\right)^2} 
        \frac{\partial}{\partial t} p_t\left(a, \z_1, \z_2\right)\right|_{t=0}
    }_{\text{(iv) fourth term}} 
\bigg\} \, d \nu(y) \, d \nu(\z_2) \, d \nu(\z_1)
\end{align*}
Each term is handled individually. For the first term:
\begin{align*}
& \int y\{\left.\frac{p\left(\z_1\right) p\left(\z_2\right)}{p\left(a, \z_1, \z_2\right)} \frac{\partial}{\partial t} p_t\left(y, a, \z_1, \z_2\right)\right|_{t=0}\}d\nu(y)d\nu(\z_2)d\nu(\z_1)\\
&= \int y\{\left.\frac{p\left(\z_2\right) p\left(\z_1\right)p_t\left(y, a, \z_1, \z_2\right)}{p\left(a, \z_1, \z_2\right)} \frac{\partial}{\partial t} \log p_t\left(y, a, \z_1, \z_2\right)\right|_{t=0}\}d\nu(y)d\nu(\z_2)d\nu(\z_1)\\
&= \int y\{\frac{p\left(\z_1\right) p\left(\z_2\right)p(o)}{p\left(a, \z_1, \z_2\right)} S(o) \} d\nu(y)d\nu(\z_2)d\nu(\z_1)\\
&= \int y\{\frac{p\left(\z_1\right) p\left(\z_2\right)p(o)}{p\left(a, \z_1, \z_2\right)} I_a(a^\prime)S(o) \} d\nu(y)d\nu(\z_2)d\nu(\z_1)d\nu(a^\prime)\\
&= \int \{\frac{I_{a}(a^\prime)p\left(\z_1\right) p\left(\z_2\right)y}{p\left(a, \z_1, \z_2\right)} S(o) p(o)\} d\nu(o)
\end{align*}
In the third equality, we extend the integration from the conditional domain over $(y, \z_1, \z_2)$ to the full domain $o = (y, a^\prime, \z_1, \z_2)$. To preserve the original conditioning on $A = a$, we introduce the indicator function \(I_a(a') := \mathbb{I}_{\{a' = a\}}\), where \(a\) denotes the fixed treatment level of interest, and \(a'\) serves as the input variable to the function, which may vary throughout the integration.

For the second term, we apply Lemma~\ref{lemma:levy}. This result allows us to express the influence of the perturbation in $p(\z_2)$ on the target parameter through a conditional expectation of the score function.
\begin{align*}
    & \int y\left\{\left.\frac{p\left(y, a, \z_1, \z_2\right) p\left(\z_1\right)}{p\left(a, \z_1, \z_2\right)} \frac{\partial}{\partial t} p_t\left(\z_2\right)\right|_{t=0}\right\}d\nu(y)d\nu(\z_2)d\nu(\z_1) \\
&= \int y\left\{\frac{p\left(y, a, \z_1, \z_2\right) p\left(\z_1\right)}{p\left(a, \z_1, \z_2\right)} \left(\E[S(O) | \z_2] - \E[S(O)]\right) p(\z_2)\right\} d\nu(y)d\nu(\z_2)d\nu(\z_1) \\
\end{align*}
We now expand the conditional expectations explicitly. The first term inside the integrand is rewritten using the definition of conditional expectation and the joint density, while the second term represents the overall mean of the score function. By computing the expectation of $Y$ given $(A=a, \Z_1, \Z_2)$ and rearranging the integration order, we ultimately express the entire term as an inner product of the score function $S(o)$ with a centered function depending on $\Z_1$ and $\Z_2$, which contributes to the influence function.
\begin{align*}
&\int y\left\{\frac{p\left(y, a, \z_1, \z_2\right) p\left(\z_1\right)}{p\left(a, \z_1, \z_2\right)} \left(\E[S(O) | \z_2] - \E[S(O)]\right) p(\z_2)\right\} d\nu(y)d\nu(\z_2)d\nu(\z_1) \\
&= \int y\left\{\frac{p\left(y, a, \z_1, \z_2\right) p\left(\z_1\right) p(\z_2)}{p\left(a, \z_2, \z_1\right)} \left( \int S(o) p(y, a, \z_1 | \z_2) d\nu(y)d\nu(a)d\nu(\z_1) \right.\right. \\
&\quad \left.\left. - \int S(o) p(o) d\nu(o) \right)\right\} d\nu(y)d\nu(\z_2)d\nu(\z_1) \\
&= \int \left\{ p(\z_1) p(\z_2) \E_Y(Y | a, \z_1, \z_2) \right\} \left( \int S(o) p(y, a, \z_1 | \z_2) d\nu(y)d\nu(a)d\nu(\z_1) \right. \\
&\quad \left. - \int S(o) p(o) d\nu(o) \right) d\nu(\z_2)d\nu(\z_1) \\
&= \int \left\{ p(\z_1) \E_Y(Y | a, \z_1, \z_2) \right\} \left( \int S(o) p(y, a, \z_1, \z_2) d\nu(y)d\nu(a)d\nu(\z_1) \right) d\nu(\z_2)d\nu(\z_1) \\
&\quad - \int T_a(P) S(o) p(o) d\nu(o)\\
&= \int \left\{ \E_{\Z_1} \left[ \E_Y[Y | A = a, \Z_1, \Z_2] \right] \right\} \left( \int S(o) p(y, a, \z_1, \z_2) d\nu(y)d\nu(a)d\nu(\z_1) \right) d\nu(\z_2) \\
&\quad - \int T_a(P) S(o) p(o) d\nu(o) \\
&= \int \left\{ \E_{\Z_1} \left[ \E_Y[Y | A = a, \Z_1, \Z_2] \right] \right\} \left( S(o) p(o) \right) d\nu(o) - \int T_a(P) S(o) p(o) d\nu(o)
\end{align*}
For the third term, we apply Lemma~\ref{lemma:levy}. This result allows us to express the influence of the perturbation in $p(\z_1)$ on the target parameter through a conditional expectation of the score function.
\begin{align*}
& \int y\left\{\left.\frac{p\left(y, a, \z_1, \z_2\right) p\left(\z_2\right)}{p\left(a, \z_1, \z_2\right)} \frac{\partial}{\partial t} p_t\left(\z_1\right)\right|_{t=0}\right\}d\nu(y)d\nu(\z_1)d\nu(\z_2) \\
&= \int y\left\{\frac{p\left(y, a, \z_1, \z_2\right) p\left(\z_2\right)}{p\left(a, \z_1, \z_2\right)} \left(\E[S(O) | \z_1] - \E[S(O)]\right) p(\z_1)\right\} d\nu(y)d\nu(\z_1)d\nu(\z_2) 
\end{align*}
The rest of the computation follows by evaluating the conditional expectation of $Y$ given $A = a, \Z_1, \Z_2$ and expressing the result in terms of an inner product between the score function and an identifying function.
\begin{align*}
&\int y\left\{\frac{p\left(y, a, \z_1, \z_2\right) p\left(\z_2\right)}{p\left(a, \z_1, \z_2\right)} \left(\E[S(O) | \z_1] - \E[S(O)]\right) p(\z_1)\right\} d\nu(y)d\nu(\z_2)d\nu(\z_1) \\
&= \int y\left\{\frac{p\left(y, a, \z_1, \z_2\right) p\left(\z_1\right) p(\z_2)}{p\left(a, \z_1, \z_2\right)} \left( \int S(o) p(y, a, \z_2 | \z_1) d\nu(y)d\nu(a)d\nu(\z_2) \right.\right. \\
&\quad \left.\left. - \int S(o) p(o) d\nu(o) \right)\right\} d\nu(y)d\nu(\z_2)d\nu(\z_1) \\
&= \int \left\{ p(\z_1) p(\z_2) \E_Y(Y | a, \z_1, \z_2) \right\} \left( \int S(o) p(y, a, \z_2 | \z_1) d\nu(y)d\nu(a)d\nu(\z_2) \right. \\
&\quad \left. - \int S(o) p(o) d\nu(o) \right) d\nu(\z_2)d\nu(\z_1) \\
&= \int \left\{ p(\z_2) \E_Y(Y | a, \z_1, \z_2) \right\} \left( \int S(o) p(y, a, \z_1, \z_2) d\nu(y)d\nu(a)d\nu(\z_2) \right) d\nu(\z_2)d\nu(\z_1) \\
&\quad - \int T_a(P) S(o) p(o) d\nu(o)\\
&= \int \left\{ \E_{\Z_2} \left[ \E_Y[Y | A = a, \Z_1, \Z_2] \right] \right\} \left( \int S(o) p(y, a, \z_1, \z_2) d\nu(y)d\nu(a)d\nu(\z_2) \right) d\nu(\z_1) \\
&\quad - \int T_a(P) S(o) p(o) d\nu(o) \\
&= \int \left\{ \E_{\Z_2} \left[ \E_Y[Y | A = a, \Z_1, \Z_2] \right] \right\} \left( S(o) p(o) \right) d\nu(o) - \int T_a(P) S(o) p(o) d\nu(o)
\end{align*}
For the fourth term, we begin by applying Lemma~\ref{lemma:levy} to express the derivative of the joint probability $p_t(a, \z_1, \z_2)$ in terms of the score function $S(O)$.
\begin{align*}
& \int y\left\{\left.\frac{p\left(y, a, \z_1, \z_2\right) p\left(\z_1\right) p\left(\z_2\right)}{p\left(a, \z_1, \z_2\right)^2} \frac{\partial}{\partial t} p_t\left(a, \z_1, \z_2\right)\right|_{t=0}\right\} d\nu(y)d\nu(\z_2)d\nu(\z_1)\\
&= \int y\left\{ \frac{p\left(y, a, \z_1, \z_2\right) p\left(\z_1\right) p\left(\z_2\right)}{p\left(a, \z_1, \z_2\right)^2} \left( \E[S(O) | a, \z_1, \z_2] - \E[S(O)] \right)p(a,\z_1,\z_2) \right\} d\nu(y)d\nu(\z_2)d\nu(\z_1)\\
\end{align*}
Next, we expand the conditional expectations as integrals over the relevant conditional distributions. The term $\E[S(O)|a,\z_1,\z_2]$ becomes an integral over $p(y|a,\z_1,\z_2)$, while $\E[S(O)]$ integrates over the full joint distribution $p(o)$. We then factor out the conditional expectation of $Y$ given $(a,\z_1,\z_2)$.
\begin{align*}
&=\int y\left\{ \frac{p\left(y, a, \z_1, \z_2\right) p\left(\z_1\right) p\left(\z_2\right)}{p\left(a, \z_1, \z_2\right)^2} \left( \E[S(O) | a, \z_1, \z_2] - \E[S(O)] \right)p(a,\z_1,\z_2) \right\} d\nu(y)d\nu(\z_2)d\nu(\z_1)\\
&= \int y \left\{ \frac{p\left(y, a, \z_1, \z_2\right) p\left(\z_1\right) p\left(\z_2\right)}{p\left(a, \z_1, \z_2\right)} \left( \int S(o) p(y |a, \z_1, \z_2) d\nu(y) - \int S(o) p(o) d\nu(o) \right) \right\} d\nu(y)d\nu(\z_2)d\nu(\z_1)\\
&= \int \left\{ p(\z_1) p(\z_2) \E_Y(Y | a, \z_1, \z_2) \right\} \left( \int S(o) p(y|a, \z_1, \z_2) d\nu(y) - \int S(o) p(o) d\nu(o) \right) d\nu(\z_1)d\nu(\z_2) \\
&= \int \left\{ p(\z_1)p(\z_2) \E_Y(Y | a, \z_1, \z_2) \right\} \left( \int S(o) \frac{p(y, a, \z_1, \z_2)}{p(a,\z_1,\z_2)} d\nu(y) \right) d\nu(\z_1)d\nu(\z_2) - \int T_a(P) S(o) p(o) d\nu(o) \\
\end{align*}
Finally, we introduce the indicator function $I_a(a^\prime)$ and combine all integrals into a single expectation with respect to $p(o)$.
\begin{align*}
& \int \left\{ p(\z_1)p(\z_2) \E_Y(Y | a, \z_1, \z_2) \right\} \left( \int S(o) \frac{p(y, a, \z_1, \z_2)}{p(a,\z_1,\z_2)} d\nu(y) \right) d\nu(\z_1)d\nu(\z_2) - \int T_a(P) S(o) p(o) d\nu(o) \\
&= \int \left\{ \frac{p(\z_1)p(\z_2)}{p(a,\z_1,\z_2)} \E_Y(Y | a, \z_1, \z_2) \right\} \left( \int S(o) {p(o)} d\nu(y) \right) d\nu(\z_1)d\nu(\z_2) - \int T_a(P) S(o) p(o) d\nu(o) \\
&= \int \left\{ \frac{I_{a}(a^\prime)p(\z_1)p(\z_2)}{p(a,\z_1,\z_2)} \E_Y(Y | a, \z_1, \z_2) \right\}  S(o) {p(o)}  d\nu(y)d\nu(\z_1)d\nu(\z_2)d\nu(a^\prime)- \int T_a(P) S(o) p(o) d\nu(o) \\
&= \int \left\{ \frac{I_{a}(a^\prime)p(\z_1)p(\z_2)}{p(a,\z_1,\z_2)} \E_Y(Y | a, \z_1, \z_2) \right\}  S(o) {p(o)}  d\nu(o)- \int T_a(P) S(o) p(o) d\nu(o) \\
\end{align*}
Summing all terms together, we obtain:
\begin{align*}
& \left.\frac{\partial T_a\left(P_t\right)}{\partial t}\right|_{t=0} \\
& =\int \{[\frac{I_{a}(a^\prime)p\left(\z_1\right) p\left(\z_2\right)(y-\E_Y(Y | a, \z_1, \z_2))}{p\left(a, \z_2, \z_1\right)}+ \E_{\Z_1} \left[ \E_Y[Y | A = a, \z_2, \z_1] \right]\\
&+ \E_{\Z_2} \left[ \E_Y[Y | A = a, \z_1, \z_2] \right]-2T_a(P)] S(o) p(o)\} d\nu(o)\\
&=\int \psi_a(o;P)S(o)p(o)d\nu(o)
\end{align*}
with
\begin{align*}
    \psi_{a}\left(Y, A, \Z_1, \Z_2 ; P\right)  = &
    \frac{ I_a(A) p(\Z_1)  p(\Z_2)}
    {p(\Z_1, \Z_2,a)} \cdot \bigg( Y - \mathbb{E}_Y[Y \mid A=a,\Z_1,\Z_2] \bigg) \notag \\
    & + \mathbb{E}_{\Z_2} \bigg[ \mathbb{E}_Y[Y \mid A=a,\Z_1,\Z_2] \bigg]   
    + \mathbb{E}_{\Z_1} \bigg[ \mathbb{E}_Y[Y \mid A=a,\Z_1,\Z_2] \bigg] - 2 T_a(P)
\end{align*}
as required.
\end{proof}

\section{Proofs of Results in Section 4}\label{appendix:sec4}
Theorem \ref{thm:optimal_adjustment} follows from the four lemmas presented in Section~\ref{sec:optimal_VAS}.
% Subsections \ref{subsec:backdoor} to \ref{subsec:overadjustment_mediator}. 
Accordingly, we first provide the proofs of Lemmas \ref{lemma:1}, \ref{lemma:2}, \ref{lemma:3}, and \ref{lemma:4} individually, and then proceed to prove the theorem.
\begin{figure}[h]
\centering

% 左图
\begin{minipage}{0.48\linewidth}
\centering
\begin{tikzpicture}[
    node distance=0.7cm and 1cm,
    every node/.style={draw, ellipse, align=center, minimum width=0.88cm, minimum height=0.6cm, font=\normalsize},
    arrow/.style={-Latex, thick} 
]
\node (A) {A};
\node[right=of A] (G1) {$G_1$};
\node[right=of G1] (Y) {$Y$};
\node[above=of G1] (B2) {$B_2$};
\node[above=of Y] (G2) {$G_2$}; 

\draw[arrow] (A) -- (G1); 
\draw[arrow] (G1) -- (Y); 
\draw[arrow] (B2) -- (A);  
\draw[arrow] (B2) -- (G1);
\draw[arrow] (B2) -- (G2); 
\draw[arrow] (G2) -- (Y);
\draw[arrow, bend right=25] (A) to (Y);
\end{tikzpicture}
\caption{A DAG illustrating Lemmas \ref{lemma:1}, \ref{lemma:2}.}
\end{minipage}
\hfill
% 右图
\begin{minipage}{0.48\linewidth}
\centering
\begin{tikzpicture}[
    node distance=0.5cm and 0.9cm,
    every node/.style={draw, ellipse, align=center, minimum width=0.88cm, minimum height=0.6cm, font=\normalsize},
    arrow/.style={-Latex, thick} 
]
\node (A) {$A$};
\node[right=of A] (B1) {$B_1$};
\node[right=of B1] (G1) {$G_1$};
\node[right=of G1] (Y) {$Y$}; 
\node[above=of G1] (G2) {$G_2$}; 

\draw[arrow] (A) -- (B1); 
\draw[arrow] (B1) -- (G1); 
\draw[arrow] (G1) -- (Y); 
\draw[arrow] (G2) -- (B1);
\draw[arrow] (G2) -- (A); 
\draw[arrow] (G2) -- (Y);
\draw[arrow, bend right=25] (A) to (Y);
\end{tikzpicture}
\caption{A DAG illustrating Lemmas \ref{lemma:3}, \ref{lemma:4}.}
\end{minipage}

\end{figure}

\subsection{Proof of Lemma \ref{lemma:1}}\label{app:proof_lemma1}
\begin{proof}(Proof of Lemma \ref{lemma:1}: Supplement of backdoor variables)\label{appendix:lemma1}

We begin by showing that the set $(\G_1, \G_2, \B_2)$ remains a VAS. We can verify this by comparing the identified WCDE functionals.
Recall that for any adjustment set $\Z=(\Z_1,\Z_2)$, the WCDE is identified as
\[
\mathrm{WCDE}_\Z
= 
\EE_{\Z_1}\!\Big\{
   \EE_{\Z_2}\!\big[\EE(Y \mid A = a, \Z_1, \Z_2)\big]
   -
   \EE_{\Z_2}\!\big[\EE(Y \mid A = a^\ast, \Z_1, \Z_2)\big]
\Big\}.
\]
For $(\G_1,\G_2,\B_2)$, consider the term corresponding to $a$:
\[
T_a(\G_1,\G_2,\B_2)
= 
\EE_{\G_1}\!\Big[
   \EE_{\G_2,\B_2}\!\big[\EE(Y \mid A=a,\G_1,\G_2,\B_2)\big]
\Big].
\]
Under the conditional independence 
\begin{equation}
\G_2 \perp\!\!\!\perp_{\mathcal{G}} A, \G_1 \mid \B_2
\label{eq:lemma1}
\end{equation} 
we have
$p(\G_2,\B_2)=p(\G_2\mid\B_2)p(\B_2)=p(\G_2\mid A=a,\G_1,\B_2)p(\B_2)$, 
and by the law of iterated expectations,
\[
\EE_{\G_2\mid\B_2}
   \!\left[
      \EE(Y \mid A=a,\G_1,\G_2,\B_2)
   \right]
= \EE(Y \mid A=a,\G_1,\B_2).
\]
Substituting this gives
\begin{align*}
&T_a(\G_1,\G_2,\B_2)
= 
\EE_{\G_1}\!\Big[\EE_{\B_2}\big[\EE_{\G_2\mid\B_2}
   \!\left[
      \EE(Y \mid A=a,\G_1,\G_2,\B_2)
   \right]\big]\Big]\\
   &=\EE_{\G_1}\!\Big[\EE_{\B_2}\big[\EE(Y \mid A=a,\G_1,\B_2)\big]\Big]
= T_a(\G_1,\B_2),
\end{align*}
and the same equality holds for $a^\ast$. Hence,
\[
\mathrm{WCDE}_{(\G_1,\G_2,\B_2)}
=\mathrm{WCDE}_{(\G_1,\B_2)}.
\]
By the uniqueness of the identified WCDE functional, any adjustment set that yields the same functional as a VAS must itself be a VAS. Therefore, $(\G_1,\G_2,\B_2)$ is also a VAS.

We now compare the asymptotic variances of two VASs, $(\G_1,\B_2)$ and $(\G_1,\G_2,\B_2)$.
 For clarity, we first focus on the first component of the $\WCDE$ in Equation~\eqref{eq:WCDE_identify},  and let $\sigma_{a,\mathbf{Z}}^2(P)$ denote its asymptotic variance. The extension to the full $\WCDE$ is straightforward.
We begin by decomposing the variance:
\begin{align*}
&\sigma^2_{a, ( \G_1, \B_2)}(P)
= \operatorname{Var}_P\left[\psi_{a}(Y,A,\G_1, \B_2; P)\right] = \operatorname{Var}_P\left[\psi_{a}(Y,A,\G_1, \G_2, \B_2; P)\right] \\
&+ \operatorname{Var}_P\left[\psi_{a}(Y,A,\G_1, \B_2; P) - \psi_{a}(Y,A,\G_1, \G_2, \B_2;P)\right] \\
& + 2\,\operatorname{Cov}_P\left(
\psi_{a}(Y,A,\G_1, \G_2, \B_2;P),
\psi_{a}(Y,A,\G_1, \B_2;P) - \psi_{a}(Y,A,\G_1, \G_2, \B_2;P)
\right).
\end{align*}
To analyze the covariance term, consider the difference in IFs evaluated at $(Y, A, \G_1, \G_2, \B_2)$:
\begin{align*}
& \psi_{a}(Y,A,\G_1,\B_2 ; P) - \psi_{a}(Y,A,\G_1,\G_2,\B_2 ; P)\\
&=\frac{ I_a(A) p(\G_1)  p(\B_2)}
    {p( \G_1, \B_2,a)} \cdot \bigg( Y - \mathbb{E}_Y[Y \mid A=a,\G_1,\B_2] \bigg) \notag \\
    & + \mathbb{E}_{\B_2} \bigg[ \mathbb{E}_Y[Y \mid A=a,\G_1,\B_2] \bigg]   
    + \mathbb{E}_{\G_1} \bigg[ \mathbb{E}_Y[Y \mid A=a,\G_1,\B_2] \bigg]\\
    &-\frac{ I_a(A) p(\G_1) p(\G_2, \B_2) }
    {p( \G_1, \G_2, \B_2,a)} \cdot \bigg( Y - \mathbb{E}_Y[Y \mid A=a,\G_1, \G_2,\B_2] \bigg) \notag \\
    & - \mathbb{E}_{\G_2, \B_2} \bigg[ \mathbb{E}_Y[Y \mid A=a,\G_1,\G_2, \B_2] \bigg]   
    - \mathbb{E}_{\G_1} \bigg[ \mathbb{E}_Y[Y \mid A=a,\G_1, \G_2, \B_2] \bigg]\\
&=\frac{I_{a}(A) \cdot p(\G_1) \cdot p(\B_2)}
    {p(A=a, \G_1,\B_2)} \cdot \bigg(  \mathbb{E}_Y[Y \mid A=a,\G_1,\G_2,\B_2] -  \mathbb{E}_Y[Y \mid A=a,\G_1,\B_2]\bigg) \notag \\
    & + \mathbb{E}_{\B_2} \bigg[ \mathbb{E}_Y[Y \mid A=a,\G_1,\B_2] \bigg]   + \mathbb{E}_{\G_1} \bigg[ \mathbb{E}_Y[Y \mid A=a,\G_1,\B_2] \bigg] \\
    & - \mathbb{E}_{\G_2,\B_2} \bigg[ \mathbb{E}_Y[Y \mid A=a,\G_1,\G_2,\B_2] \bigg]   - \mathbb{E}_{\G_1}\bigg[ \mathbb{E}_Y[Y \mid A=a,\G_1,\G_2,\B_2] \bigg] \\
\end{align*}
The second equality above holds under the conditional independence assumption \ref{eq:lemma1}, from which it follows that
\[
p(\G_2 \mid \B_2) = p(\G_2 \mid \G_1, \B_2, A=a),
\]
which implies:
\[
\frac{ p(\G_2, \B_2) p(\G_1) }{ p(\G_1, \G_2, \B_2, A=a) }
= \frac{ p(\B_2) p(\G_1) }{ p(\G_1, \B_2, A=a) } \cdot \frac{ p(\G_2 \mid \B_2) }{ p(\G_2 \mid \B_2, \G_1, A=a) }
= \frac{ p(\B_2) p(\G_1) }{ p(A=a, \G_1, \B_2) }.
\]
Therefore, the first terms in the two influence functions cancel out. To simplify the final term, note that:
\begin{align*}
&\mathbb{E}_{\G_2,\B_2}\bigg[ \mathbb{E}_Y[Y \mid A=a,\G_1,\G_2,\B_2] \bigg]=\mathbb{E}_{\B_2}\bigg[\E_{\G_2\mid\B_2}\bigg[ \mathbb{E}_Y[Y \mid A=a,\G_1,\G_2,\B_2] 
\bigg]\bigg]\\&=\mathbb{E}_{\B_2}\bigg[\E_{\G_2\mid\G_1,\B_2,A=a}\bigg[ \mathbb{E}_Y[Y \mid A=a,\G_1,\G_2,\B_2] \bigg]\bigg]=\mathbb{E}_{\B_2} \bigg[ \mathbb{E}_Y[Y \mid A=a,\G_1,\B_2] \bigg]
\end{align*}
this equality follows from the conditional independence assumption \ref{eq:lemma1} and the law of iterated expectations. Substituting back, the difference simplifies to:
\begin{align*}
& \frac{I_{a}(A) \cdot p(\G_1)\cdot p(\B_2) }
    {p(A=a, \G_1,\B_2)} \cdot \bigg(  \mathbb{E}_Y[Y \mid A=a,\G_1,\G_2,\B_2] -  \mathbb{E}_Y[Y \mid A=a,\G_1,\B_2]\bigg) \notag \\
    & + \mathbb{E}_{\B_2} \bigg[ \mathbb{E}_Y[Y \mid A=a,\G_1,\B_2] \bigg]   + \mathbb{E}_{\G_1} \bigg[ \mathbb{E}_Y[Y \mid A=a,\G_1,\B_2] \bigg] \\
    & - \mathbb{E}_{\G_2,\B_2} \bigg[ \mathbb{E}_Y[Y \mid A=a,\G_1,\G_2,\B_2] \bigg]   - \mathbb{E}_{\G_1}\bigg[ \mathbb{E}_Y[Y \mid A=a,\G_1,\G_2,\B_2] \bigg] \\
    &= \frac{I_{a}(A)\cdot p(\G_1) \cdot p(\B_2) }
    {p(A=a, \G_1, \B_2)} \cdot \bigg(  \mathbb{E}_Y[Y \mid A=a,\G_1,\G_2,\B_2] -  \mathbb{E}_Y[Y \mid A=a,\G_1,\B_2]\bigg) \notag \\
    & + \mathbb{E}_{\G_1} \bigg[ \mathbb{E}_Y[Y \mid A=a,\G_1,\B_2] \bigg] - \mathbb{E}_{\G_1} \bigg[ \mathbb{E}_Y[Y \mid A=a,\G_1,\G_2,\B_2] \bigg]
\end{align*}
Therefore, we arrive at the simplified expression for their difference.

We then construct a parametric submodel that perturbs only the conditional law \( P(A, \G_1 \mid \G_2, \B_2) \), while keeping all other components of the distribution fixed. We show that under the  conditional independence assumption \ref{eq:lemma1}, the difference between the two IFs is orthogonal to the influence function of the larger adjustment set:
\[
\mathbb{E}_P\left[\psi_a(Y, A, \G_1, \G_2, \B_2;P) \cdot \left(
\psi_a(Y, A, \G_1, \B_2;P) - \psi_a(Y, A, \G_1, \G_2, \B_2;P)
\right)\right] = 0.
\]
Hence, the covariance term vanishes. Substituting this into the variance decomposition yields:
\begin{align*}
   & \operatorname{Var}_P[\psi_a(Y, A, \G_1, \B_2;P)]\\
&= \operatorname{Var}_P[\psi_a(Y, A, \G_1, \G_2, \B_2;P)] 
+ \operatorname{Var}_P[\psi_a(Y, A, \G_1, \B_2;P) - \psi_a(Y, A, \G_1, \G_2, \B_2;P)],
\end{align*}

where the second term is nonnegative. To formalize this orthogonality, we appeal to two key results from semiparametric theory~\citep{tsiatis2006semiparametric}:
\begin{enumerate}
    \item If \( P_t \) is a smooth parametric submodel with \( P_0 = P \) and score function \( g \), then the influence function \( \psi_a \) satisfies:
    \[
    \left.\frac{d}{dt} T_a(P_t) \right|_{t=0} = \mathbb{E}_P[\psi_a \cdot g].
    \]
    \item In a nonparametric model, any function \( g \) such that
    \[
    \mathbb{E}_P[g(A, \G_1, \G_2, \B_2) \mid \G_2, \B_2] = 0
    \]
    is a valid score function for the conditional law \( P(A, \G_1 \mid \G_2, \B_2) \).
\end{enumerate}
Let us now consider any distribution \( P \) and define the function
\[
g(A, \G_1, \G_2, \B_2) := \psi_a(Y, A, \G_1, \B_2; P) - \psi_a(Y, A, \G_1, \G_2, \B_2; P),
\]
which satisfies the second condition above. Using this function, we define the following submodel for sufficiently small \( t \geq 0 \):
\[
P_t(Y, A, \G_1, \G_2, \B_2) := P(Y \mid A, \G_1, \G_2, \B_2) \cdot P_t(A, \G_1 \mid \G_2, \B_2) \cdot P(\G_2, \B_2),
\]
where
\[
P_t(A, \G_1 \mid \G_2, \B_2) = P(A, \G_1 \mid \G_2, \B_2)(1 + t \cdot g(A, \G_1, \G_2, \B_2)).
\]
This submodel perturbs only the conditional law of \( (A, \G_1) \) given \( (\G_2, \B_2) \), leaving the remaining components of the joint distribution fixed. The score function of this model at \( t = 0 \) is
\[
\left.\frac{\partial}{\partial t} \log p_t(Y, A, \G_1, \G_2, \B_2)\right|_{t=0}
= \left.\frac{\partial}{\partial t} \log \{1 + t \cdot g(A, \G_1, \G_2, \B_2)\}\right|_{t=0}
= g(A, \G_1, \G_2, \B_2).
\]
Applying result (1), we obtain:
\[
\left.\frac{d}{dt} T_a(P_t) \right|_{t=0} = \mathbb{E}_P[\psi_a \cdot g].
\]
Now consider the functional:
\[
T_a(P) =\sum_{\G_1} \sum_{\G_2,\B_2} \left[\mathbb{E}[Y \mid a, \G_1, \G_2, \B_2]\right] p(\G_2, \B_2) p(\G_1).
\]
This functional depends only on the joint distribution of \( (\G_2, \B_2) \), \( \G_1 \), and the conditional distribution \( P(Y \mid A, \G_1, \G_2, \B_2) \). If the function \( g \) satisfies
\begin{equation}
\mathbb{E}_P[g(A, \G_1, \G_2, \B_2)\mid \G_2, \B_2] = 0 
\quad \text{and} \quad 
\mathbb{E}_P[g(A, \G_1, \G_2, \B_2) \mid \G_1] = 0,
\label{eq:orthogonality_conditions1}
\end{equation}
then perturbations in \( P_t(Y, A, \G_1, \G_2, \B_2) \) do not affect those distributions.
That is, the submodel is locally ancillary for the functional \( T_a \), and therefore:
\[
\left.\frac{d}{dt} T_a(P_t) \right|_{t=0} = 0.
\]
Combining with the earlier result gives:
\[
\mathbb{E}_P[\psi_a g] = 0
\]
Then we obtain the orthogonality:
\begin{align*}
&\E_P\left[\boldsymbol{\psi}_{a}(Y,A,\G_1, \G_2,\B_2 ; P) g(A, \G_1, \G_2,\B_2)\right]
=0 
\end{align*}
We now verify that this function \( g(A, \G_1, \G_2, \B_2) \) satisfies the two orthogonality conditions in Condition \ref{eq:orthogonality_conditions1}. First, consider the conditional expectation given \( \G_2, \B_2 \):
\begin{align*}
 &\E_P[\psi_{a}(Y,A,\G_1,\B_2 ; P) - \psi_{a}(Y,A,\G_1,\G_2,\B_2 ; P) \mid \G_2,\B_2]\\
&=\E_P\Bigg\{ 
 \frac{I_{a}(A) \cdot p(\B_2) \cdot p(\G_1)}
    {p(A=a, \G_1, \B_2)} \cdot \bigg( \mathbb{E}_Y[Y \mid A=a,\G_1,\G_2,\B_2] 
    -  \mathbb{E}_Y[Y \mid A=a,\G_1,\B_2] \bigg) \\
& + \mathbb{E}_{\G_1} \bigg[ \mathbb{E}_Y[Y \mid A=a,\G_1,\B_2] \bigg] 
    - \mathbb{E}_{\G_1} \bigg[ \mathbb{E}_Y[Y \mid A=a,\G_1,\G_2,\B_2] \bigg]\mid \G_2,\B_2
\Bigg\}\\ 
&=\E_P\Bigg\{ 
 \frac{ p(\G_2,\B_2) \cdot p(\G_1)}
    {p(\G_1, \G_2, \B_2)} \cdot \bigg( \mathbb{E}_Y[Y \mid A=a,\G_1,\G_2,\B_2] 
    -  \mathbb{E}_Y[Y \mid A=a,\G_1,\B_2] \bigg) \\
& + \mathbb{E}_{\G_1} \bigg[ \mathbb{E}_Y[Y \mid A=a,\G_1,\B_2] \bigg] 
    - \mathbb{E}_{\G_1} \bigg[ \mathbb{E}_Y[Y \mid A=a,\G_1,\G_2,\B_2] \bigg]\mid \G_2,\B_2
\Bigg\}\\  
&=\E_P\Bigg\{ 
 \frac{ p(\G_2,\B_2) \cdot p(\G_1)}
    {p(\G_1, \G_2, \B_2)} \cdot \bigg( \mathbb{E}_Y[Y \mid A=a,\G_1,\G_2,\B_2] 
    -  \mathbb{E}_Y[Y \mid A=a,\G_1,\B_2] \bigg) \mid \G_2,\B_2
\Bigg\}\\  
& + \mathbb{E}_{\G_1} \bigg[ \mathbb{E}_Y[Y \mid A=a,\G_1,\B_2] \bigg] 
    - \mathbb{E}_{\G_1} \bigg[ \mathbb{E}_Y[Y \mid A=a,\G_1,\G_2,\B_2] \bigg]
\end{align*}
In the second equality, we integrate over \(A\) conditional on \((\G_2, \B_2)\). Note that the last two expectation terms are fixed once \((\G_2, \B_2)\) are given, so we can move them outside the outer expectation accordingly.

We now rewrite the first expectation as an integral over $\G_1$ conditional on $(\G_2, \B_2)$:
\begin{align*}
&\E_P\Bigg\{ 
 \frac{ p(\G_2,\B_2) \cdot p(\G_1)}
    {p(\G_1, \G_2, \B_2)} \cdot \bigg( \mathbb{E}_Y[Y \mid A=a,\G_1,\G_2,\B_2] 
    -  \mathbb{E}_Y[Y \mid A=a,\G_1,\B_2] \bigg) \,\Big|\, \G_2,\B_2
\Bigg\} \\[5pt]
&= \int \frac{ p(\G_2,\B_2)\cdot p(\g_1)}
{p(\g_1, \G_2, \B_2)} 
\left(\mathbb{E}_Y[Y \mid A=a,\g_1, \G_2, \B_2]-\mathbb{E}_Y[Y \mid A=a,\g_1,\B_2]\right)
p(\g_1 \mid \G_2,\B_2)\, d\g_1 \\[5pt]
&= \int \left(\mathbb{E}_Y[Y \mid A=a,\g_1,\G_2,\B_2]-\mathbb{E}_Y[Y \mid A=a,\g_1,\B_2]\right)
p(\g_1)\, d\g_1 \\[5pt]
    &= \mathbb{E}_{\G_1} \bigg[ \mathbb{E}_Y[Y \mid A=a,\G_1,\G_2,\B_2] \bigg] 
    - \mathbb{E}_{\G_1} \bigg[ \mathbb{E}_Y[Y \mid A=a,\G_1,\B_2] \bigg]  \\[5pt]
\end{align*}
Substituting this expression back into the original conditional expectation, we obtain:
\begin{align*}
 &\E_P[\psi_{a}(Y,A,\G_1,\B_2 ; P) - \psi_{a}(Y,A,\G_1,\G_2,\B_2 ; P) \mid \G_2,\B_2]\\
    &= \mathbb{E}_{\G_1} \bigg[ \mathbb{E}_Y[Y \mid A=a,\G_1,\G_2,\B_2] \bigg] 
    - \mathbb{E}_{\G_1} \bigg[ \mathbb{E}_Y[Y \mid A=a,\G_1,\B_2] \bigg]  \\[5pt]
& + \mathbb{E}_{\G_1} \bigg[ \mathbb{E}_Y[Y \mid A=a,\G_1,\B_2] \bigg] 
    - \mathbb{E}_{\G_1} \bigg[ \mathbb{E}_Y[Y \mid A=a,\G_1,\G_2,\B_2] \bigg]\\
    &=0
\end{align*}
This establishes that the function \( g(A,  \G_1, \G_2,\B_2) \) satisfies the first orthogonality condition in Condition~\ref{eq:orthogonality_conditions1}, $\mathbb{E}_P[g(A, \G_1, \G_2, \B_2)\mid \G_2, \B_2] = 0$

To verify that it satisfies the second orthogonality condition, we proceed similarly by conditioning on \( \G_1 \):
\begin{align*}
 &\E_P[\psi_{a}(Y,A,\G_1,\B_2 ; P) - \psi_{a}(Y,A,\G_1,\G_2,\B_2 ; P) \mid \G_1] \notag \\
&=\E_P\Bigg\{ 
 \frac{I_{a}(A)\cdot p(\G_1) \cdot p(\B_2) }
    {p(A=a, \G_1, \B_2)} \cdot \bigg( \mathbb{E}_Y[Y \mid A=a,\G_1,\G_2,\B_2] 
    -  \mathbb{E}_Y[Y \mid A=a,\G_1,\B_2] \bigg) \notag \\
& \quad + \mathbb{E}_{\G_1} \bigg[ \mathbb{E}_Y[Y \mid A=a,\G_1,\B_2] \bigg] 
    - \mathbb{E}_{\G_1} \bigg[ \mathbb{E}_Y[Y \mid A=a,\G_1,\G_2,\B_2] \bigg]\mid \G_1
\Bigg\} \notag \\
&=\E_P\Bigg\{ 
 \frac{ p(\G_1) \cdot p(\G_2,\B_2)}
    {p(\G_1, \G_2, \B_2)} \cdot \bigg( \mathbb{E}_Y[Y \mid A=a,\G_1,\G_2,\B_2] 
    -  \mathbb{E}_Y[Y \mid A=a,\G_1,\B_2] \bigg) \notag \\
&\quad + \mathbb{E}_{\G_1} \bigg[ \mathbb{E}_Y[Y \mid A=a,\G_1,\B_2] \bigg] 
    - \mathbb{E}_{\G_1} \bigg[ \mathbb{E}_Y[Y \mid A=a,\G_1,\G_2,\B_2] \bigg]\mid \G_1
\Bigg\}
\end{align*}
In the second equality, we integrate over \(A\) conditional on \(\G_1\). We now rewrite the first term as an integral over $(\G_2, \B_2)$ with respect to $P(\G_2, \B_2 \mid \G_1)$, and then simplify.
\begin{align*}
&\E_P\Bigg\{ 
 \frac{ p(\G_1) \cdot p(\G_2,\B_2)}
    {p(\G_1, \G_2, \B_2)} \cdot \bigg( \mathbb{E}_Y[Y \mid A=a,\G_1,\G_2,\B_2] 
    -  \mathbb{E}_Y[Y \mid A=a,\G_1,\B_2] \bigg) \notag \\
&\quad + \mathbb{E}_{\G_1} \bigg[ \mathbb{E}_Y[Y \mid A=a,\G_1,\B_2] \bigg] 
    - \mathbb{E}_{\G_1} \bigg[ \mathbb{E}_Y[Y \mid A=a,\G_1,\G_2,\B_2] \bigg]\mid \G_1
\Bigg\}\\
&= \int \frac{  p(\G_1) \cdot p(\G_2,\B_2)}
{p(\G_1, \g_2, \b_2)} 
\left(\mathbb{E}_Y[Y \mid A=a,\G_1,\g_2,\b_2]-\mathbb{E}_Y[Y \mid A=a,\G_1,\b_2]\right)
p(\g_2,\b_2 \mid \G_1)\, d\g_2\, d\b_2 \\[5pt]
&\quad + \mathbb{E}_P\Bigg\{ 
 \mathbb{E}_{\G_1} \bigg[ \mathbb{E}_Y[Y \mid A=a,\G_1,\B_2] \bigg] 
    - \mathbb{E}_{\G_1} \bigg[ \mathbb{E}_Y[Y \mid A=a,\G_1,\G_2,\B_2] \bigg] \,\Big|\, \G_1
\Bigg\} \\[5pt]
&= \int p(\g_2,\b_2)
\left(\mathbb{E}_Y[Y \mid A=a,\G_1,\g_2,\b_2]-\mathbb{E}_Y[Y \mid A=a,\G_1,\b_2]\right)
\, d\g_2\, d\b_2 \\[5pt]
&\quad + \mathbb{E}_P\Bigg\{ 
 \mathbb{E}_{\G_1} \bigg[ \mathbb{E}_Y[Y \mid A=a,\G_1,\B_2] \bigg] 
    - \mathbb{E}_{\G_1} \bigg[ \mathbb{E}_Y[Y \mid A=a,\G_1,\G_2,\B_2] \bigg] \,\Big|\, \G_1
\Bigg\} \\[5pt]
\end{align*}
Next, we cancel the two components of the expression by invoking the conditional independence assumption \ref{eq:lemma1}.
\begin{align*}
&\int p(\g_2,\b_2)
\left(\mathbb{E}_Y[Y \mid A=a,\G_1,\g_2,\b_2]-\mathbb{E}_Y[Y \mid A=a,\G_1,\b_2]\right)
\, d\g_2\, d\b_2 \\[5pt]
&\quad + \mathbb{E}_P\Bigg\{ 
 \mathbb{E}_{\G_1} \bigg[ \mathbb{E}_Y[Y \mid A=a,\G_1,\B_2] \bigg] 
    - \mathbb{E}_{\G_1} \bigg[ \mathbb{E}_Y[Y \mid A=a,\G_1,\G_2,\B_2] \bigg] \,\Big|\, \G_1
\Bigg\} \\[5pt]
&= \int
\left(\mathbb{E}_Y[Y \mid A=a,\G_1,\g_2,\b_2]-\mathbb{E}_Y[Y \mid A=a,\G_1,\b_2]\right) p(\g_2\mid \b_2, A=a, \G_1)p(\b_2)
\, d\g_2\, d\b_2 \\[5pt]
&\quad + \mathbb{E}_P\Bigg\{ 
 \mathbb{E}_{\G_1} \bigg[ \mathbb{E}_Y[Y \mid A=a,\G_1,\B_2] \bigg] 
    - \mathbb{E}_{\G_1} \bigg[ \mathbb{E}_Y[Y \mid A=a,\G_1,\G_2,\B_2] \bigg] \,\Big|\, \G_1
\Bigg\} \\[5pt]
&= \mathbb{E}_P\Bigg\{ 
 \mathbb{E}_{\G_1} \bigg[ \mathbb{E}_Y[Y \mid A=a,\G_1,\B_2] \bigg] 
    - \mathbb{E}_{\G_1} \bigg[ \mathbb{E}_Y[Y \mid A=a,\G_1,\G_2,\B_2] \bigg] \,\Big|\, \G_1
\Bigg\}
\end{align*}
The second equality holds because  we have the conditional independence assumption \ref{eq:lemma1}, so:
\[
p(\g_2, \b_2) = p(\g_2 \mid \b_2) \cdot p(\b_2) = p(\g_2 \mid \b_2, A=a, \G_1)p(\b_2),
\]
which justifies replacing the joint density in the integral. Furthermore, by the law of iterated expectations,
\[
\int \mathbb{E}_Y[Y \mid A=a, \G_1, \g_2, \b_2] \cdot p(\g_2 \mid \b_2, A=a, \G_1) \, d\g_2 = \mathbb{E}_Y[Y \mid A=a,\G_1, \b_2].
\]
As a result, the two integrals cancel out, yielding the final equality. To simplify this expression, we compute both terms as iterated integrals.
\begin{align*}
&\E_P\Bigg\{ 
 \mathbb{E}_{\G_1} \bigg[ \mathbb{E}_Y[Y \mid A=a,\G_1,\B_2] \bigg] 
    - \mathbb{E}_{\G_1} \bigg[ \mathbb{E}_Y[Y \mid A=a,\G_1,\G_2,\B_2] \bigg]\mid \G_1
\Bigg\}\\
&= \iiint y \cdot \left[ p(y \mid A=a, \g_1, \b_2) \right] \, dy \, p(\g_1) \, d\g_1 \cdot p(\b_2 \mid \G_1) \, d\b_2 \\
&\quad - \iiiint y \cdot \left[ p(y \mid A=a, \g_1, \g_2, \b_2) \right] \, dy \, p(\g_1) \, d\g_1 \cdot p(\g_2, \b_2 \mid \G_1) \, d\g_2 \, d\b_2 \\
\end{align*}
Based on Condition~\ref{eq:lemma1}, we can simplify the second expression through the following derivation. First, we apply Condition~\ref{eq:lemma1} ($p(\g_2\mid \b_2)=p(\g_2\mid A=a,\G_1, \b_2)$) to expand the conditioning set:
\begin{align*}
& \iiiint y \cdot \left[ p(y \mid A=a, \g_1, \g_2, \b_2) \right] \, dy \, p(\g_1) \, d\g_1 \cdot p(\g_2, \b_2 \mid \G_1) \, d\g_2 \, d\b_2 \\
&=\iiiint y \cdot p(y \mid A=a, \g_1, \g_2, \b_2) \, dy \, p(\g_1) d\g_1 \cdot p(\g_2 \mid \b_2) d\g_2 \cdot p(\b_2 \mid \mathbf{   G}_1) d\b_2 \\
&= \iiiint y \cdot p(y \mid A=a, \g_1, \g_2, \b_2) \, dy \, p(\g_1) d\g_1 \cdot p(\g_2 \mid A=a, \g_1, \b_2) d\g_2 \cdot p(\b_2 \mid \G_1) d\b_2
\end{align*}
Next, we express the integrand as a joint probability distribution over $y$ and $\g_2$:
\begin{align*}
&= \iiiint y \cdot p(y, \g_2 \mid A=a, \g_1, \b_2) \, dy d\g_2 \, p(\g_1) d\g_1 \cdot p(\b_2 \mid \G_1) d\b_2
\end{align*}
Using the chain rule of probability, we factorize the joint probability:
\begin{align*}
&= \iiiint y \cdot p(\g_2 \mid y, A=a, \g_1, \b_2) p(y \mid A=a, \g_1, \b_2) \, d\g_2 dy \, p(\g_1) d\g_1 \cdot p(\b_2 \mid \G_1) d\b_2
\end{align*}
Finally, we marginalize over $\g_2$ by integrating it out (since $\int p(\g_2 \mid y, A=a, \g_1, \b_2) d\g_2 = 1$), yielding the simplified expression:
\begin{align*}
&= \iiint y \cdot p(y \mid A=a, \g_1, \b_2) \, dy \, p(\g_1) d\g_1 \cdot p(\b_2 \mid \G_1) d\b_2
\end{align*}
This shows that the second term equals the first one. Thus, we conclude:
\begin{align*}
\E_P\Bigg\{ 
 \mathbb{E}_{\G_1} \bigg[ \mathbb{E}_Y[Y \mid A=a,\G_1,\B_2] \bigg] 
    - \mathbb{E}_{\G_1} \bigg[ \mathbb{E}_Y[Y \mid A=a,\G_1,\G_2,\B_2] \bigg]\mid \G_1
\Bigg\}=0
\end{align*}
So far we have shown that  the function \( g(A, \G_1, \G_2, \B_2) \) satisfies Condition~\ref{eq:orthogonality_conditions1}, so we obtain the orthogonality:
\[
\mathbb{E}_P\left[\psi_a(Y, A, \G_1, \G_2, \B_2;P) \cdot \left(
\psi_a(Y, A, \G_1, \B_2;P) - \psi_a(Y, A, \G_1, \G_2, \B_2;P)
\right)\right] = 0.
\]
Therefore, we conclude that the variance decomposes as:
\begin{align*}
& \sigma_{a,(\G_1,\B_2)}^2(P) = \operatorname{var}_P\left[\psi_{ a}(Y,A,\G_1,\B_2 ; P)\right] \\
& =\operatorname{var}_P\left[\psi_{a}(Y,A,\G_1, \G_2,\B_2 ; P)\right]+\operatorname{var}_P\left[\psi_{a}(Y,A, \G_1,\B_2 ; P)-\psi_{ a}(Y,A,\G_1,\G_2,\B_2 ;P)\right]\\
&\geq \sigma_{a, (\G_1, \G_2, \B_2)}^2(P)
\end{align*}
Now extend this to the vectorized $\WCDE$. Define: $\mathbf{c} \equiv (c_a)_{a \in \mathcal{A}}$, 
$\mathbf{Q} \equiv (Q_a)_{a \in \mathcal{A}}$ where 
\[
Q_a= \psi_{a}(Y,A,\G_1,\B_2 ;P) - \psi_{a}(Y,A,\G_1,\G_2,\B_2 ; P)
\]
For any $\mathbf{Z}$, define $\boldsymbol{\psi}(Y,A,\mathbf{Z}; P) 
\equiv (\psi_{ a}(Y,A,\mathbf{Z};P))_{a \in \mathcal{A}}$.  
Then, writing $\sum_{a \in \mathcal{A}} c_a \psi_{a}(Y,A,\mathbf{Z};P) 
= \mathbf{c}^\top \boldsymbol{\psi}(Y,A,\mathbf{Z}; P)$  
We notice that:
\[
 \E_P[\Q\mid \G_2,\B_2]=\mathbf{0},
\]
\[
 \E_P[\Q \mid \G_1]=\mathbf{0},
\]
Then we obtain the orthogonality:
\begin{align*}
&\E_P\left[\Q^\top\boldsymbol{\psi}(Y,A,\G_1, \G_2,\B_2 ; P) \right]
=\mathbf{0} 
\end{align*}
Therefore, we conclude that the variance decomposes as:
\begin{align*}
    &\sigma^2_{ (\G_1, \B_2)}(P) = \operatorname{var}_P \left[ \mathbf{c}^\top \boldsymbol{\psi}(Y,A,\G_1,\B_2; P) \right] \\
=& \operatorname{var}_P \left[ \mathbf{c}^\top \boldsymbol{\psi}(Y,A,\G_1, \G_2,\B_2; P) \right] 
+ \operatorname{var}_P \left[ \mathbf{c}^\top \mathbf{Q} \right] 
= \sigma^2_{(\G_1,\G_2,\B_2)}(P) + \mathbf{c}^\top \operatorname{var}_P(\mathbf{Q}) \mathbf{c}\\
\geq &  \sigma^2_{(\G_1,\G_2,\B_2)}(P) 
\end{align*}
This completes the proof.
\end{proof}

\subsection{Proof of Lemma \ref{lemma:2}}\label{app:proof_lemma2}
\begin{proof}(Proof of Lemma \ref{lemma:2}: Deletion of overadjustment backdoor variables)\label{appendix:lemma2}

We begin by showing that the set $(\G_1,\G_2)$ remains a VAS. We can verify this by comparing the identified WCDE functionals.
Recall that for any adjustment set $\Z=(\Z_1,\Z_2)$, the WCDE is identified as
\[
\mathrm{WCDE}_\Z
= 
\EE_{\Z_1}\!\Big\{
   \EE_{\Z_2}\!\big[\EE(Y \mid A = a, \Z_1, \Z_2)\big]
   -
   \EE_{\Z_2}\!\big[\EE(Y \mid A = a^\ast, \Z_1, \Z_2)\big]
\Big\}.
\]
For $(\G_1,\G_2,\B_2)$, consider the term corresponding to $a$:
\[
T_a(\G_1,\G_2,\B_2)
= 
\EE_{\G_1}\!\Big[
   \EE_{\G_2,\B_2}\!\big[\EE(Y \mid A=a,\G_1,\G_2,\B_2)\big]
\Big].
\]
Under the conditional independence
\begin{equation}
Y \perp\!\!\!\perp_{\mathcal{G}} \B_2 \mid \G_1,\G_2,A,
\label{eq:lemma2}
\end{equation}
the inner expectation simplifies as
\[
\EE\!\left[Y \mid A=a,\G_1,\G_2,\B_2\right]
=\EE\!\left[Y \mid A=a,\G_1,\G_2\right],
\]
Substituting this gives
\begin{align*}
&T_a(\G_1,\G_2,\B_2)
= 
\EE_{\G_1}\!\Big[\EE_{\G_2, \B_2}\!\big[\EE(Y \mid A=a,\G_1,\G_2)\big]\Big]\\
&=\EE_{\G_1}\!\Big[\EE_{\G_2}\!\big[\EE(Y \mid A=a,\G_1,\G_2)\big]\Big]
= T_a(\G_1,\G_2),
\end{align*}
and the same equality holds for $a^\ast$. Hence,
\[
\mathrm{WCDE}_{(\G_1,\G_2,\B_2)}
=\mathrm{WCDE}_{(\G_1,\G_2)}.
\]
By the uniqueness of the identified WCDE functional, any adjustment set that yields the same functional as a VAS must itself be a VAS. Therefore, $(\G_1,\G_2)$ is also a VAS.

We now compare the asymptotic variances of two VASs, $(\G_1,\G_2)$ and $(\G_1,\G_2,\B_2)$.
 For clarity, we first focus on the first component of the $\WCDE$ in Equation~\eqref{eq:WCDE_identify},  and let $\sigma_{a,\mathbf{Z}}^2(P)$ denote its asymptotic variance. The extension to the full $\WCDE$ is straightforward.

We begin by applying the law of total variance to decompose the variance under the larger adjustment set $(\G_1, \G_2, \B_2)$:
\begin{align*}
\operatorname{Var}[\psi_a(Y, A, \G_1, \G_2, \B_2;P)] 
&= \operatorname{Var}[\mathbb{E}[\psi_a(Y, A, \G_1, \G_2, \B_2;P) \mid A, Y, \G_1, \G_2]] \\
&\quad + \mathbb{E}\left[\operatorname{Var}[\psi_a(Y, A, \G_1, \G_2, \B_2;P) \mid A, Y, \G_1, \G_2]\right].
\end{align*}
Next, we will show that under the conditional independence assumption \ref{eq:lemma2}  
the IF based on the adjustment set $(\G_1, \G_2, \B_2)$ satisfies  
\[
\mathbb{E}_P[\psi_a(Y, A, \G_1, \G_2, \B_2;P) \mid A, Y, \G_1, \G_2]
= \psi_a(Y, A, \G_1, \G_2;P).
\]  
This equality allows us to relate the variances of the two IFs via the law of total variance. We now compute the conditional expectation of $\psi_a(Y, A, \G_1, \G_2, \B_2;P)$:
\begin{align*}
& \E_P\left[\psi_{a}(Y,A,\G_1, \G_2, \B_2 ; P) \mid A, Y, \G_1,\G_2\right] = \\
& \underbrace{
    I_{a}(A)\left\{Y - \E_Y[Y \mid A=a,\G_1,\G_2]\right\} 
    \cdot \E_P\left[\left.\frac{p(\G_1) \cdot  p(\G_2,\B_2)}{p(A=a,\G_1,\G_2,\B_2)} \right\rvert\, A=a, Y, \G_1,\G_2\right]
}_{\text{(i) first term}} \\[5pt]
& + \underbrace{
    \E_{P} \left[ \E_{\G_2,\B_2} \left[ \E_Y[Y \mid A=a,\G_1,\G_2,\B_2] \right] \,\middle|\, \G_1 \right]
}_{\text{(ii) second term}} \\[5pt]
& + \underbrace{
    \E_{P} \left[ \E_{\G_1} \left[ \E_Y[Y \mid A=a,\G_1,\G_2,\B_2] \right] \,\middle|\, \G_2 \right]
}_{\text{(iii) third term}} 
- 2 T_a(P)
\end{align*}
This equality holds because, given the conditional independence relation \ref{eq:lemma2}, it implies the key simplification:
\[
\mathbb{E}_Y(Y \mid A=a, \G_1, \G_2, \B_2) = \mathbb{E}_Y(Y \mid A=a, \G_1, \G_2).
\]
For the first conditional expectation term, we perform the detailed computation:
\begin{align*}
    &\mathbb{E}_P \left[ \frac{p(\G_2, \B_2) \cdot p(\G_1)}{p(A, \G_1, \G_2, \B_2)} \,\Big|\, A = a, Y, \G_1, \G_2 \right] \\
    &=\; \mathbb{E}_P \left[ \frac{p(\G_1) \cdot p(\G_2, \B_2) }{p(\B_2 \mid A = a, \G_1, \G_2) \cdot p(A=a, \G_1, \G_2)} \,\Big|\, A = a, Y, \G_1, \G_2 \right] \\
    &=\; \frac{1}{p(A=a, \G_1, \G_2)} \cdot \mathbb{E}_P \left[ \frac{p(\G_1) \cdot p(\G_2, \B_2)}{p(\B_2 \mid A = a, \G_1, \G_2)} \,\Big|\, A = a, Y, \G_1, \G_2 \right] \\
       &=\; \frac{1}{p(A=a, \G_1, \G_2)} \cdot \mathbb{E}_P \left[ \frac{p(\G_1) \cdot p(\G_2, \B_2)}{p(\B_2 \mid A = a, \G_1, \G_2)} \,\Big|\, A = a, Y, \G_1, \G_2 \right] \\
    &=\; \frac{1}{p(A=a, \G_1, \G_2)} \cdot \int_{\b_2} \frac{p(\G_1) \cdot p(\G_2, \b_2)}{p(\b_2 \mid A = a, \G_1, \G_2)} \cdot p(\b_2 \mid A = a, Y, \G_1, \G_2) \, d\b_2 \\
    &=\; \frac{1}{p(A=a, \G_1, \G_2)} \cdot \int_{\b_2}  p(\G_1)\cdot p(\G_2, \b_2) \cdot \frac{p(\b_2 \mid A = a, Y, \G_1, \G_2)}{p(\b_2 \mid A = a, \G_1, \G_2)} \, d\b_2 \\
\end{align*}
In the last line above, we expand the conditional expectation as an integral over \(\B_2\). Now, under the conditional independence assumption \ref{app:proof_lemma2}, it follows that
\[
p(\b_2 \mid A = a, Y, \G_1, \G_2) = p(\b_2 \mid A = a, \G_1, \G_2).
\]
Substituting this into the integrand, the ratio simplifies to 1:
\begin{align*}
    &\; \frac{1}{p(A=a, \G_1, \G_2)} \cdot \int_{\b_2} p(\G_2, \b_2) \cdot p(\G_1) \cdot \frac{p(\b_2 \mid A = a, Y, \G_1, \G_2)}{p(\b_2 \mid A = a, \G_1, \G_2)} \, d\b_2 \\
    &=\; \frac{1}{p(A=a, \G_1, \G_2)} \cdot \int_{\b_2} p(\G_2, \b_2) \cdot p(\G_1) \, d\b_2 \\
    &=\; \frac{1}{p(A=a, \G_1, \G_2)} \cdot \int_{\b_2} p(\G_2 \mid \b_2) \cdot p(\G_1) \cdot p(\G_2) \, d\b_2 \\
    &=\; \frac{p(\G_1) \cdot p(\G_2)}{p(A=a, \G_1, \G_2)} \cdot \int_{\b_2} p(\b_2 \mid \G_2) \, d\b_2 \\
    &=\; \frac{p(\G_1) \cdot p(\G_2)}{p(A=a, \G_1, \G_2)}
\end{align*}
Since $\mathbb{E}_Y(Y \mid A=a, \G_1, \G_2, \B_2) = \mathbb{E}_Y(Y \mid A=a, \G_1,\G_2)$, the second and third terms simplify as follows:
\begin{align*}
& \E_{P}  \bigg[E_{\G_2,\B_2} [ E_Y[Y \mid A=a,\G_1,\G_2,\B_2] ]\mid \G_1\bigg]\\
&= \E_{P}  \bigg[E_{\G_2,\B_2} [ E_Y[Y \mid A=a,\G_1,\G_2] ]\mid \G_1\bigg]\\
&= \E_{\G_2} \bigg[ E_Y[Y \mid A=a,\G_1,\G_2] \bigg]
\end{align*}
\begin{align*}
&\E_{P}  \bigg[E_{\G_1} [ E_Y[Y \mid A=a,\G_1,\G_2,\B_2] ]\mid \G_2\bigg]\\
&=\E_{P}  \bigg[E_{\G_1} [ E_Y[Y \mid A=a,\G_1,\G_2] ]\mid \G_2\bigg]\\
&= \E_{\G_1} \bigg[ E_Y[Y \mid A=a,\G_1,\G_2] \bigg]
\end{align*}
Combining these results, we obtain:
\begin{align*}
&\E_P\left[\psi_{ a}(Y,A,\G_1,\G_2, \B_2 ;P) \mid A, Y, \G_1,\G_2\right]\\
&=\frac{I_{a}(A) \cdot p(\G_1) \cdot p(\G_2)}
    {p(A=a, \G_1, \G_2)} \cdot \bigg( Y - \E_Y[Y \mid A=a,\G_1,\G_2] \bigg) \notag \\
    & + \E_{\G_2} \bigg[ E_Y[Y \mid A=a, \G_1, \G_2] \bigg]   + \E_{\G_1} \bigg[ E_Y[Y \mid A=a,\G_1,\G_2] \bigg] - 2 T_a(P)\\
    &=\psi_{ a}(Y,A,\G_1, \G_2 ; P) 
\end{align*}
That is, the conditional expectation of the influence function under the larger adjustment set equals the influence function under the smaller set.

Finally, applying the law of total variance, we establish the variance dominance relationship:
\begin{align*}
\sigma_{a, (\G_1,\G_2, \B_2)}^2(P) & =\mathrm{var}_P\left[\psi_{a}(Y,A,\G_1,\G_2, \B_2 ; P)\right] \\
& = \mathrm{var}_P\left[\psi_{a}(Y,A,\G_1,\G_2 ;P)\right] + \E_P\left[\mathrm{var}_P\left[\psi_{ a}(Y,A,\G_1,\G_2, \B_2 ; P) \mid A, Y, \G_1,\G_2\right]\right] \\
& \geq \sigma_{a, (\G_1,\G_2)}^2(P)
\end{align*}
The inequality follows because the conditional variance term is non-negative.

Now for the full $\WCDE$ estimator, define: $\mathbf{c} \equiv (c_a)_{a \in \mathcal{A}}$, for any $\mathbf{Z}$, define $\boldsymbol{\psi}(Y,A,\mathbf{Z}; P) 
\equiv (\psi_{ a}(Y,A,\mathbf{Z};P))_{a \in \mathcal{A}}$.  
Then, writing $\sum_{a \in \mathcal{A}} c_a \psi_{a}(Y,A,\mathbf{Z};P) 
= \mathbf{c}^\top \boldsymbol{\psi}(Y,A,\mathbf{Z}; P)$  

note that:
\[
\mathbb{E}_P \left[ \mathbf{c}^\top \boldsymbol{\psi}(Y,A,\G_1,\G_2,\B_2; P) \mid A,Y,\G_1,\G_2\right] =\mathbf{c}^\top \boldsymbol{\psi} (Y,A,\G_1,\G_2;P) 
\]
So again by variance decomposition:
\begin{align*}
&\sigma^2_{(\G_1,\G_2,\B_2)}(P)=\operatorname{var}_P \left[ \mathbf{c}^\top \boldsymbol{\psi}(Y,A,\G_1,\G_2,\B_2; P) \right] \\
&= \mathrm{var}_P[\mathbb{E}_P \left[ \mathbf{c}^\top \boldsymbol{\psi}(Y,A,\G_1,\G_2,\B_2; P) \mid A,Y,\G_1,\G_2\right]] + \mathbb{E}_P \left[ \mathrm{var}_P \left[ \mathbf{c}^\top \boldsymbol{\psi}(Y,A,\G_1,\G_2,\B_2; P)  \mid A, Y, \G_1,\G_2 \right] \right] \\
&= \mathrm{var}_P \left[ \mathbf{c}^\top \boldsymbol{\psi} (Y,A,\G_1,\G_2;P) \right] + \mathbf{c}^\top \mathbb{E}_P \left[ \mathrm{var}_P \left[\boldsymbol{\psi} (Y,A,\G_1,\G_2,\B_2; P) \mid A, Y, \G_1,\G_2\right] \right] \mathbf{c} \\
&= \sigma^2_{(\G_1,\G_2)}(P) +\mathbf{c}^\top \mathbb{E}_P \left[ \mathrm{var}_P \left[\boldsymbol{\psi} (Y,A,\G_1,\G_2,\B_2; P) \mid A, Y, \G_1,\G_2 \right] \right] \mathbf{c}\\
&\geq \sigma^2_{(\G_1,\G_2)}(P)
\end{align*}
This completes the proof.

\end{proof}

\subsection{Proof of Lemma \ref{lemma:3}}\label{app:proof_lemma3}
\begin{proof}(Proof of Lemma \ref{lemma:3}: Supplement of mediator variables)\label{appendix:lemma3}

We begin by showing that the set $(\G_1,\B_1,\G_2)$ remains a VAS.
We can verify this by comparing the identified WCDE functionals.
Recall that for any adjustment set $\Z=(\Z_1,\Z_2)$, the WCDE is identified as
\[
\mathrm{WCDE}_\Z
= 
\EE_{\Z_1}\!\Big\{
   \EE_{\Z_2}\!\big[\EE(Y \mid A = a, \Z_1, \Z_2)\big]
   -
   \EE_{\Z_2}\!\big[\EE(Y \mid A = a^\ast, \Z_1, \Z_2)\big]
\Big\}.
\]
For $(\G_1,\B_1,\G_2)$, consider the $a$-term:
\begin{align*}
T_a(\G_1,\B_1,\G_2)
&=
\EE_{\G_1,\B_1}\!\Bigg[
  \EE_{\G_2}\!\Big[
    \EE\!\big[Y \mid A=a,\G_1,\B_1,\G_2\big]
  \Big]
\Bigg] \\
&= 
\EE_{\B_1}\!\Bigg[
  \EE_{\G_1\mid \B_1}\!\Big[
    \EE_{\G_2}\!\Big[
      \EE\!\big[Y \mid A=a,\G_1,\B_1,\G_2\big]
    \Big]
  \Big]
\Bigg].
\end{align*}

By the conditional independence assumption:
\begin{equation}
\G_1 \perp\!\!\!\perp_{\mathcal{G}} A, \G_2 \mid \B_1
\label{eq:lemma3}
\end{equation}
we have
$p(\G_1,\B_1)=p(\G_1\mid \B_1)p(\B_1)=p(\G_1\mid A=a,\G_2,\B_1)p(\B_1)$.
Applying the law of iterated expectations 
\begin{align*}
&\EE_{\G_1\mid \B_1}\!\Big[
    \EE_{\G_2}\!\Big[
      \EE\!\big[Y \mid A=a,\G_1,\B_1,\G_2\big]
    \Big]
  \Big]
=
\EE_{\G_2}\!\Big[
\EE_{\G_1\mid \B_1}
 \!\left[\EE(Y \mid A=a,\G_1,\B_1,\G_2)\right]\Big]\\
 &=
\EE_{\G_2}\!\Big[
\EE_{\G_1\mid \B_1,\G_2,A=a}
 \!\left[\EE(Y \mid A=a,\G_1,\B_1,\G_2)\right]\Big]\\
 &=
\EE_{\G_2}
 \!\left[\EE(Y \mid A=a, \B_1,\G_2)\right].
\end{align*}
Therefore
\[
T_a(\G_1,\B_1,\G_2)
=
\EE_{\B_1}\!\Big[\EE_{\G_2}\!\big[\EE(Y \mid A=a,\B_1,\G_2)\big]\Big]
= T_a(\B_1,\G_2),
\]
and the same equality holds for $a^\ast$. Hence,
\[
\mathrm{WCDE}_{(\G_1,\B_1,\G_2)}
=
\mathrm{WCDE}_{(\B_1,\G_2)}.
\]
By the uniqueness of the identified WCDE functional, any adjustment set that yields the same functional as a VAS must itself be a VAS. Therefore, $(\G_1,\B_1,\G_2)$ is also a VAS.

We now compare the asymptotic variances of two VASs, $(\B_1,\G_2)$ and $(\G_1,\B_1,\G_2)$.
 For clarity, we first focus on the first component of the $\WCDE$ in Equation~\eqref{eq:WCDE_identify},  and let $\sigma_{a,\mathbf{Z}}^2(P)$ denote its asymptotic variance. The extension to the full $\WCDE$ is straightforward.
We begin by decomposing the variance:
\begin{align*}
&\sigma^2_{a, ( \B_1, \G_2)}(P)
= \operatorname{Var}_P\left[\psi_{a}(Y,A,\B_1, \G_2; P)\right] = \operatorname{Var}_P\left[\psi_{a}(Y,A,\G_1, \B_1, \G_2; P)\right] \\
&+ \operatorname{Var}_P\left[\psi_{a}(Y,A,\B_1, \G_2; P) - \psi_{a}(Y,A,\G_1, \B_1, \G_2;P)\right] \\
& + 2\,\operatorname{Cov}_P\left(
\psi_{a}(Y,A,\G_1, \B_1, \G_2;P),
\psi_{a}(Y,A,\B_1, \G_2;P) - \psi_{a}(Y,A,\G_1, \B_1, \G_2;P)
\right).
\end{align*}
To analyze the covariance term, consider the difference in IFs evaluated at $(Y, A, \G_1, \B_1, \G_2)$:
\begin{align*}
& \psi_{a}(Y,A,\B_1,\G_2 ; P) - \psi_{a}(Y,A,\G_1,\B_1,\G_2 ; P)\\
&=\frac{ I_a(A) p(\B_1)  p(\G_2)}
    {p( \B_1, \G_2,a)} \cdot \bigg( Y - \mathbb{E}_Y[Y \mid A=a,\B_1,\G_2] \bigg) \notag \\
    & + \mathbb{E}_{\G_2} \bigg[ \mathbb{E}_Y[Y \mid A=a,\B_1,\G_2] \bigg]   
    + \mathbb{E}_{\B_1} \bigg[ \mathbb{E}_Y[Y \mid A=a,\B_1,\G_2] \bigg]\\
    &-\frac{ I_a(A) p(\G_1, \B_1)  p(\G_2)}
    {p( \G_1, \B_1, \G_2,a)} \cdot \bigg( Y - \mathbb{E}_Y[Y \mid A=a,\G_1, \B_1,\G_2] \bigg) \notag \\
    & - \mathbb{E}_{\G_2} \bigg[ \mathbb{E}_Y[Y \mid A=a,\G_1,\B_1, \G_2] \bigg]   
    - \mathbb{E}_{\G_1,\B_1} \bigg[ \mathbb{E}_Y[Y \mid A=a,\G_1, \B_1, \G_2] \bigg]\\
&=  \frac{I_{a}(A) \cdot p(\B_1) \cdot p(\G_2)}
    {p(A=a, \B_1,\G_2)} \cdot \bigg(  \mathbb{E}_Y[Y \mid A=a,\G_1,\B_1,\G_2] -  \mathbb{E}_Y[Y \mid A=a,\B_1,\G_2]\bigg) \notag \\
    & + \mathbb{E}_{\G_2} \bigg[ \mathbb{E}_Y[Y \mid A=a,\B_1,\G_2] \bigg]   + \mathbb{E}_{\B_1} \bigg[ \mathbb{E}_Y[Y \mid A=a,\B_1,\G_2] \bigg] \\
    & - \mathbb{E}_{\G_2} \bigg[ \mathbb{E}_Y[Y \mid A=a,\G_1,\B_1,\G_2] \bigg]   - \mathbb{E}_{\G_1,\B_1}\bigg[ \mathbb{E}_Y[Y \mid A=a,\G_1,\B_1,\G_2] \bigg] \\
\end{align*}
The second equality above holds under the conditional independence assumption\ref{eq:lemma3}, from which it follows that
\[
p(\G_1 \mid \B_1) = p(\G_1 \mid \B_1, \G_2, A=a),
\]
which implies:
\[
\frac{ p(\G_1, \B_1) p(\G_2) }{ p(\G_1, \B_1, \G_2, A=a) }
= \frac{ p(\B_1) p(\G_2) }{ p(\B_1, \G_2, A=a) } \cdot \frac{ p(\G_1 \mid \B_1) }{ p(\G_1 \mid \B_1, \G_2, A=a) }
= \frac{ p(\B_1) p(\G_2) }{ p(A=a, \B_1, \G_2) }.
\]
Therefore, the first terms in the two influence functions cancel out. To simplify the final term, note that:
\begin{align*}
&\mathbb{E}_{\G_1,\B_1}\bigg[ \mathbb{E}_Y[Y \mid A=a,\G_1,\B_1,\G_2] \bigg]=\mathbb{E}_{\B_1}\bigg[\E_{\G_1\mid\B_1}\bigg[ \mathbb{E}_Y[Y \mid A=a,\G_1,\B_1,\G_2] 
\bigg]\bigg]\\&=\mathbb{E}_{\B_1}\bigg[\E_{\G_1\mid\B_1,\G_2,A=a}\bigg[ \mathbb{E}_Y[Y \mid A=a,\G_1,\B_1,\G_2] \bigg]\bigg]=\mathbb{E}_{\B_1} \bigg[ \mathbb{E}_Y[Y \mid A=a,\B_1,\G_2] \bigg]
\end{align*}
this equality follows from the conditional independence assumption \ref{eq:lemma3} and the law of iterated expectations. Substituting back, the difference simplifies to:
\begin{align*}
& \frac{I_{a}(A) \cdot p(\B_1) \cdot p(\G_2)}
    {p(A=a, \B_1,\G_2)} \cdot \bigg(  \mathbb{E}_Y[Y \mid A=a,\G_1,\B_1,\G_2] -  \mathbb{E}_Y[Y \mid A=a,\B_1,\G_2]\bigg) \notag \\
    & + \mathbb{E}_{\G_2} \bigg[ \mathbb{E}_Y[Y \mid A=a,\B_1,\G_2] \bigg]   + \mathbb{E}_{\B_1} \bigg[ \mathbb{E}_Y[Y \mid A=a,\B_1,\G_2] \bigg] \\
    & - \mathbb{E}_{\G_2} \bigg[ \mathbb{E}_Y[Y \mid A=a,\G_1,\B_1,\G_2] \bigg]   - \mathbb{E}_{\B_1,\G_1}\bigg[ \mathbb{E}_Y[Y \mid A=a,\G_1,\B_1,\G_2] \bigg] \\
    &= \frac{I_{a}(A) \cdot p(\B_1) \cdot p(\G_2)}
    {p(A=a, \B_1, \G_2)} \cdot \bigg(  \mathbb{E}_Y[Y \mid A=a,\G_1,\B_1,\G_2] -  \mathbb{E}_Y[Y \mid A=a,\B_1,\G_2]\bigg) \notag \\
    & + \mathbb{E}_{\G_2} \bigg[ \mathbb{E}_Y[Y \mid A=a,\B_1,\G_2] \bigg] - \mathbb{E}_{\G_2} \bigg[ \mathbb{E}_Y[Y \mid A=a,\G_1,\B_1,\G_2] \bigg]
\end{align*}
Therefore, we arrive at the simplified expression for their difference.

We then construct a parametric submodel that perturbs only the conditional law \( P(A, \G_2 \mid \G_1, \B_1) \), while keeping all other components of the distribution fixed. We show that under the assumption \ref{eq:lemma3}, the difference between the two IFs is orthogonal to the influence function of the larger adjustment set:
\[
\mathbb{E}_P\left[\psi_a(Y, A, \G_1, \B_1, \G_2;P) \cdot \left(
\psi_a(Y, A, \B_1, \G_2;P) - \psi_a(Y, A, \G_1, \B_1, \G_2;P)
\right)\right] = 0.
\]
Hence, the covariance term vanishes. Substituting this into the variance decomposition yields:
\begin{align*}
&\operatorname{Var}_P[\psi_a(Y, A, \B_1, \G_2;P)]\\
&= \operatorname{Var}_P[\psi_a(Y, A, \G_1, \B_1, \G_2;P)] 
+ \operatorname{Var}_P[\psi_a(Y, A, \B_1, \G_2;P) - \psi_a(Y, A, \G_1, \B_1, \G_2;P)],
\end{align*}
where the second term is nonnegative. To formalize this orthogonality, we appeal to two key results from semiparametric theory~\citep{tsiatis2006semiparametric}:
\begin{enumerate}
    \item If \( P_t \) is a smooth parametric submodel with \( P_0 = P \) and score function \( g \), then the influence function \( \psi_a \) satisfies:
    \[
    \left.\frac{d}{dt} T_a(P_t) \right|_{t=0} = \mathbb{E}_P[\psi_a \cdot g].
    \]
    \item In a nonparametric model, any function \( g \) such that
    \[
    \mathbb{E}_P[g(A, \G_1, \B_1, \G_2) \mid \G_1, \B_1] = 0
    \]
    is a valid score function for the conditional law \( P(A, \G_2 \mid \G_1, \B_1) \).
\end{enumerate}
Let us now consider any distribution \( P \) and define the function
\[
g(A, \G_1, \B_1, \G_2) := \psi_a(Y, A, \B_1, \G_2; P) - \psi_a(Y, A, \G_1, \B_1, \G_2; P),
\]
which satisfies the second condition above. Using this function, we define the following submodel for sufficiently small \( t \geq 0 \):
\[
P_t(Y, A, \G_1, \B_1, \G_2) := P(Y \mid A, \G_1, \B_1, \G_2) \cdot P_t(A, \G_2 \mid \G_1, \B_1) \cdot P(\G_1, \B_1),
\]
where
\[
P_t(A, \G_2 \mid \G_1, \B_1) = P(A, \G_2 \mid \G_1, \B_1)(1 + t \cdot g(A, \G_1, \B_1, \G_2)).
\]
This submodel perturbs only the conditional law of \( (A, \G_2) \) given \( (\G_1, \B_1) \), leaving the remaining components of the joint distribution fixed.
The score function of this model at \( t = 0 \) is
\[
\left.\frac{\partial}{\partial t} \log p_t(Y, A, \G_1, \B_1, \G_2)\right|_{t=0}
= \left.\frac{\partial}{\partial t} \log \{1 + t \cdot g(A, \G_1, \B_1, \G_2)\}\right|_{t=0}
= g(A, \G_1, \B_1, \G_2).
\]
Applying result (1), we obtain:
\[
\left.\frac{d}{dt} T_a(P_t) \right|_{t=0} = \mathbb{E}_P[\psi_a \cdot g].
\]
Now consider the functional:
\[
T_a(P) =\sum_{\mathbf{\G_1,\B_1}} \sum_{\mathbf{\G_2}} \left[\mathbb{E}[Y \mid a, \G_1, \B_1, \G_2]\right] p(\G_2) p(\G_1,\B_1).
\]
This functional depends only on the joint distribution of \( (\G_1, \B_1) \), \( \G_2 \), and the conditional distribution \( P(Y \mid A, \G_1, \B_1, \G_2) \). If the function \( g \) satisfies
\begin{equation}
\mathbb{E}_P[g(A, \G_1, \B_1, \G_2)\mid \G_1, \B_1] = 0 \quad \text{and} \quad \mathbb{E}_P[g(A, \G_1, \B_1, \G_2) \mid \G_2] = 0,
\label{eq:orthogonality_conditions2}
\end{equation}
then perturbations in \( P_t(Y, A, \G_1, \B_1, \G_2) \) do not affect those distributions. That is, the submodel is locally ancillary for the functional \( T_a \), and therefore:
\[
\left.\frac{d}{dt} T_a(P_t) \right|_{t=0} = 0.
\]
Combining with the earlier result gives:
\[
\mathbb{E}_P[\psi_a g] = 0
\]
Then we obtain the orthogonality:
\begin{align*}
&\E_P[{\psi}_{a}(Y,A,\G_1, \B_1,\G_2 ; P) g(A, \G_1, \B_1,\G_2)]
=0 
\end{align*}
We now verify that this function \( g(A, \G_1, \B_1, \G_2) \) satisfies the two orthogonality conditions in Condition \ref{eq:orthogonality_conditions2}. First, consider the conditional expectation given \( \G_1, \B_1 \):
\begin{align*}
 &\E_P[\psi_{a}(Y,A,\B_1,\G_2 ; P) - \psi_{a}(Y,A,\G_1,\B_1,\G_2 ; P) \mid \G_1,\B_1]\\
&=\E_P\Bigg\{ 
 \frac{I_{a}(A) \cdot p(\G_2) \cdot p(\B_1)}
    {p(A=a, \B_1, \G_2)} \cdot \bigg( \mathbb{E}_Y[Y \mid A=a,\G_1,\B_1,\G_2] 
    -  \mathbb{E}_Y[Y \mid A=a,\B_1,\G_2] \bigg) \\
& + \mathbb{E}_{\G_2} \bigg[ \mathbb{E}_Y[Y \mid A=a,\B_1,\G_2] \bigg] 
    - \mathbb{E}_{\G_2} \bigg[ \mathbb{E}_Y[Y \mid A=a,\G_1,\B_1,\G_2] \bigg]\mid \G_1,\B_1
\Bigg\}\\ 
&=\E_P\Bigg\{ 
 \frac{ p(\G_1,\B_1) \cdot p(\G_2)}
    {p(\G_1, \B_1, \G_2)} \cdot \bigg( \mathbb{E}_Y[Y \mid A=a,\G_1,\B_1,\G_2] 
    -  \mathbb{E}_Y[Y \mid A=a,\B_1,\G_2] \bigg) \\
& + \mathbb{E}_{\G_2} \bigg[ \mathbb{E}_Y[Y \mid A=a,\B_1,\G_2] \bigg] 
    - \mathbb{E}_{\G_2} \bigg[ \mathbb{E}_Y[Y \mid A=a, \G_1, \B_1, \G_2] \bigg]\mid \G_1, \B_1
\Bigg\}\\  
&=\E_P\Bigg\{ 
 \frac{ p(\G_1,\B_1) \cdot p(\G_2)}
    {p(\G_1, \B_1, \G_2)} \cdot \bigg( \mathbb{E}_Y[Y \mid A=a,  \G_1, \B_1,\G_2] 
    -  \mathbb{E}_Y[Y \mid A=a,  \B_1,\G_2] \bigg) \mid \B_1,\G_1
\Bigg\}\\  
& + \mathbb{E}_{\G_2} \bigg[ \mathbb{E}_Y[Y \mid A=a, \B_1, \G_2] \bigg] 
    - \mathbb{E}_{\G_2} \bigg[ \mathbb{E}_Y[Y \mid A=a, \G_1, \B_1, \G_2] \bigg]\\
\end{align*}
In the second equality, we integrate over \(A\) conditional on \((\G_1, \B_1)\). Note that the last two expectation terms are fixed once \((\G_1, \B_1)\) are given, so we can move them outside the outer expectation accordingly.

We now rewrite the first expectation as an integral over $\G_2$ conditional on $(\G_1, \B_1)$:
\begin{align*}
&\E_P\Bigg\{ 
 \frac{ p(\G_1,\B_1) \cdot p(\G_2)}
    {p(\G_1, \B_1,\G_2)} \cdot \bigg( \mathbb{E}_Y[Y \mid A=a, \G_1, \B_1, \G_2] 
    -  \mathbb{E}_Y[Y \mid A=a, \B_1, \G_2] \bigg) \mid \G_1,\B_1
\Bigg\}\\  \\
 &= \int \frac{ p(\G_1,\B_1)\cdot p(\g_2)}
{p(\G_1, \B_1,\g_2)} 
\left(\mathbb{E}_Y[Y|A=a,\G_1,\B_1, \g_2]-\mathbb{E}_Y[Y|A=a,\B_1,\g_2]\right)
p(\g_2| \G_1, \B_1)\, d\g_2 \\[5pt]
&= \int \left(\mathbb{E}_Y[Y \mid A=a,\G_1,\B_1,\g_2]-\mathbb{E}_Y[Y \mid A=a,\B_1,\g_2]\right)
p(\g_2)\, d\g_2\\[5pt]
&= \mathbb{E}_{\G_2} \bigg[ \mathbb{E}_Y[Y \mid A=a, \G_1, \B_1, \G_2] \bigg] 
    - \mathbb{E}_{\G_2} \bigg[ \mathbb{E}_Y[Y \mid A=a, \B_1, \G_2] \bigg]  \\
\end{align*}
Substituting this expression back into the original conditional expectation, we obtain:
\begin{align*}
 &\E_P[\psi_{a}(Y,A,\B_1,\G_2 ; P) - \psi_{a}(Y,A,\G_1,\B_1,\G_2 ; P) \mid \G_1,\B_1]\\
    &= \mathbb{E}_{\G_2} \bigg[ \mathbb{E}_Y[Y \mid A=a, \G_1, \B_1, \G_2] \bigg] 
    - \mathbb{E}_{\G_2} \bigg[ \mathbb{E}_Y[Y \mid A=a, \B_1, \G_2] \bigg]  \\
& + \mathbb{E}_{\G_2} \bigg[ \mathbb{E}_Y[Y \mid A=a, \B_1, \G_2] \bigg] 
    - \mathbb{E}_{\G_2} \bigg[ \mathbb{E}_Y[Y \mid A=a, \G_1, \B_1, \G_2] \bigg]\\
    &=0
\end{align*}
This establishes that the function \( g(A, \G_1, \B_1, \G_2) \) satisfies the first orthogonality condition in Condition~\ref{eq:orthogonality_conditions1}, $\mathbb{E}_P[g(A, \G_1, \B_1, \G_2)\mid \G_1, \B_1] = 0 $.

To verify that it satisfies the second orthogonality condition, we proceed similarly by conditioning on \( \G_2 \):
\begin{align*}
&\E_P[\psi_{a}(Y,A,\B_1,\G_2 ; P) - \psi_{a}(Y,A,\G_1,\B_1,\G_2 ; P) \mid \G_2] \notag \\
&=\E_P\Bigg\{ 
 \frac{I_{a}(A) \cdot p(\G_2) \cdot p(\B_1)}
    {p(A=a, \B_1, \G_2)} \cdot \bigg( \mathbb{E}_Y[Y \mid A=a,\G_1,\B_1,\G_2] 
    -  \mathbb{E}_Y[Y \mid A=a,\B_1,\G_2] \bigg) \notag \\
&\quad + \mathbb{E}_{\G_2} \bigg[ \mathbb{E}_Y[Y \mid A=a,\B_1,\G_2] \bigg] 
    - \mathbb{E}_{\G_2} \bigg[ \mathbb{E}_Y[Y \mid A=a,\G_1,\B_1,\G_2] \bigg] \mid \G_2 \Bigg\} \notag \\
&=\E_P\Bigg\{ 
 \frac{ p(\G_1,\B_1) \cdot p(\G_2)}
    {p(\G_1, \B_1, \G_2)} \cdot \bigg( \mathbb{E}_Y[Y \mid A=a,\G_1, \B_1,\G_2] 
    -  \mathbb{E}_Y[Y \mid A=a,\B_1,\G_2] \bigg) \notag \\
&\quad + \mathbb{E}_{\G_2} \bigg[ \mathbb{E}_Y[Y \mid A=a,\B_1,\G_2] \bigg] 
    - \mathbb{E}_{\G_2} \bigg[ \mathbb{E}_Y[Y \mid A=a,\G_1,\B_1,\G_2] \bigg]\mid \G_2 \Bigg\}
\end{align*}
In the second equality, we integrate over \(A\) conditional on \(\G_2\). We now rewrite the first term as an integral over $(\G_1, \B_1)$ with respect to $P(\G_1, \B_1 \mid \G_2)$, and then simplify.
\begin{align*}
&\E_P\Bigg\{ 
 \frac{ p(\G_1,\B_1) \cdot p(\G_2)}
    {p(\G_1, \B_1, \G_2)} \cdot \bigg( \mathbb{E}_Y[Y \mid A=a,\G_1, \B_1,\G_2] 
    -  \mathbb{E}_Y[Y \mid A=a,\B_1,\G_2] \bigg) \notag \\
&\quad + \mathbb{E}_{\G_2} \bigg[ \mathbb{E}_Y[Y \mid A=a,\B_1,\G_2] \bigg] 
    - \mathbb{E}_{\G_2} \bigg[ \mathbb{E}_Y[Y \mid A=a,\G_1,\B_1,\G_2] \bigg]\mid \G_2 \Bigg\}\\
&= \int \frac{ p(\g_1,\b_1)\cdot p(\G_2)}
{p( \g_1, \b_1,\G_2)} 
\left(\mathbb{E}_Y[Y \mid A=a,\g_1,\b_1,\G_2]-\mathbb{E}_Y[Y \mid A=a,\b_1,\G_2]\right)
p(\g_1,\b_1 \mid \G_2)\, d\b_1\, d\g_1 \\[5pt]
&\quad + \mathbb{E}_P\Bigg\{ 
 \mathbb{E}_{\G_2} \bigg[ \mathbb{E}_Y[Y \mid A=a,\B_1,\G_2] \bigg] 
    - \mathbb{E}_{\G_2} \bigg[ \mathbb{E}_Y[Y \mid A=a,\G_1,\B_1,\G_2] \bigg] \,\Big|\, \G_2
\Bigg\} \\[5pt]
&= \int p(\b_1,\g_1)
\left(\mathbb{E}_Y[Y \mid A=a,\g_1,\b_1,\G_2]-\mathbb{E}_Y[Y \mid A=a,\b_1,\G_2]\right)
\, d\b_1\, d\g_1 \\[5pt]
&\quad + \mathbb{E}_P\Bigg\{ 
 \mathbb{E}_{\G_2} \bigg[ \mathbb{E}_Y[Y \mid A=a,\B_1,\G_2] \bigg] 
    - \mathbb{E}_{\G_2} \bigg[ \mathbb{E}_Y[Y \mid A=a,\G_1,\B_1,\G_2] \bigg] \,\Big|\, \G_2
\Bigg\} \\[5pt]
\end{align*}
Next, we cancel the two components of the expression by invoking the conditional independence assumption \ref{eq:lemma3}.
\begin{align*}
& \int p(\g_1,\b_1)
\left(\mathbb{E}_Y[Y \mid A=a,\g_1,\b_1,\G_2]-\mathbb{E}_Y[Y \mid A=a,\b_1,\G_2]\right)
\, d\b_1\, d\g_1 \\[5pt]
&\quad + \mathbb{E}_P\Bigg\{ 
 \mathbb{E}_{\G_2} \bigg[ \mathbb{E}_Y[Y \mid A=a,\B_1,\G_2] \bigg] 
    - \mathbb{E}_{\G_2} \bigg[ \mathbb{E}_Y[Y \mid A=a,\G_1,\B_1,\G_2] \bigg] \,\Big|\, \G_2
\Bigg\} \\[5pt]
&= \int 
\left(\mathbb{E}_Y[Y \mid A=a,\g_1,\b_1,\G_2]-\mathbb{E}_Y[Y \mid A=a,\b_1,\G_2]\right)p(\g_1\mid \b_1, A=a, \G_2)p(\b_1)
\, d\g_1\, d\b_1 \\[5pt]
&\quad + \mathbb{E}_P\Bigg\{ 
 \mathbb{E}_{\G_2} \bigg[ \mathbb{E}_Y[Y \mid A=a,\B_1,\G_2] \bigg] 
    - \mathbb{E}_{\G_2} \bigg[ \mathbb{E}_Y[Y \mid A=a,\G_1,\B_1,\G_2] \bigg] \,\Big|\, \G_2
\Bigg\} \\[5pt]
&= \mathbb{E}_P\Bigg\{ 
 \mathbb{E}_{\G_2} \bigg[ \mathbb{E}_Y[Y \mid A=a,\B_1,\G_2] \bigg] 
    - \mathbb{E}_{\G_2} \bigg[ \mathbb{E}_Y[Y \mid A=a,\G_1,\B_1,\G_2] \bigg] \,\Big|\, \G_2
\Bigg\}
\end{align*}
The second equality holds because  we have the conditional independence assumption \ref{eq:lemma3}, so:
\[
p(\g_1, \b_1) = p(\g_1 \mid \b_1) \cdot p(\b_1) = p(\g_1 \mid \b_1, A=a, \G_2)p(\b_1),
\]
which justifies replacing the joint density in the integral. Furthermore, by the law of iterated expectations,
\[
\int \mathbb{E}_Y[Y \mid A=a, \g_1, \b_1, \G_2] \cdot p(\g_1 \mid \b_1, A=a, \G_2) \, d\g_2 = \mathbb{E}_Y[Y \mid A=a, \b_1, \G_2].
\]
As a result, the two integrals cancel out, yielding the final equality. To simplify this expression, we compute both terms as iterated integrals.
\begin{align*}
& \mathbb{E}_P\Bigg\{ 
 \mathbb{E}_{\G_2} \bigg[ \mathbb{E}_Y[Y \mid A=a,\B_1,\G_2] \bigg] 
    - \mathbb{E}_{\G_2} \bigg[ \mathbb{E}_Y[Y \mid A=a,\G_1,\B_1,\G_2] \bigg] \,\Big|\, \G_2
\Bigg\}\\
&= \iiint y \cdot \left[ p(y \mid A=a, \b_1, \g_2) \right] \, dy \, p(\g_2) \, d\g_2 \cdot p(\b_1 \mid \G_2) \, d\b_1\\
&-\iiiint y \cdot \left[p(y \mid A=a, \g_1, \b_1, \g_2) \right] \, dy \, p(\g_2) \, d\g_2  \cdot p(\g_1, \b_1 \mid \G_2) \, d\g_1 d\b_1\\
\end{align*}
Based on Condition~\ref{eq:lemma3}, we can simplify the second expression through the following derivation. First, we apply Condition~\ref{eq:lemma3} ($p(\g_1\mid \b_1)=p(\g_1\mid A=a, \b_1,\G_2)$) to expand the conditioning set:
\begin{align*}
&\iiiint y \cdot \left[p(y \mid A=a, \g_1, \b_1, \g_2) \right] \, dy \, p(\g_2) \, d\g_2  \cdot p(\g_1, \b_1 \mid \G_2) \, d\g_1 d\b_1\\
&=\iiiint y \cdot p(y \mid A=a, \g_1, \b_1, \g_2) \, dy \, p(\g_2) d\g_2 \cdot p(\g_1 \mid \b_1) d\g_1 \cdot p(\b_1 \mid \G_2) d\b_1 \\
&= \iiiint y \cdot p(y \mid A=a, \g_1, \b_1, \g_2) \, dy \, p(\g_2) d\g_2 \cdot p(\g_1 \mid A=a, \b_1, \g_2) d\g_1 \cdot p(\b_1 \mid \G_2) d\b_1
\end{align*}
Next, we express the integrand as a joint probability distribution over $y$ and $\g_2$:
\begin{align*}
&= \iiiint y \cdot p(y, \g_1 \mid A=a, \b_1, \g_2) \, dy d\g_1 \, p(\g_2) d\g_2 \cdot p(\b_1 \mid \G_2) d\b_1
\end{align*}
Using the chain rule of probability, we factorize the joint probability:
\begin{align*}
&= \iiiint y \cdot p(\g_1 \mid y, A=a, \b_1, \g_2) p(y \mid A=a, \b_1, \g_2) \, d\g_1 dy \, p(\g_2) d\g_2 \cdot p(\b_1 \mid \G_2) d\b_1
\end{align*}
Finally, we marginalize over $\g_1$ by integrating it out (since $\int p(\g_1 \mid y, A=a, \b_1, \g_2) d\g_1 = 1$), yielding the simplified expression:
\begin{align*}
&= \iiint y \cdot p(y \mid A=a, \b_1, \g_2) \, dy \, p(\g_2) d\g_2 \cdot p(\b_1 \mid \G_2) d\b_1
\end{align*}
This shows that the second term equals the first one. Thus, we conclude:
\begin{align*}
\mathbb{E}_P\Bigg\{ 
 \mathbb{E}_{\G_2} \bigg[ \mathbb{E}_Y[Y \mid A=a,\B_1,\G_2] \bigg] 
    - \mathbb{E}_{\G_2} \bigg[ \mathbb{E}_Y[Y \mid A=a,\G_1,\B_1, \G_2] \bigg] \,\Big|\, \G_2
\Bigg\} =0
\end{align*}
So far we have shown that  the function \( g(A, \G_1, \B_1, \G_2) \) satisfies Condition~\ref{eq:orthogonality_conditions2}, so we obtain the orthogonality:
\[
\mathbb{E}_P\left[\psi_a(Y, A, \G_1, \B_1, \G_2;P) \cdot \left(
\psi_a(Y, A, \B_1, \G_2;P) - \psi_a(Y, A, \G_1, \B_1, \G_2;P)
\right)\right] = 0.
\]
Therefore, we conclude that the variance decomposes as:
\begin{align*}
& \sigma_{a, (\B_1,\G_2)}^2(P) = \operatorname{var}_P\left[\psi_{a}(Y,A,\B_1,\G_2 ; P)\right] \\
& =\operatorname{var}_P\left[\psi_{a}(Y,A,\G_1, \B_1,\G_2 ; P)\right]+\operatorname{var}_P\left[\psi_{a}(Y,A,\G_1,\B_1 ;P)-\psi_{a}(Y,A,\G_1,\B_1,\G_2 ; P)\right]\\
&\geq \sigma_{a, (\G_1, \B_1,\G_2)}^2(P)
\end{align*}
Now extend this to the vectorized $\WCDE$. Define: $\mathbf{c} \equiv (c_a)_{a \in \mathcal{A}}$, 
$\mathbf{Q} \equiv (Q_a)_{a \in \mathcal{A}}$ where 
\[
Q_a= \psi_{a}(Y,A,\B_1, \G_2 ;P) - \psi_{a}(Y,A,\G_1,\B_1,\G_2 ; P)
\]
For any $\mathbf{Z}$, define $\boldsymbol{\psi}(Y,A,\mathbf{Z}; P) 
\equiv (\psi_{ a}(Y,A,\mathbf{Z};P))_{a \in \mathcal{A}}$.  
Then, writing $\sum_{a \in \mathcal{A}} c_a \psi_{a}(Y,A,\mathbf{Z};P) 
= \mathbf{c}^\top \boldsymbol{\psi}(Y,A,\mathbf{Z}; P)$  
We notice that:
\[
 \E_P[\Q\mid \G_1,\B_1]=\mathbf{0},
\]
\[
 \E_P[\Q \mid \G_2]=\mathbf{0},
\]
Then we obtain the orthogonality:
\begin{align*}
&\E_P[\Q^\top\boldsymbol{\psi}(Y,A,\G_1, \B_1,\G_2 ; P)]
=\mathbf{0} 
\end{align*}
Therefore, we conclude that the variance decomposes as:
\begin{align*}
    &\sigma^2_{ (\B_1, \G_2)}(P) = \operatorname{var}_P \left[ \mathbf{c}^\top\boldsymbol{\psi}(Y,A,\B_1,\G_2; P) \right] \\
=&\operatorname{var}_P \left[ \mathbf{c}^\top \boldsymbol{\psi}(Y,A,\G_1, \B_1,\G_2; P) \right] 
+ \operatorname{var}_P \left[ \mathbf{c}^\top \mathbf{Q} \right] 
= \sigma^2_{(\G_1,\B_1,\G_2)}(P) + \mathbf{c}^\top \operatorname{var}_P(\mathbf{Q}) \mathbf{c} \\
\geq& \sigma^2_{(\G_1,\B_1,\G_2)}(P)
\end{align*}
This completes the proof.

\end{proof}

\subsection{Proof of Lemma \ref{lemma:4}}\label{app:proof_lemma4}
\begin{proof}(Proof of Lemma \ref{lemma:4}: Deletion of overadjustment mediator variables)\label{appendix:lemma4}

We begin by showing that the set $(\G_1,\G_2)$ remains a valid adjustment set (VAS).
We can verify this by comparing the identified WCDE functionals.
For any adjustment set $\Z=(\Z_1,\Z_2)$, the WCDE is identified as
\[
\mathrm{WCDE}_\Z
= 
\EE_{\Z_1}\!\Big\{
   \EE_{\Z_2}\!\big[\EE(Y \mid A = a, \Z_1, \Z_2)\big]
   -
   \EE_{\Z_2}\!\big[\EE(Y \mid A = a^\ast, \Z_1, \Z_2)\big]
\Big\}.
\]
For $(\G_1,\B_1,\G_2)$, consider the $a$-term:
\[
T_a(\G_1,\B_1,\G_2)
=
\EE_{\G_1, \B_1}\!\Big[
  \EE_{\G_2}\!\big[\EE(Y \mid A=a,\G_1,\B_1,\G_2)\big]
\Big].
\]
From the conditional independence
\begin{equation}
Y \perp\!\!\!\perp_{\mathcal{G}} \B_1 \mid \G_1,\G_2,A,
\label{eq:lemma4}
\end{equation}
it follows that
\[
\EE\!\big[Y \mid A=a,\G_1,\B_1,\G_2\big]
= \EE\!\big[Y \mid A=a,\G_1,\G_2\big],
\]
Substituting this gives
\begin{align*}
&T_a(\G_1,\B_1,\G_2)
= 
\EE_{\G_1, \B_1}\!\Big[\EE_{\G_2}\!\big[\EE(Y \mid A=a,\G_1,\G_2)\big]\Big]\\
&=\EE_{\G_1}\!\Big[\EE_{\G_2}\!\big[\EE(Y \mid A=a,\G_1,\G_2)\big]\Big]
= T_a(\G_1,\G_2),
\end{align*}
and the same equality holds for $a^\ast$. Therefore
\[
\mathrm{WCDE}_{(\G_1,\B_1,\G_2)}
= \mathrm{WCDE}_{(\G_1,\G_2)}.
\]
By the uniqueness of the identified WCDE functional, any adjustment set that yields the same functional as a VAS must itself be a VAS. Hence $(\G_1,\G_2)$ is a VAS.

We now compare the asymptotic variances of two VASs, $(\G_1,\G_2)$ and $(\G_1, \B_1,\G_2)$.
 For clarity, we first focus on the first component of the $\WCDE$ in Equation~\eqref{eq:WCDE_identify},  and let $\sigma_{a,\mathbf{Z}}^2(P)$ denote its asymptotic variance. The extension to the full $\WCDE$ is straightforward.

We begin by applying the law of total variance to decompose the variance under the larger adjustment set $(\G_1,\B_1, \G_2)$:
\begin{align*}
\operatorname{Var}[\psi_a(Y, A, \G_1, \B_1, \G_2;P)] 
&= \operatorname{Var}[\mathbb{E}[\psi_a(Y, A, \G_1, \B_1, \G_2;P) \mid A, Y, \G_1, \G_2]] \\
&\quad + \mathbb{E}\left[\operatorname{Var}[\psi_a(Y, A, \G_1, \B_1, \G_2;P) \mid A, Y, \G_1, \G_2]\right].
\end{align*}
Next, we will show that under the conditional independence assumption \ref{eq:lemma4}
, the IF based on the adjustment set $(\G_1, \B_1, \G_2)$ satisfies  
\[
\mathbb{E}_P[\psi_a(Y, A, \G_1, \B_1, \G_2;P) \mid A, Y, \G_1, \G_2]
= \psi_a(Y, A, \G_1, \G_2;P).
\]  
This equality allows us to relate the variances of the two IFs via the law of total variance. We now compute the conditional expectation of $\psi_a(Y, A, \G_1, \B_1, \G_2;P)$:
\begin{align*}
& \E_P\left[\psi_{a}(Y,A,\G_1,\B_1, \G_2 ; P) \mid A, Y, \G_1,\G_2\right] = \\
& \underbrace{
    I_{a}(A)\left\{Y - \E_Y[Y \mid A=a,\G_1,\G_2]\right\} 
    \cdot \E_P\left[\left.\frac{p(\G_2)\cdot p(\G_1,\B_1)}{p(A=a,\G_1,\B_1,\G_2)} \right\rvert\, A=a, Y, \G_1,\G_2\right]
}_{\text{(i) first term}} \\[5pt]
& + \underbrace{
    \E_{P} \left[ \E_{\G_2} \left[ \E_Y[Y \mid A=a,\G_1,\B_1,\G_2] \right] \,\middle|\, \G_1, \B_1 \right]
}_{\text{(ii) second term}} \\[5pt]
& + \underbrace{
    \E_{P} \left[ \E_{\G_1, \B_1} \left[ \E_Y[Y \mid A=a,\G_1,\B_1,\G_2] \right] \,\middle|\, \G_2 \right]
}_{\text{(iii) third term}} 
- 2 T_a(P)
\end{align*}
This equality holds because, given the conditional independence relation \ref{eq:lemma4}, it implies the key simplification:
\[
\mathbb{E}_Y(Y \mid A=a, \G_1,\B_1,\G_2) = \mathbb{E}_Y(Y \mid A=a, \G_1,\G_2).
\]
For the first conditional expectation term, we perform the detailed computation:
\begin{align*}
    &\mathbb{E}_P \left[ \frac{p(\G_1,\B_1) \cdot p(\G_2)}{p(A, \G_1, \B_1, \G_2)} \,\Big|\, A = a, Y, \G_1, \G_2 \right] \\
    &=\; \mathbb{E}_P \left[ \frac{ p(\G_1,\B_1) \cdot p(\G_2) }{p(\B_1 \mid A = a, \G_1, \G_2) \cdot p(A=a, \G_1, \G_2)} \,\Big|\, A = a, Y, \G_1, \G_2 \right] \\
    &=\; \frac{1}{p(A=a, \G_1, \G_2)} \cdot \mathbb{E}_P \left[ \frac{ p(\G_1,\B_1) \cdot p(\G_2)}{p(\B_1 \mid A = a, \G_1, \G_2)} \,\Big|\, A = a, Y, \G_1, \G_2 \right] \\
       &=\; \frac{1}{p(A=a, \G_1, \G_2)} \cdot \mathbb{E}_P \left[ \frac{ p(\G_1,\B_1) \cdot p(\G_2) }{p(\B_1 \mid A = a, \G_1, \G_2)} \,\Big|\, A = a, Y, \G_1, \G_2 \right] \\
    &=\; \frac{1}{p(A=a, \G_1, \G_2)} \cdot \int_{\b_1} \frac{p(\G_1, \b_1)\cdot p(\G_2)  }{p(\b_1 \mid A = a, \G_1, \G_2)} \cdot p(\b_1 \mid A = a, Y, \G_1, \G_2) \, d\b_1 \\
    &=\; \frac{1}{p(A=a,\G_1, \G_2)} \cdot \int_{\b_1}p(\G_1, \b_1) \cdot p(\G_2)\cdot \frac{p(\b_1 \mid A = a, Y, \G_1, \G_2)}{p(\b_1 \mid A = a, \G_1, \G_2)} \, d\b_1 \\
\end{align*}
In the last line above, we expand the conditional expectation as an integral over \(\B_1\). Now, under the conditional independence assumption \ref{eq:lemma4}, it follows that
\[
p(\b_1 \mid A = a, Y, \G_1, \G_2) = p(\b_1 \mid A = a, \G_1, \G_2).
\]
Substituting this into the integrand, the ratio simplifies to 1:
\begin{align*}
    &\; \frac{1}{p(A=a, \G_1, \G_2)} \cdot \int_{\b_1} p(\G_2) \cdot p(\G_1,\b_1) \cdot \frac{p(\b_1 \mid A = a, Y, \G_1, \G_2)}{p(\b_1 \mid A = a, \G_1, \G_2)} \, d\b_1 \\
    &=\; \frac{1}{p(A=a, \G_1, \G_2)} \cdot \int_{\b_1} p(\G_2) \cdot p(\G_1, \b_1) \, d\b_1 \\
    &=\; \frac{1}{p(A=a, \G_1, \G_2)} \cdot \int_{\b_1} p(\b_1 \mid \G_1) \cdot p(\G_1) \cdot p(\G_2) \, d\b_1 \\
    &=\; \frac{p(\G_1) \cdot p(\G_2)}{p(A=a, \G_1, \G_2)} \cdot \int_{\b_1} p(\b_1 \mid \G_1) \, d\b_1 \\
    &=\; \frac{p(\G_1) \cdot p(\G_2)}{p(A=a, \G_1, \G_2)}
\end{align*}
Since $\mathbb{E}_Y(Y \mid A=a, \G_1,\B_1, \G_2) = \mathbb{E}_Y(Y \mid A=a, \G_1,\G_2)$, so the second and third terms simplify as follows:
\begin{align*}
& \E_{P}  \bigg[E_{\G_2} [ E_Y[Y \mid A=a,\G_1,\B_1,\G_2] ]\mid \G_1,\B_1\bigg]\\
&= \E_{P}  \bigg[E_{\G_2} [ E_Y[Y \mid A=a,\G_1,\G_2] ]\mid \G_1\bigg]\\
&= \E_{\G_2} \bigg[ E_Y[Y \mid A=a,\G_1,\G_2] \bigg]
\end{align*}
\begin{align*}
&\E_{P}  \bigg[E_{\B_1,\G_1} [ E_Y[Y \mid A=a,\G_1,\B_1,\G_2] ]\mid \G_2\bigg]\\
&=\E_{P}  \bigg[E_{\B_1,\G_1} [ E_Y[Y \mid A=a,\G_1,\G_2] ]\mid \G_2\bigg]\\
&= \E_{\G_1} \bigg[ E_Y[Y \mid A=a,\G_1,\G_2] \bigg]
\end{align*}
Combining these results, we obtain:
\begin{align*}
&\E_P\left[\psi_{ a}(Y,A,\G_1,\B_1, \G_2 ;P) \mid A, Y, \G_1,\G_2\right]\\
&=\frac{I_{a}(A) \cdot p(\G_2) \cdot p(\G_1)}
    {p(A=a, \G_1, \G_2)} \cdot \bigg( Y - \E_Y[Y \mid A=a,\G_1,\G_2] \bigg) \notag \\
    & + \E_{\G_2} \bigg[ E_Y[Y \mid A=a, \G_1, \G_2] \bigg]   + \E_{\G_1} \bigg[ E_Y[Y \mid A=a,\G_1,\G_2] \bigg] - 2 T_a(P)\\
    &=\psi_{ a}(Y,A,\G_1, \G_2 ; P) 
\end{align*}
That is, the conditional expectation of the influence function under the larger adjustment set equals the influence function under the smaller set.

Finally, applying the law of total variance, we establish the variance dominance relationship:
\begin{align*}
\sigma_{a, (\G_1,\B_1, \G_2)}^2(P) & =\mathrm{var}_P\left[\psi_{a}(Y,A,\G_1,\B_1, \G_2 ; P)\right] \\
& = \mathrm{var}_P\left[\psi_{a}(Y,A,\G_1,\G_2 ;P)\right] + \E_P\left[\mathrm{var}_P\left[\psi_{ a}(Y,A,\G_1,\B_1, \G_2 ; P) \mid A, Y, \G_1,\G_2\right]\right] \\
& \geq \sigma_{a, (\G_1,\G_2)}^2(P)
\end{align*}
The inequality follows because the conditional variance term is non-negative.

Now for the full $\WCDE$ estimator, define: $\mathbf{c} \equiv (c_a)_{a \in \mathcal{A}}$, for any $\mathbf{Z}$, define $\boldsymbol{\psi}(Y,A,\mathbf{Z}; P) 
\equiv (\psi_{ a}(Y,A,\mathbf{Z};P))_{a \in \mathcal{A}}$.  
Then, writing $\sum_{a \in \mathcal{A}} c_a \psi_{a}(Y,A,\mathbf{Z};P) 
= \mathbf{c}^\top \boldsymbol{\psi}(Y,A,\mathbf{Z}; P)$  

note that:
\[
\mathbb{E}_P \left[ \mathbf{c}^\top \boldsymbol{\psi}(Y,A,\G_1,\B_1,\G_2; P) \mid A,Y,\G_1,\G_2\right] =\mathbf{c}^\top \boldsymbol{\psi} (Y,A,\G_1,\G_2;P) 
\]
So again by variance decomposition:
\begin{align*}
&\sigma^2_{(\G_1,\B_1,\G_2)}(P)=\operatorname{var}_P \left[ \mathbf{c}^\top \boldsymbol{\psi}(Y,A,\G_1,\B_1,\G_2; P) \right] \\
&= \mathrm{var}_P[\mathbb{E}_P \left[ \mathbf{c}^\top \boldsymbol{\psi}(Y,A,\G_1,\B_1,\G_2; P) \mid A,Y,\G_1,\G_2\right]] + \mathbb{E}_P \left[ \mathrm{var}_P \left[ \mathbf{c}^\top \boldsymbol{\psi}(Y,A,\G_1,\B_1,\G_2; P)  \mid A, Y, \G_1,\G_2 \right] \right] \\
&= \mathrm{var}_P \left[ \mathbf{c}^\top \boldsymbol{\psi} (Y,A,\G_1,\G_2;P) \right] + \mathbf{c}^\top \mathbb{E}_P \left[ \mathrm{var}_P \left[\boldsymbol{\psi} (Y,A,\G_1,\B_1,\G_2; P) \mid A, Y, \G_1,\G_2 \right] \right] \mathbf{c} \\
&= \sigma^2_{(\G_1,\G_2)}(P) +\mathbf{c}^\top \mathbb{E}_P \left[ \mathrm{var}_P \left[\boldsymbol{\psi} (Y,A,\G_1,\B_1,\G_2; P) \mid A, Y, \G_1,\G_2 \right] \right] \mathbf{c}\\
&\geq \sigma^2_{(\G_1,\G_2)}(P)
\end{align*}
This completes the proof.
\end{proof}

\subsection{Proof of Theorem \ref{thm:optimal_adjustment}}\label{app:proof_thm_optimal}
\begin{proof}(Proof of Theorem \ref{thm:optimal_adjustment})

Let \(\mathbf{Z} = (\mathbf{Z}_1, \mathbf{Z}_2)\) be a valid adjustment set, where \(\mathbf{Z}_1 = \mathbf{Z} \cap \M\) contains mediating variables and \(\mathbf{Z}_2 = \mathbf{Z} \setminus \mathbf{Z}_1\) contains non-mediators (typically backdoor covariates).

We aim to construct an adjustment set that minimizes the asymptotic variance of the $\WCDE$ estimator by augmenting and pruning \(\mathbf{Z}_1\) and \(\mathbf{Z}_2\) based on conditional independencies in the DAG.

We prove the result through a sequence of variance-reducing steps.
\subsubsection*{Step 1: Augmenting \(\mathbf{Z}_2\) with Non-mediating Parents of \(Y\)}
% \Lry{
According to Condition \ref{condition:mediator_path} and Condition \ref{condition:cde}, all backdoor paths for $\{A,Y\}$ and $\{\Z_1,Y\}$ are blocked by $\Z_2$, so we have the conditional independence:
\[
(\mathbf{X}_{1 \in \Pa(Y)} \cup \mathbf{X}_{4 \in \Pa(Y)}\setminus \Z_2) \perp\!\!\!\perp_{\mathcal{G}} A, \mathbf{Z}_1 \mid \mathbf{Z}_2.
\]
% }
Then by Lemma~\ref{lemma:1}, augmenting \(\mathbf{Z}_2\) with \(\mathbf{X}_{1 \in \Pa(Y)}\) and \(\mathbf{X}_{4 \in \Pa(Y)}\) reduces the variance:
\[
\sigma^2_{(\mathbf{Z}_1,\, \mathbf{Z}_2 \cup \mathbf{X}_{1 \in \Pa(Y)} \cup \mathbf{X}_{4 \in \Pa(Y)})}(P) \leq \sigma^2_{(\mathbf{Z}_1, \mathbf{Z}_2)}(P).
\]
\subsubsection*{Step 2: Augmenting \(\mathbf{Z}_1\) with Mediating Parents of \(Y\)}
According to Condition \ref{condition:unique_WCDE},  $\{\mathbf{M}^{\prime}
    \bigsetminus \mathbf{Z_1}\} \perp\!\!\!\perp_{\mathcal{G}} \{\Pa(Y)\bigsetminus \mathbf{M ^{\prime}}\}\mid \mathbf{Z_1}$, so we have the conditional independence:
\[
(\mathbf{X}_{3 \in \Pa(Y)} \setminus\Z_1)\perp\!\!\!\perp_{\mathcal{G}} {A}, \mathbf{Z}_2 \cup \mathbf{X}_{1 \in \Pa(Y)} \cup \mathbf{X}_{4 \in \Pa(Y)} \mid \mathbf{Z}_1.
\]
Then by Lemma~\ref{lemma:3}, adding \(\mathbf{X}_{3 \in \Pa(Y)}\) further reduces variance:
\[
\sigma^2_{(\mathbf{Z}_1 \cup \mathbf{X}_{3 \in \Pa(Y)},\, \mathbf{Z}_2 \cup \mathbf{X}_{1 \in \Pa(Y)} \cup \mathbf{X}_{4 \in \Pa(Y)})}(P) \leq \sigma^2_{(\mathbf{Z}_1,\, \mathbf{Z}_2 \cup \mathbf{X}_{1 \in \Pa(Y)} \cup \mathbf{X}_{4 \in \Pa(Y)})}(P).
\]

\subsubsection*{Step 3: Pruning Redundant Backdoor Variables from \(\mathbf{Z}_2\)}
Since \(\mathbf{X}_{1 \in \Pa(Y)} \cup \mathbf{X}_{4 \in \Pa(Y)}\), \(\mathbf{Z}_1 \cup \mathbf{X}_{3 \in \Pa(Y)}\), and \({A}\) together form the full set of parents of \(Y\),  according to the \textit{local Markov property}, \(Y\) is conditionally independent of all non-descendant non-parent variables given its parents. Therefore, we have the following conditional independence:
\[
Y \perp\!\!\!\perp_{\mathcal{G}} \mathbf{Z}_2 \setminus (\mathbf{X}_{1 \in \Pa(Y)} \cup \mathbf{X}_{4 \in \Pa(Y)}) \mid \mathbf{X}_{1 \in \Pa(Y)} \cup \mathbf{X}_{4 \in \Pa(Y)}, \mathbf{Z}_1 \cup \mathbf{X}_{3 \in \Pa(Y)}, {A}.
\]
Then by Lemma~\ref{lemma:2}, removing extraneous elements of \(\mathbf{Z}_2\) does not increase variance:
\[
\sigma^2_{(\mathbf{Z}_1 \cup \mathbf{X}_{3 \in \Pa(Y)},\, \mathbf{X}_{1 \in \Pa(Y)} \cup \mathbf{X}_{4 \in \Pa(Y)})}(P) \leq \sigma^2_{(\mathbf{Z}_1 \cup \mathbf{X}_{3 \in \Pa(Y)},\, \mathbf{Z}_2 \cup \mathbf{X}_{1 \in \Pa(Y)} \cup \mathbf{X}_{4 \in \Pa(Y)})}(P).
\]

\subsubsection*{Step 4: Pruning Redundant Mediating Variables from \(\mathbf{Z}_1\)}

Note that \(\mathbf{X}_{1 \in \Pa(Y)} \cup \mathbf{X}_{4 \in \Pa(Y)}\), together with \(\mathbf{Z}_1 \cup \mathbf{X}_{3 \in \Pa(Y)}\) and \({A}\), constitute the full set of parents of \(Y\). According to the \textit{local Markov property}, \(Y\) is conditionally independent of all non-descendant non-parent variables given its parents, so we have the conditional independence:

\[
Y \perp\!\!\!\perp_{\mathcal{G}} \mathbf{Z}_1 \setminus \mathbf{X}_{3 \in \Pa(Y)} \mid \mathbf{X}_{1 \in \Pa(Y)} \cup \mathbf{X}_{4 \in \Pa(Y)}, \mathbf{X}_{3 \in \Pa(Y)}, {A}.
\]
Then by Lemma~\ref{lemma:4}, variables in \(\mathbf{Z}_1\) outside \(\mathbf{X}_{3 \in \Pa(Y)}\) can be excluded:
\[
\sigma^2_{(\mathbf{X}_{3 \in \Pa(Y)},\, \mathbf{X}_{1 \in \Pa(Y)} \cup \mathbf{X}_{4 \in \Pa(Y)})}(P) \leq \sigma^2_{(\mathbf{Z}_1 \cup \mathbf{X}_{3 \in \Pa(Y)},\, \mathbf{X}_{1 \in \Pa(Y)} \cup \mathbf{X}_{4 \in \Pa(Y)})}(P).
\]

\subsubsection*{Conclusion}
Through these steps, we reduce the adjustment set to the minimal form that preserves validity and achieves minimal asymptotic variance. The final optimal adjustment set is:
\[
\mathbf{O} := \mathbf{X}_{1 \in \Pa(Y)} \cup \mathbf{X}_{3 \in \Pa(Y)} \cup \mathbf{X}_{4 \in \Pa(Y)}.
\]
\end{proof}

\section{Experimental Details}
\subsection{Synthetic Experiments Data-Generating Process}
\label{app:synthetic_DGP}

\textbf{Edge coefficients.} For each DAG edge, coefficients are sampled independently: with $50\%$ probability from $\mathrm{Uniform}[-1.5, -0.5]$, otherwise from $\mathrm{Uniform}[0.5, 1.5]$, ensuring non-negligible effects.

\textbf{Structural equations.} Each variable $X_i$ in the DAG is generated recursively according to a structural equation model:
\[
  X_i \;=\; \sum_{j \in \mathrm{Pa}(i)} w_{ji}\, f_j(X_j) \;+\; \varepsilon_i,
\]
where $w_{ji}$ is the sampled edge coefficient from parent $j$ to $i$. The function $f_j(\cdot)$ is applied independently to each parent variable $X_j$ before combination and is selected uniformly at random from:
\[
\text{Identity: } f(x) = x, \quad
\text{Sine: } f(x) = \sin(x), \quad
\text{Cosine: } f(x) = \cos(x).
\]

\textbf{Noise.} The additive noise term $\varepsilon_i$ is sampled independently from $\mathcal{N}(0, 1/4)$ for each node with parents. Root nodes (i.e., nodes with no parents) are sampled from $\mathcal{N}(0, 1)$.

\textbf{Outcome transformation.} For each simulation run, we randomly select one of the three nonlinear functions $f$ to transform the outcome. The final outcome $Y$ is generated by applying the chosen $f$ to its parent variables.

\textbf{Replications.}
We repeat this process for 50 randomly generated coefficient sets to assess the robustness of WCDE estimation.
For each fixed DAG, adjustment set, and sample size, we additionally perform a Monte Carlo experiment with size $100$
to estimate the empirical variance of the estimator across independent replications.

\subsection{MILDEW Network}\label{app:realworld_DAG}

\begin{table}[H]
\centering
\caption{Variance and MSE of WCDE estimates on the MILDEW network (5 representative VASs) under different sample sizes. 
Variable names are abbreviated as follows: $m_1$: meldug\_3, $m_2$: middel\_3, $k$: mikro\_3, $k_2$: mikro\_2, $l_1$: lai\_1, $l_2$: lai\_2, $l_3$: lai\_3.}
\label{tab:mildew_combined}
\resizebox{\textwidth}{!}{
\begin{tabular}{lccccccccccc}
\toprule
& & \multicolumn{2}{c}{$n=250$} & \multicolumn{2}{c}{$n=500$} & \multicolumn{2}{c}{$n=1000$} & \multicolumn{2}{c}{$n=4000$} & \multicolumn{2}{c}{$n=10000$} \\
\cmidrule(lr){3-4} \cmidrule(lr){5-6} \cmidrule(lr){7-8} \cmidrule(lr){9-10} \cmidrule(lr){11-12}
\textbf{Adj. Set} & & Var & MSE & Var & MSE & Var & MSE & Var & MSE & Var & MSE \\
\midrule
% ------------------- Optimal -------------------
{[}$\mathbf{m}_1, \mathbf{m}_2, \mathbf{k}${]} & & \textbf{0.00466} & \textbf{0.00470} & \textbf{0.00269} & \textbf{0.00269} & \textbf{0.00104} & \textbf{0.00106} & \textbf{0.00020} & \textbf{0.00021} & \textbf{0.00010} & \textbf{0.00011} \\
% ------------------- Reference -------------------
{[}$l_2, l_3, m_1, m_2, k${]} & & 0.00512 & 0.00513 & 0.00301 & 0.00302 & 0.00114 & 0.00114 & 0.00027 & 0.00028 & 0.00012 & 0.00012 \\
% ------------------- Largest -------------------
{[}$l_1$, $l_2$, $l_3$, $m$, $m_1$, $m_2$, $k_2$, $k${]} &  & 0.00609 & 0.00609 & 0.00422 & 0.00422 & 0.00248 & 0.00255 & 0.00057 & 0.00058 & 0.00026 & 0.00026 \\
% ------------------- Smallest -------------------
{[}$m_1, k${]} & & 0.01445 & 0.01521 & 0.00810 & 0.00810 & 0.00451 & 0.00455 & 0.00084 & 0.00084 & 0.00042 & 0.00042 \\
{[}$l_2, m_1${]} & & 0.01292 & 0.01400 & 0.00873 & 0.00873 & 0.00450 & 0.00460 & 0.00082 & 0.00083 & 0.00046 & 0.00046 \\
{[}$l_3, m_1${]} & & 0.01409 & 0.01527 & 0.00888 & 0.00888 & 0.00463 & 0.00472 & 0.00081 & 0.00081 & 0.00047 & 0.00047 \\
\bottomrule
\end{tabular}
}
\end{table}

The MILDEW network exhibits a highly complex structure with a large number of nodes and edges, resulting in an enormous space of potential valid adjustment sets. To maintain both interpretability and representativeness, we adopted the following construction strategy. 
First, we identified all nodes that appear on the mediator paths between the treatment variable \texttt{mikro\_1} and the outcome variable \texttt{meldug\_4}. 
We note that this particular treatment outcome pair admits no backdoor path on MILDEW network (potentially explaining the close performance of top performing VASs.)
Next, we selected relevant nodes from the parents of both the treatment and outcome. 
We then systematically enumerated all possible combinations of these nodes and filtered them using the adjustment criterion to ensure validity. 
This process yielded a total of \textbf{122} valid adjustment sets.

\begin{table}[H]
\centering
\caption[Variance and MSE on MILDEW (Top-50 by $n=10000$)]{Variance and MSE of WCDE estimates on the MILDEW network (top-50 VAS ranked by $n=10000$). \textbf{Abbreviations:} $k$ (\texttt{mikro\_3}), $k_2$ (\texttt{mikro\_2}), $l_1$ (\texttt{lai\_1}), $l_2$ (\texttt{lai\_2}), $l_3$ (\texttt{lai\_3}), $m_1$ (\texttt{meldug\_3}), $m_2$ (\texttt{middel\_3}), $m$ (\texttt{meldug\_2}).}
\label{tab:mildew_top50_combined_allabbr}
\resizebox{\linewidth}{!}{
\begin{tabular}{lccccccccccc}
\toprule
& & \multicolumn{2}{c}{$n=250$} & \multicolumn{2}{c}{$n=500$} & \multicolumn{2}{c}{$n=1000$} & \multicolumn{2}{c}{$n=4000$} & \multicolumn{2}{c}{$n=10000$} \\
\cmidrule(lr){3-4} \cmidrule(lr){5-6} \cmidrule(lr){7-8} \cmidrule(lr){9-10} \cmidrule(lr){11-12}
\textbf{Adj. Set} & & Var & MSE & Var & MSE & Var & MSE & Var & MSE & Var & MSE \\
\midrule
{[}$m_1$, $m_2$, $k${]} &  & 0.00466 & 0.00470 & 0.00269 & 0.00269 & 0.00104 & 0.00106 & 0.00020 & 0.00021 & 0.00010 & 0.00011 \\
{[}$l_2$, $m_1$, $m_2$, $k_2$, $k${]} &  & 0.00465 & 0.00465 & 0.00275 & 0.00276 & 0.00125 & 0.00125 & 0.00026 & 0.00026 & 0.00011 & 0.00011 \\
{[}$l_3$, $m_1$, $m_2$, $k_2$, $k${]} &  & 0.00470 & 0.00471 & 0.00270 & 0.00271 & 0.00123 & 0.00124 & 0.00025 & 0.00025 & 0.00011 & 0.00011 \\
{[}$l_2$, $m_1$, $m_2$, $k${]} &  & 0.00534 & 0.00534 & 0.00301 & 0.00302 & 0.00112 & 0.00112 & 0.00026 & 0.00027 & 0.00011 & 0.00012 \\
{[}$l_3$, $m_1$, $m_2$, $k${]} &  & 0.00518 & 0.00520 & 0.00300 & 0.00301 & 0.00119 & 0.00119 & 0.00024 & 0.00024 & 0.00011 & 0.00012 \\
{[}$l_2$, $l_3$, $m_1$, $m_2$, $k_2$, $k${]} &  & 0.00436 & 0.00436 & 0.00272 & 0.00272 & 0.00120 & 0.00120 & 0.00027 & 0.00027 & 0.00011 & 0.00012 \\
{[}$m_1$, $m_2$, $k_2$, $k${]} &  & 0.00443 & 0.00453 & 0.00244 & 0.00245 & 0.00117 & 0.00121 & 0.00023 & 0.00023 & 0.00012 & 0.00012 \\
{[}$l_2$, $m$, $m_1$, $m_2$, $k${]} &  & 0.00616 & 0.00622 & 0.00319 & 0.00320 & 0.00121 & 0.00121 & 0.00026 & 0.00027 & 0.00012 & 0.00012 \\
{[}$l_2$, $l_3$, $m_1$, $m_2$, $k${]} &  & 0.00512 & 0.00513 & 0.00301 & 0.00302 & 0.00114 & 0.00114 & 0.00027 & 0.00028 & 0.00012 & 0.00012 \\
{[}$l_3$, $m$, $m_1$, $m_2$, $k${]} &  & 0.00578 & 0.00587 & 0.00299 & 0.00299 & 0.00131 & 0.00131 & 0.00027 & 0.00028 & 0.00012 & 0.00012 \\
{[}$m$, $m_1$, $m_2$, $k${]} &  & 0.00603 & 0.00608 & 0.00270 & 0.00271 & 0.00121 & 0.00121 & 0.00025 & 0.00027 & 0.00012 & 0.00013 \\
{[}$l_2$, $l_3$, $m$, $m_1$, $m_2$, $k${]} &  & 0.00555 & 0.00563 & 0.00307 & 0.00308 & 0.00134 & 0.00134 & 0.00028 & 0.00029 & 0.00012 & 0.00013 \\
{[}$l_3$, $m$, $m_1$, $m_2$, $k_2$, $k${]} &  & 0.00476 & 0.00484 & 0.00285 & 0.00285 & 0.00143 & 0.00143 & 0.00026 & 0.00026 & 0.00012 & 0.00013 \\
{[}$l_2$, $l_3$, $m$, $m_1$, $m_2$, $k_2$, $k${]} &  & 0.00446 & 0.00449 & 0.00272 & 0.00273 & 0.00133 & 0.00134 & 0.00027 & 0.00027 & 0.00012 & 0.00013 \\
{[}$m$, $m_1$, $m_2$, $k_2$, $k${]} &  & 0.00531 & 0.00548 & 0.00269 & 0.00271 & 0.00130 & 0.00130 & 0.00027 & 0.00028 & 0.00012 & 0.00013 \\
{[}$l_2$, $m$, $m_1$, $m_2$, $k_2$, $k${]} &  & 0.00469 & 0.00474 & 0.00282 & 0.00282 & 0.00132 & 0.00132 & 0.00025 & 0.00026 & 0.00013 & 0.00013 \\
{[}$l_1$, $m_1$, $m_2$, $k${]} &  & 0.00554 & 0.00555 & 0.00306 & 0.00307 & 0.00118 & 0.00119 & 0.00030 & 0.00030 & 0.00013 & 0.00013 \\
{[}$l_2$, $m_1$, $m_2$, $k_2${]} &  & 0.00518 & 0.00523 & 0.00325 & 0.00326 & 0.00155 & 0.00159 & 0.00031 & 0.00031 & 0.00013 & 0.00013 \\
{[}$l_2$, $l_3$, $m_1$, $m_2$, $k_2${]} &  & 0.00548 & 0.00552 & 0.00319 & 0.00321 & 0.00152 & 0.00154 & 0.00030 & 0.00030 & 0.00013 & 0.00013 \\
{[}$l_3$, $m_1$, $m_2${]} &  & 0.00534 & 0.00540 & 0.00303 & 0.00304 & 0.00135 & 0.00138 & 0.00030 & 0.00030 & 0.00013 & 0.00013 \\
{[}$l_2$, $l_3$, $m_1$, $m_2${]} &  & 0.00548 & 0.00550 & 0.00322 & 0.00322 & 0.00147 & 0.00149 & 0.00032 & 0.00032 & 0.00013 & 0.00014 \\
{[}$l_2$, $m_1$, $m_2${]} &  & 0.00524 & 0.00525 & 0.00330 & 0.00330 & 0.00148 & 0.00151 & 0.00031 & 0.00031 & 0.00014 & 0.00014 \\
{[}$l_3$, $m_1$, $m_2$, $k_2${]} &  & 0.00558 & 0.00566 & 0.00302 & 0.00305 & 0.00142 & 0.00145 & 0.00029 & 0.00029 & 0.00014 & 0.00014 \\
{[}$l_1$, $m_1$, $m_2$, $k_2$, $k${]} &  & 0.00537 & 0.00538 & 0.00272 & 0.00275 & 0.00129 & 0.00131 & 0.00029 & 0.00029 & 0.00014 & 0.00014 \\
{[}$l_1$, $m$, $m_1$, $m_2$, $k_2$, $k${]} &  & 0.00558 & 0.00559 & 0.00275 & 0.00276 & 0.00128 & 0.00128 & 0.00029 & 0.00031 & 0.00014 & 0.00015 \\
{[}$l_3$, $m$, $m_1$, $m_2${]} &  & 0.00693 & 0.00705 & 0.00335 & 0.00335 & 0.00165 & 0.00165 & 0.00033 & 0.00033 & 0.00015 & 0.00015 \\
{[}$l_1$, $m$, $m_1$, $m_2$, $k${]} &  & 0.00640 & 0.00640 & 0.00314 & 0.00314 & 0.00115 & 0.00117 & 0.00031 & 0.00033 & 0.00015 & 0.00016 \\
{[}$l_2$, $m$, $m_1$, $m_2${]} &  & 0.00655 & 0.00667 & 0.00342 & 0.00343 & 0.00163 & 0.00163 & 0.00031 & 0.00032 & 0.00015 & 0.00015 \\
{[}$l_2$, $l_3$, $m$, $m_1$, $m_2${]} &  & 0.00630 & 0.00642 & 0.00318 & 0.00319 & 0.00166 & 0.00166 & 0.00031 & 0.00032 & 0.00015 & 0.00016 \\
{[}$l_3$, $m$, $m_1$, $m_2$, $k_2${]} &  & 0.00550 & 0.00566 & 0.00309 & 0.00309 & 0.00162 & 0.00163 & 0.00033 & 0.00034 & 0.00015 & 0.00016 \\
{[}$l_2$, $m$, $m_1$, $m_2$, $k_2${]} &  & 0.00528 & 0.00543 & 0.00329 & 0.00329 & 0.00169 & 0.00169 & 0.00031 & 0.00031 & 0.00016 & 0.00016 \\
{[}$l_2$, $l_3$, $m$, $m_1$, $m_2$, $k_2${]} &  & 0.00526 & 0.00538 & 0.00309 & 0.00309 & 0.00163 & 0.00163 & 0.00031 & 0.00032 & 0.00016 & 0.00016 \\
{[}$l_1$, $l_3$, $m$, $m_1$, $m_2$, $k${]} &  & 0.00648 & 0.00649 & 0.00385 & 0.00385 & 0.00184 & 0.00190 & 0.00041 & 0.00042 & 0.00018 & 0.00018 \\
{[}$l_1$, $l_3$, $m$, $m_1$, $m_2$, $k_2$, $k${]} &  & 0.00585 & 0.00585 & 0.00332 & 0.00332 & 0.00203 & 0.00207 & 0.00040 & 0.00041 & 0.00018 & 0.00018 \\
{[}$l_1$, $l_2$, $m_1$, $m_2$, $k_2$, $k${]} &  & 0.00884 & 0.00885 & 0.00423 & 0.00425 & 0.00267 & 0.00275 & 0.00055 & 0.00056 & 0.00020 & 0.00020 \\
{[}$l_1$, $l_3$, $m_1$, $m_2$, $k${]} &  & 0.00767 & 0.00767 & 0.00367 & 0.00368 & 0.00183 & 0.00190 & 0.00038 & 0.00038 & 0.00020 & 0.00020 \\
{[}$l_1$, $l_3$, $m_1$, $m_2$, $k_2$, $k${]} &  & 0.00694 & 0.00694 & 0.00335 & 0.00335 & 0.00187 & 0.00194 & 0.00038 & 0.00039 & 0.00020 & 0.00020 \\
{[}$l_1$, $l_2$, $m_1$, $m_2$, $k${]} &  & 0.00955 & 0.00955 & 0.00634 & 0.00640 & 0.00258 & 0.00267 & 0.00052 & 0.00053 & 0.00020 & 0.00020 \\
{[}$l_1$, $l_2$, $l_3$, $m_1$, $m_2$, $k_2$, $k${]} &  & 0.00867 & 0.00870 & 0.00427 & 0.00428 & 0.00243 & 0.00255 & 0.00057 & 0.00057 & 0.00021 & 0.00021 \\
{[}$l_1$, $l_2$, $l_3$, $m_1$, $m_2$, $k${]} &  & 0.01021 & 0.01021 & 0.00593 & 0.00597 & 0.00251 & 0.00265 & 0.00054 & 0.00054 & 0.00023 & 0.00023 \\
{[}$l_1$, $l_3$, $m$, $m_1$, $m_2$, $k_2${]} &  & 0.00765 & 0.00766 & 0.00447 & 0.00448 & 0.00248 & 0.00249 & 0.00053 & 0.00054 & 0.00024 & 0.00025 \\
{[}$l_1$, $l_3$, $m_1$, $m_2$, $k_2${]} &  & 0.00749 & 0.00751 & 0.00425 & 0.00425 & 0.00273 & 0.00275 & 0.00046 & 0.00046 & 0.00024 & 0.00024 \\
{[}$l_1$, $l_3$, $m$, $m_1$, $m_2${]} &  & 0.00940 & 0.00940 & 0.00454 & 0.00455 & 0.00308 & 0.00310 & 0.00051 & 0.00052 & 0.00025 & 0.00026 \\
{[}$l_1$, $l_2$, $m$, $m_1$, $m_2$, $k_2$, $k${]} &  & 0.00686 & 0.00686 & 0.00443 & 0.00444 & 0.00262 & 0.00269 & 0.00055 & 0.00057 & 0.00025 & 0.00025 \\
{[}$l_1$, $l_2$, $l_3$, $m$, $m_1$, $m_2$, $k_2$, $k${]} &  & 0.00609 & 0.00609 & 0.00422 & 0.00422 & 0.00248 & 0.00255 & 0.00057 & 0.00058 & 0.00026 & 0.00026 \\
{[}$l_1$, $l_3$, $m_1$, $m_2${]} &  & 0.00943 & 0.00944 & 0.00407 & 0.00409 & 0.00302 & 0.00304 & 0.00046 & 0.00046 & 0.00026 & 0.00026 \\
{[}$l_1$, $l_2$, $l_3$, $m$, $m_1$, $m_2$, $k${]} &  & 0.00755 & 0.00757 & 0.00520 & 0.00520 & 0.00270 & 0.00276 & 0.00057 & 0.00058 & 0.00026 & 0.00027 \\
{[}$l_1$, $l_2$, $m$, $m_1$, $m_2$, $k${]} &  & 0.00799 & 0.00799 & 0.00548 & 0.00548 & 0.00322 & 0.00335 & 0.00053 & 0.00055 & 0.00027 & 0.00027 \\
{[}$l_1$, $l_2$, $m_1$, $m_2$, $k_2${]} &  & 0.01119 & 0.01125 & 0.00732 & 0.00739 & 0.00443 & 0.00446 & 0.00068 & 0.00069 & 0.00027 & 0.00027 \\
{[}$l_1$, $l_2$, $l_3$, $m_1$, $m_2$, $k_2${]} &  & 0.01017 & 0.01036 & 0.00659 & 0.00660 & 0.00397 & 0.00401 & 0.00064 & 0.00066 & 0.00028 & 0.00028 \\
\bottomrule
\end{tabular}
}
\end{table}

\end{document}